\documentclass[pra,aps,reprint,twocolumn,superscriptaddress,10pt,nofootinbib,longbibliography,floatfix]{revtex4-2}
\pdfoutput=1

\usepackage[utf8]{inputenc}
\usepackage[T1]{fontenc}
\usepackage[english]{babel}
\usepackage{newtxtext} 
\usepackage{graphicx,epsfig}
\usepackage{color}
\usepackage[usenames,dvipsnames]{xcolor}

\usepackage{longtable}

\usepackage{physics}
\usepackage{amsmath,amssymb,amsthm}
\usepackage{bbm,dsfont,bm} 
\usepackage{stmaryrd}
\usepackage{enumitem}
\usepackage[caption=false]{subfig}
\usepackage[normalem]{ulem}

\definecolor{linkcolor}{rgb}{0,0,0.6}	
\usepackage[colorlinks=true,pdfstartview=FitV,linkcolor=linkcolor,citecolor=linkcolor,urlcolor=linkcolor,hyperindex=true,hyperfigures=false]{hyperref}

\usepackage{numprint}

\usepackage[bb=boondox]{mathalfa}

\usepackage{upgreek}

\newcommand{\kket}[1]{{|#1 \rangle \!\rangle}}
\newcommand{\bbra}[1]{{\langle \!\langle #1 |}}
\renewcommand{\ket}[1]{{|#1 \rangle}}
\renewcommand{\bra}[1]{{\langle #1 |}}
\renewcommand{\ketbra}[2]{{|#1 \rangle \!\langle #2|}}
\newcommand{\kketbra}[2]{{|#1 \rangle \!\rangle \langle \!\langle #2|}}

\newcommand{\id}{\mathbbm{1}}
\newcommand{\HS}{\mathcal{H}}
\newcommand{\W}{\textup{\textsf{\textbf{W}}}}
\DeclareMathOperator{\bmTr}{Tr}

\begin{document}

\title{Reassessing the advantage of indefinite causal orders for quantum metrology}

\author{Raphaël Mothe}
\thanks{Corresponding author:\\
raphael.mothe@neel.cnrs.fr}
\affiliation{Univ.\ Grenoble Alpes, CNRS, Grenoble INP\footnote{Institute of Engineering Univ. Grenoble Alpes}, Institut N\'eel, 38000 Grenoble, France}
\affiliation{Univ.\ Grenoble Alpes, Inria, 38000 Grenoble, France}

\author{Cyril Branciard}
\affiliation{Univ.\ Grenoble Alpes, CNRS, Grenoble INP\footnote{Institute of Engineering Univ. Grenoble Alpes}, Institut N\'eel, 38000 Grenoble, France}

\author{Alastair A.\ Abbott}
\affiliation{Univ.\ Grenoble Alpes, Inria, 38000 Grenoble, France}

\date{April 3, 2024}


\begin{abstract}
The quantum switch, the canonical example of a process with indefinite causal order, has been claimed to provide various advantages over processes with definite causal orders for some particular tasks in the field of quantum metrology. In this work, we argue that some of these advantages in fact do not hold if a fairer comparison is made. To this end, we consider a framework that allows for a proper comparison between the performance, quantified by the quantum Fisher information, of different classes of indefinite causal order processes and that of causal strategies on a given metrological task. More generally, by considering the recently proposed classes of circuits with classical or quantum control of the causal order, we come up with different examples where processes with indefinite causal order offer (or not) an advantage over processes with definite causal order, qualifying the interest of indefinite causal order regarding quantum metrology. As it turns out, for a range of examples, the class of quantum circuits with quantum control of causal order, which are known to be physically realizable, is shown to provide a strict advantage over causally ordered quantum circuits as well as over the class of quantum circuits with causal superposition. Thus, the consideration of this class provides new evidence that indefinite causal order strategies can strictly outperform definite causal order strategies in quantum metrology.\end{abstract}


\maketitle


\section{Introduction}

In the usual formalism of quantum mechanics, the causal order between different operations, i.e., the order in which the operations are applied to a target system, is not subject to any kind of quantum indefiniteness. Yet, an extension of quantum mechanical laws encapsulated in the process matrix formalism was proposed in Ref.~\cite{oreshkov12}, which allows for processes with so-called indefinite causal order. For instance, by controlling the causal order between two operations with a quantum degree of freedom placed in a superposition, it is possible to generate superpositions of---and thus indefinite---causal orders.
This phenomenon is at the heart of the canonical example of a process with indefinite causal order, the quantum switch~\cite{chirib13}. In addition to the fundamental questions raised regarding quantum causality~\cite{hardy05,oreshkov12,zych19}, these indefinite causal order processes have attracted attention through the various advantages they can provide in different subfields of quantum information over processes with well-defined causal orders~\cite{chiribella12,ebler,feix15,araujo_witnessing_2015,guerin16,Chiribella_2021}.

Quantum metrology is one such field, with several studies examining the potential for the quantum switch to provide advantages over processes with well-defined causal orders~\cite{mukhopadhyay,frey,zhao,Frey21,Altherr,Chapeau-Blondeau,Chapeau-Blondeau21coherentcontrol,chapeau-blondeau22,liu,chiribella22,kurdzialek,goldberg23,Chapeau_Blondeau_23}. 
A basic scenario in quantum metrology is that of channel parameter estimation, where the goal is to maximize the precision of an estimate of an unknown parameter of a quantum channel, given $N$ uses (or copies) of the channel and with access to different kinds of quantum resources such as quantum superposition or entanglement. 
Thanks to the quantum Cramér-Rao bound~\cite{Braunstein94,Braunstein96}, one can quantify the possible precision of an estimate using the quantum Fisher information (QFI).

Several studies over the last few years have attempted to show that the quantum switch can provide advantages in this simplest metrological scenario of estimating the parameter of rather natural families of quantum channels, including depolarizing channels~\cite{frey} and thermalizing channels~\cite{mukhopadhyay}, amongst others~\cite{Frey21,Chapeau-Blondeau,Chapeau-Blondeau21coherentcontrol,chapeau-blondeau22,Chapeau_Blondeau_23}.
These works all show that using the quantum switch to sequentially apply these channels in a superposition of different orders provides an advantage over the simple sequential, causally definite, composition of the channels, and conclude, perhaps questionably, that indefinite causal order is a resource for quantum metrology.
Indeed, one may wonder whether the quantum switch indeed outperforms any standard adaptive, fixed-order strategy in these tasks, and similar questions have been raised over supposed advantages in thermodynamic tasks~\cite{capela23}.
This motivates us to analyze here this question in more detail and, more generally, to investigate the potential advantages of different types of causally indefinite strategies in quantum metrology~\cite{liu}.
We note that another result~\cite{zhao}, which was recently demonstrated experimentally~\cite{yin23}, showed an interesting and rigorous advantage using the quantum switch in a rather different kind of (multi-parameter) metrological task; here, however, we focus on the simple framework described above and studied in Refs.~\cite{mukhopadhyay,frey,Frey21,Chapeau-Blondeau,Chapeau-Blondeau21coherentcontrol,chapeau-blondeau22,Chapeau_Blondeau_23}. 

Rather than studying the performance of particular processes or strategies, such as the quantum switch or sequential composition, we look here to understand the capabilities of different families or classes of processes.
We thereby build on the framework on Refs.~\cite{Altherr,liu} to optimize the QFI, and thus bound the precision of estimation, over several such pertinent strategies.
Indeed, Ref.~\cite{liu} already considered two classes that include indefinite causal order: quantum circuits with causal superposition (QC-Sup), which includes the quantum switch as well as some other circuits with indefinite causal order; and the whole general class of process matrices, which includes all valid indefinite causal order processes (Gen).
By comparing these classes to standard sequential quantum circuits (quantum circuits with fixed causal order; QC-FO), and parallel circuits with entanglement (QC-Par), they established hierarchies of performance in estimating certain families of channels.
The particular interest devoted to the quantum switch in metrological settings is due to its position as one of the few causally indefinite processes with a clear physical interpretation and potential realization~\cite{rubino17,goswami18,Wei2019,Goswami2020,guo20,rubino20,rubino21}.
It is known, however, that physically meaningful processes beyond the quantum switch, and more generally beyond QC-Sup, exist~\cite{wechs21}; indeed, there are also causally indefinite processes that go beyond QC-FO by exploiting dynamical causal orders.

Here, we seek to study the physical limits of both causally definite and indefinite quantum metrology.
We thereby consider two further classes of processes introduced in Ref.~\cite{wechs21}: quantum circuits with classical control of causal order (QC-CC), and the quantum counterpart of QC-CCs, namely quantum circuits with quantum control of causal order (QC-QC).
We show how the framework of Ref.~\cite{liu} can be adapted to these classes to bound---in some cases tightly---the QFI, and thus metrological performance, in the metrological tasks considered.
Revisiting the results of Refs.~\cite{frey,mukhopadhyay}, we find the claims therein to be unfounded, with no advantage attainable with processes exhibiting indefinite causal orders.
Focusing on the $N=2,3$ cases, we detail the different possible hierarchies and show that the potential advantage of indefinite causal order strategies crucially depends on the family of quantum channels whose parameter is being estimated.
Several situations where indefinite causal order strategies outperform causally definite strategies could be identified, showing that indefinite causal order can indeed be seen as a resource in quantum metrology. 
A strict advantage of QC-QCs over QC-Sups could also be identified in various settings, showing that this class pushes further
the underlying physical limits of quantum metrology.

This paper is organized as follows.
In Sec.~\ref{sec:prelims} we give a brief overview of quantum metrology and causally indefinite matrices using the process matrix framework.
In Sec.~\ref{sec:reassessing} we study in more detail the purported advantages of Refs.~\cite{frey,mukhopadhyay} and show that the quantum switch can be outperformed by causally definite strategies.
In Sec.~\ref{sec:SDPs} we show how the metrological task can be solved using semidefinite programming over different classes of strategies, which we detail in Sec.~\ref{sec:classes}.
In Sec.~\ref{sec:results} we present a comparison of the performance of different classes of strategies when estimating different channel families' parameters, before presenting our conclusions in Sec.~\ref{sec:conclusion}.


\section{Preliminaries}
\label{sec:prelims}


\subsection{Quantum metrology}

\begin{figure}[t]
    \centering
    \includegraphics[width=0.6\columnwidth]{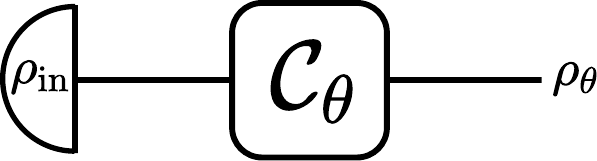}
    \caption{The basic typical scenario of channel parameter estimation. A probe state $\rho_\text{in}$ is sent through a quantum channel $\mathcal{C}_\theta$ specified by an unknown parameter $\theta$. The parameter can be estimated by measuring the output state of the channel, $\rho_\theta$.}
    \label{fig:quantum_metrology}
\end{figure}

In its simplest version, with one use of the channel, the metrological problem of channel parameter estimation can be described as follows (see Fig.~\ref{fig:quantum_metrology}).
Consider a quantum channel $\mathcal{C}_\theta$ parametrized (in a smooth enough manner) by an unknown continuous parameter $\theta$ that we wish to estimate. 
We can probe the channel by applying it to a quantum state $\rho_\text{in}$ to obtain the output state $\rho_\theta = \mathcal{C}_\theta(\rho_\text{in})$ that hence acquires a dependence on the parameter $\theta$. By measuring this output state, information about the parameter can be obtained, allowing one to estimate the parameter.

The uncertainty, quantified as the root-mean-square error, $\delta\hat{\theta}$ of an unbiased estimator $\hat{\theta}$ of $\theta$ can be bounded by the quantum Cramér-Rao bound~\cite{helstrom67,helstrom69} as 
\begin{equation}
    \delta\hat{\theta} \geq \frac{1}{\sqrt{J_\theta(\rho_\theta)}},
\end{equation}
where $J_\theta(\rho_\theta)$ is the quantum Fisher information (QFI) of the output state $\rho_\theta$ with respect to the parameter $\theta$. 
For single-parameter estimation, which is the scope of this work, the quantum Cramér-Rao bound is known to be achievable~\cite{helstrom67,helstrom69}, making the QFI a central quantity in order to quantify the ability to estimate an unknown parameter. 
The QFI $J_\theta(\rho_\theta)$ of the output state $\rho_\theta$ (which is assumed to be continuously differentiable) can be written as~\cite{paris09}
\begin{equation}
    J_\theta(\rho_\theta) := 2\!\! \sum_{\substack{i,j : \\ \lambda_{\theta,i} + \lambda_{\theta,j} > 0}}\!\! \frac{\abs{\bra{\lambda_{\theta,j}}\partial_\theta \rho_\theta \ket{\lambda_{\theta,i}}}^2}{\lambda_{\theta,i} + \lambda_{\theta,j}},
    \label{eq:QFI_simple}
\end{equation}
where the $\lambda_{\theta,i}$'s and $\ket{\lambda_{\theta,i}}$'s are the eigenvalues and corresponding (normalized) eigenvectors of $\rho_\theta$, so that $\rho_\theta = \sum_i \lambda_{\theta,i} \ketbra{\lambda_{\theta,i}}{\lambda_{\theta,i}}$. 
In this work we shall use another expression of the QFI introduced in Ref.~\cite{fujiwara08},
\begin{equation}
	J_\theta(\rho_\theta)=4 \min_{\{\ket{\psi_{\theta,j}}\}_{j=1}^q}\Tr\left(\sum_{j=1}^q\ketbra{\dot{\psi}_{\theta,j}}{\dot{\psi}_{\theta,j}}\right),
	\label{eq:QFI_Fujiwara}
\end{equation}
where $q$ can be any fixed value no smaller than the rank of $\rho_\theta$, and where the minimization is over all sets $\{\ket{\psi_{\theta,j}}\}_{j=1}^q$ of $q$ (potentially unnormalized, not necessarily orthogonal) vectors that form a so-called ensemble decomposition of size $q$ of $\rho_\theta$~\cite{fujiwara08}, such that $\rho_\theta = \sum_{j=1}^q\ketbra{\psi_{\theta,j}}{\psi_{\theta,j}}$, and where all $\ket{\psi_{\theta,j}}$'s are themselves continuously differentiable.
The dot denotes the derivation with respect to $\theta$, $\ket{\dot{\psi}_{\theta,j}} = \partial_\theta\ket{\psi_{\theta,j}}$.

\begin{figure}[t]
     \centering
     \includegraphics[width=0.9\columnwidth]{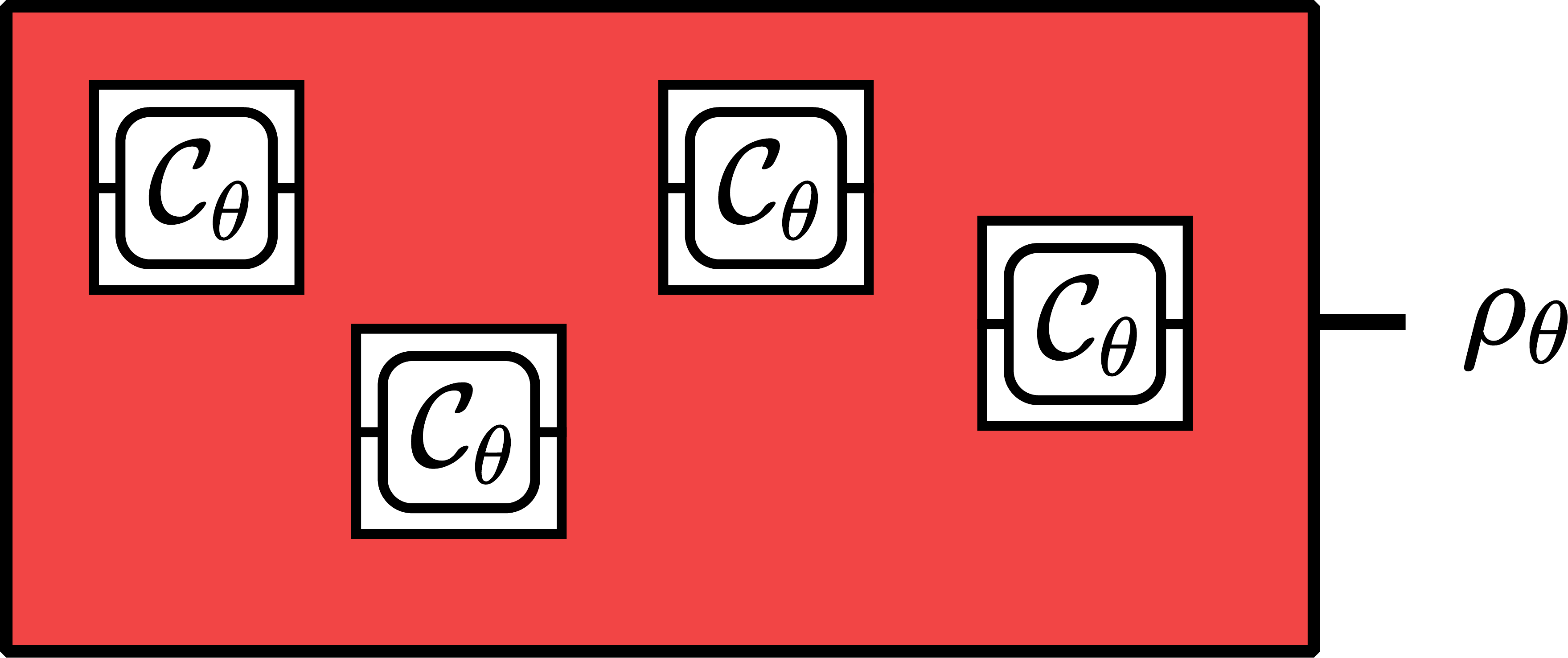}
     \caption{Abstract representation of a strategy (depicted in red) that connects $N=4$ instances of the quantum channel $\mathcal{C}_\theta$ in order to produce the output state $\rho_\theta$. In this paper we study the precision of estimating $\theta$ from $\rho_\theta$ given access to different classes of such strategies.}
    \label{fig:strategy}
\end{figure}

In the simple scenario of Fig.~\ref{fig:quantum_metrology}, the channel $\mathcal{C}_\theta$ is probed only once and it is only possible to play with the initial probe state $\rho_{\text{in}}$ to maximize the QFI of the output state, and thus the precision of estimation. 
However, as soon as we consider scenarios in which $\mathcal{C}_\theta$ can be used, or probed, several times, more possibilities arise. 
If one independently probes $N$ copies of the channel $\mathcal{C}_\theta$ one obtains the output state $\rho_\theta^{\otimes N}$, for which $J_\theta(\rho_\theta^{\otimes N}) = N J_\theta(\rho_\theta)$, so that the Cramér-Rao bound gives $\delta\hat{\theta} \geq \frac{1}{\sqrt{N J_\theta(\rho_\theta)}}$.
In addition to playing with the initial probe state, however, one can also change the structure of the probing strategy in order to maximize the QFI of the output state. 
For instance, it is well-known that, by entangling several probes, one can increase the precision of an estimate beyond the classical square-root scaling in $N$~\cite{Huang16}, even achieving Heisenberg scaling---a quadratic advantage in scaling. Compared with the strategy shown in Fig.~\ref{fig:quantum_metrology}, it can likewise be advantageous to consider probes that are entangled with additional ancillary systems that the channel is not applied to~\cite{Fujiwara01,Fujiwara04}.
More generally, given $N$ uses of, or ``queries'' to, the quantum channel $\mathcal{C}_\theta$, one may wonder what the best circuit architecture, or strategy, is that maximizes the precision of estimation of $\theta$; this scenario is depicted abstractly in Fig.~\ref{fig:strategy}.

This question and the underlying metrological task, which we will formalize further in Sec.~\ref{sec:SDPs}, is the central object of study in this paper.
Thus far, most approaches to quantum metrology have considered the most general strategy to be described by a standard, causally ordered quantum circuit. However, one can also envisage---as we will do here---more general compositional strategies using indefinite causal order, and study how useful this new resource is for improving the precision of parameter estimation.


\subsection{The process matrix formalism}

In order to describe more general metrological strategies, we will make use of the process matrix formalism that was introduced in Ref.~\cite{oreshkov12} to describe processes with potentially indefinite causal order. 
This formalism generalizes quantum theory by relaxing the assumption of a well-defined global causal order between different operations.
Standard, causally definite quantum theory is instead only assumed to hold locally: operations are assigned to local closed and spatially separated laboratories described by quantum laws, but no assumption of a predefined global causal structure is made.

In the general scenario described by the process matrix formalism, $N$ quantum operations $\mathcal{A}_k$ (with $1\le k\le N$) are considered.
Each operation is performed once and only once on the physical system received in its input Hilbert space $\mathcal{H}^{A_k^I}$, and the resulting system is sent out in the output Hilbert space $\mathcal{H}^{A_k^O}$.%
\footnote{All Hilbert spaces are assumed here to be finite-dimensional. For simplicity we shall typically consider that the operations input and output Hilbert spaces are of the same dimension, $d$.}
The operations are described by completely positive (CP) maps $\mathcal{A}_k:\mathcal{L}(\mathcal{H}^{A_k^I})\to \mathcal{L}(\mathcal{H}^{A_k^O})$, where $\mathcal{L}(\mathcal{H}^{X})$ denotes the space of linear operators on $\mathcal{H}^X$. 
The overall process describes how the $N$ operations are related, be it in a definite or indefinite causal order, and outputs a system in the global causal future of all the operations, described by some Hilbert space $\mathcal{H}^F$.%
\footnote{Although the original formulation of the process matrix framework did not involve the output system in the space $\mathcal{H}^F$, it is rather easy to extend the formalism so as to include it~\cite{Araujo2017purification}. It is also possible to introduce some space $\mathcal{H}^P$ in which some input system could be prepared, in the global causal past of all other operations. As in Refs.~\cite{Altherr,liu}, however, we do not define such a global past, since any fixed initial state (as one has in the metrological scenario we consider) can be incorporated directly into the process itself.\\
Note also that in general the operations $\mathcal{A}_k$ can be extended to also act on auxiliary spaces $\mathcal{H}^{A_k^{I'/O'}}$, i.e., the supermap can be applied to CP maps $\mathcal{A}_k:\mathcal{L}(\mathcal{H}^{A_k^IA_k^{I'}})\to \mathcal{L}(\mathcal{H}^{A_k^OA_k^{O'}})$.}
This process can equivalently be seen as a higher-order map, or supermap, that maps the $N$ operations $\mathcal{A}_k$ to the state output in the global future. 
By considering its Choi–Jamiołkowski representation~\cite{Choi}, the process can be uniquely described by a matrix $W \in \mathcal{L}(\mathcal{H}^{A_{\mathcal{N}}^{IO}F})$, its \emph{process matrix}, where we use the shorthand notation $\mathcal{H}^{XY}:=\mathcal{H}^{X} \otimes \mathcal{H}^{Y}$, $\mathcal{H}^{A_k^{IO}} := \mathcal{H}^{A_k^IA_k^O}$, and $\mathcal{H}^{A_\mathcal{K}^{IO}}:=\bigotimes_{k\in\mathcal{K}}\mathcal{H}^{A_k^{IO}}$, for any $\mathcal{K}\subseteq\mathcal{N}:=\{ 1,\ldots,N\}$. 
The fact that the overall process must yield a proper, well-defined supermap, but without making any assumption of a predefined global causal order between the operations, imposes some constraints on the process matrix $W$ that must be satisfied for it to be valid; we will detail these in Sec.~\ref{sec:classes}.
Within the set of all valid process matrices one can identify specific relevant classes of process matrices (for instance, those compatible with a definite causal order) obtained by imposing further constraints on $W$, which will also be detailed in Sec.~\ref{sec:classes}.
We shall denote by \textup{\textsf{\textbf{W}}} a generic class of processes.

When considering the scenario of Fig.~\ref{fig:strategy}, we are typically interested in expressing the output state $\rho_\theta$ as a function of the channels $\mathcal{C}_\theta$ and the process matrix $W$ describing a given strategy.
To this end, let us first introduce the Choi matrix~\cite{Choi} $C_\theta$ of the channel $\mathcal{C}_\theta:\mathcal{L}(\mathcal{H}^{A_k^I})\to\mathcal{L}(\mathcal{H}^{A_k^O})$, that connects an input space $\mathcal{H}^{A_{k}^I}$ to an output space $\mathcal{H}^{A_{k}^O}$, as
\begin{equation}
    C_\theta := \sum_{i,j} \ketbra{i}{j}^{A_{k}^I}\otimes \mathcal{C}_\theta(\ketbra{i}{j}^{A_{k}^I}) \in \mathcal{L}(\mathcal{H}^{A_{k}^{IO}}),\label{eq:Choi_def}
\end{equation}
where $\{\ket{i}^X\}_i$ denotes a fixed orthonormal basis, which we refer to as the computational basis of $\mathcal{H}^{X}$, and where we use the short-hand notation $\ketbra{i}{j}^X := \ket{i}^{X}\bra{j}^X$. 
Throughout this paper we will refer to the channel $\mathcal{C}_\theta$ and its Choi matrix $C_\theta$ indifferently, as Eq.~\eqref{eq:Choi_def} describes an isomorphism that can be inverted to express $\mathcal{C}_\theta$ in terms of $C_\theta$.

When using $N$ copies of the channel, as in Fig.~\ref{fig:strategy}, one can then rewrite the output state $\rho_\theta$ using the so-called link product as
\begin{align}
    \rho_\theta = C_\theta^{\otimes N} * W.
\end{align}
The link product between two operators $A\in \mathcal{L}(\mathcal{H}^{XY})$ and $B\in \mathcal{L}(\mathcal{H}^{YZ})$ that share a common system $Y$ associated to $\mathcal{H}^Y$ is defined as~\cite{Chiribella08,Chiribella2009}
\begin{equation}
   A * B := \Tr_Y[(A^{T_Y}\otimes \id^Z)(\id^X \otimes B)] \in \mathcal{L}(\mathcal{H}^{XZ}), \label{eq:def_link_prod}
\end{equation}
where $\Tr_Y$ and ${}^{T_Y}$ denote, respectively, the partial trace and partial transpose (in the computational basis) with respect to the Hilbert space $\mathcal{H}^Y$, and $\id$ is the identity operator in the appropriate space (indicated here by the superscripts).


\subsection{Example: the quantum switch}


\subsubsection{Description of the dynamics}

The quantum switch~\cite{chirib13} is the canonical example of a process with indefinite causal order, and can be described as follows. Two operations $\mathcal{A}_1$ and $\mathcal{A}_2$ are applied locally to a target system initially in the state $\ket{\psi_{\text{in}}}^t\in\HS^t$ in an order that depends on a control qubit in the space $\HS^c$, after which both the target system and the control qubit are sent to a region corresponding to the global future $F$ with Hilbert space $\HS^F:=\HS^{F^c}\otimes \HS^{F^t}$. 
If the control qubit is in the state $\ket{0}^c$, then $\mathcal{A}_1$ is applied to the target system before $\mathcal{A}_2$, while if the control qubit is in the state $\ket{1}^c$, then $\mathcal{A}_2$ is applied before $\mathcal{A}_1$.
By initializing the control qubit in the superposed state $\ket{+}^c=\frac{1}{\sqrt{2}}(\ket{0}^c+\ket{1}^c)$, and in the case where $\mathcal{A}_i$ (for $i=1,2$) is a unitary operation $\rho \mapsto U_{A_i}\rho U_{A_i}^\dagger$, with $U_{A_i}$ a unitary, the dynamics of the quantum switch is given by
\begin{align}
    & \!\ket{+}^c\otimes \ket{\psi_{\text{in}}}^t \notag \\
    & \!\to \frac{1}{\sqrt{2}}\big(\ket{0}^c \otimes U_{A_2}U_{A_1} \ket{\psi_{\text{in}}}^t+\ket{1}^c \otimes U_{A_1}U_{A_2} \ket{\psi_{\text{in}}}^t\big).
    \label{eq:QS}
\end{align}
The switch can thus be interpreted as a coherent control of the two possible causal orders between $\mathcal{A}_1$ and $\mathcal{A}_2$.
Although its expression is slightly more involved, the dynamics can also be written for nonunitary operations $\mathcal{A}_i$, with a similar interpretation~\cite{chirib13}.


\subsubsection{Process matrix description}

The process matrix for the quantum switch process can be we written as~\cite{araujo_witnessing_2015} 
\begin{align}
    & W^{\text{QS}}=\ketbra{w}{w} \nonumber\\
    & \text{with} \quad \ket{w}=\frac{1}{\sqrt{2}}\big(\ket{\psi_{\text{in}}}^{A_1^I}|\mathbbm{1} \rangle \! \rangle^{A_1^OA_2^I}|\mathbbm{1} \rangle \! \rangle^{A_2^OF^t}\ket{0}^{F^c} \nonumber\\[-2mm]
    & \hspace{25mm} + \ket{\psi_{\text{in}}}^{A_2^I}|\mathbbm{1} \rangle \! \rangle^{A_2^OA_1^I}|\mathbbm{1} \rangle \! \rangle^{A_1^OF^t}\ket{1}^{F^c}\big),
    \label{eq:PM_QS}
\end{align}
with the tensor products left implicit for readability, and where $\kket{\id}^{XY}:=\sum_i\ket{i}^X\otimes\ket{i}^Y$ is defined for two isomorphic spaces $\mathcal{H}^X$ and $\mathcal{H}^Y$, whose computational bases $\{\ket{i}^X\}_i$ and $\{\ket{i}^Y\}_i$ are in one-to-one correspondence.  
The terms of the form $|\mathbbm{1} \rangle \! \rangle^{XY}$ can be understood as representing an identity channel between $\mathcal{H}^X$ and $\mathcal{H}^Y$.
Eq.~\eqref{eq:PM_QS} involves the coherent superposition of two terms that correspond to the two possible orders in which $\mathcal{A}_1$ comes before $\mathcal{A}_2$ and in which $\mathcal{A}_2$ comes before $\mathcal{A}_1$, in agreement with Eq.~\eqref{eq:QS}. 

The quantum switch has been shown to provide an advantage over processes with definite causal order for different tasks in several subfields of quantum information, including quantum computation~\cite{chiribella12,araujo14} and quantum communication~\cite{guerin16,ebler}.
Recently, several attempts to demonstrate advantages in quantum metrology using the quantum switch have been presented.
Some of these fit within the standard framework of quantum metrology we described above~\cite{mukhopadhyay,frey,Frey21,Chapeau-Blondeau,Chapeau-Blondeau21coherentcontrol,chapeau-blondeau22,Chapeau_Blondeau_23}, while others have considered more particular scenarios involving, for example, products of position and momentum displacement operators~\cite{zhao}.
These works, particularly the former ones in the more standard quantum metrology scenario, motivate our work, and we will review some of them further in what follows.


\section{Re-examining the literature}
\label{sec:reassessing}

Here we examine more carefully some of the supposed advantages in quantum metrology obtained from indefinite causal order.
Refs.~\cite{frey,mukhopadhyay} exemplify the attempts to use the quantum switch for metrology in the standard metrological setting, and by studying these examples we will argue that it is misleading to claim that indefinite causal order would be a resource for quantum metrology with respect to the evidence they provide.


\subsection{Review of some previous work}

In Ref.~\cite{frey} the author considers the $d$-dimensional depolarizing channel $\mathcal{D}_\theta$ defined as
\begin{equation}
    \mathcal{D}_\theta(\rho) = \theta \rho + (1-\theta)\frac{\id}{d},
    \label{eq:depo_channel}
\end{equation}
for any (normalized) $d$-dimensional density matrix $\rho$, and where $\theta\in [0,1]$ parametrizes the channel.
They study the following metrological comparison: given two instances of $\mathcal{D}_\theta$, compare the QFI of the output state produced by the quantum switch,%
\footnote{Both Refs.~\cite{frey,mukhopadhyay} actually consider a quantum switch with a control qubit initialized in a more general state of the form $\sqrt{p}\ket{0}^c+\sqrt{1-p}\ket{1}^c$. The claimed advantages are found to be the largest for $p=1/2$; for simplicity we restrict here the discussion to that case.}
$J_\theta(D_\theta^{\otimes 2}*W^{\text{QS}})$, and that produced by the sequential quantum circuit that composes the instances of $\mathcal{D}_\theta$ one after the other (described by the process matrix $W^{\text{Seq}}$; see Appendix~\ref{app:Frey}), $J_\theta(D_\theta^{\otimes 2}*W^{\text{Seq}})$. 
On the grounds that $J_\theta(D_\theta^{\otimes 2}*W^{\text{QS}})> J_\theta(D_\theta^{\otimes 2}*W^{\text{Seq}})$ for all $\theta<1$, the author concludes that the quantum switch is an aid for channel probing, in the sense that causal indefiniteness leads to higher QFI, and hence better performance. 
For the comparison of these two specific processes, the quantum switch $W^\text{QS}$ and the sequential circuit $W^\text{seq}$, this statement is strictly true, but insofar as it concerns causal indefiniteness more generally it appears rather limited or even misleading.
One may wonder, for instance, whether the quantum switch outperforms all strategies with a definite causal order---as one would like if one is to claim indefinite causal order is a resource for channel probing---or simply the specific sequential strategy considered.

The same kind of comparison was made in Ref.~\cite{mukhopadhyay}, where this time the quantum switch and the sequential strategy are compared in a metrological task related to quantum thermometry. 
The same scenario as in Ref.~\cite{frey} is considered, with the depolarizing channel $\mathcal{D}_\theta$ replaced by a qubit thermalizing channel $\mathcal{T}_{\theta}$, where $\theta>0$ is here the temperature of the bath the qubit is probing. The channel $\mathcal{T}_{\theta}$ is defined by the following Kraus operators: 
\begin{align}
&K_{\theta,0}=\sqrt{p_{\theta}}
    \begin{pmatrix}
1 & 0 \\
0 & \sqrt{1-\lambda}
\end{pmatrix}
\!,\, K_{\theta,1}=\sqrt{p_{\theta}}
    \begin{pmatrix}
0 & \sqrt{\lambda} \\
0 & 0
\end{pmatrix}
\!, \nonumber \\
&K_{\theta,2}=\sqrt{1{-}p_{\theta}}
    \begin{pmatrix}
\sqrt{1-\lambda} & 0 \\
0 & 1
\end{pmatrix}
\!,\, K_{\theta,3}=\sqrt{1{-}p_{\theta}}
    \begin{pmatrix}
0 & 0 \\
\sqrt{\lambda} & 0
\end{pmatrix}\!,
\label{eq:therma_channel}
\end{align}
so that $\mathcal{T}_{\theta}(\rho) = \sum_{i=0}^3 K_{\theta,i}\, \rho\, K_{\theta,i}^\dagger$, where $p_{\theta}=1/(1+e^{-\epsilon/\theta})$, $\lambda = 1 - e^{-t/\tau}$, with $\epsilon >0$ the energy gap of the probe, $t >0$ the interaction time between the probe and the bath, and $\tau >0$ the relaxation time of the bath. 
In the limit $t\to \infty$, for which $\lambda \to 1$, any input state is thermalized so that the thermalizing channel becomes constant: $\mathcal{T}_\theta(\rho)=p_{\theta}\ketbra{0}{0} + (1-p_{\theta}) \ketbra{1}{1}$, $\forall\, \rho$ (with $\Tr \rho = 1$). 
In this limit, by showing that $J_\theta(T_\theta^{\otimes 2}*W^{\text{QS}})> J_\theta(T_\theta^{\otimes 2}*W^{\text{Seq}})$ for all $\theta$ (and for the input state of the quantum switch $\ket{\psi_{\text{in}}} = \ket{0}$, see Appendix~\ref{app:Mukhopadhyay}), the authors conclude again that indefinite causal order is a resource in quantum thermometry. 
As before, the scope of this comparison is rather limited by considering only a single causally-definite strategy for comparison with the quantum switch.


\subsection{Comment and critique}
\label{sec:critique}

\begin{figure}
     \centering
     \subfloat[\label{fig:2TQOD}]{%
        \includegraphics[width=0.47\columnwidth]{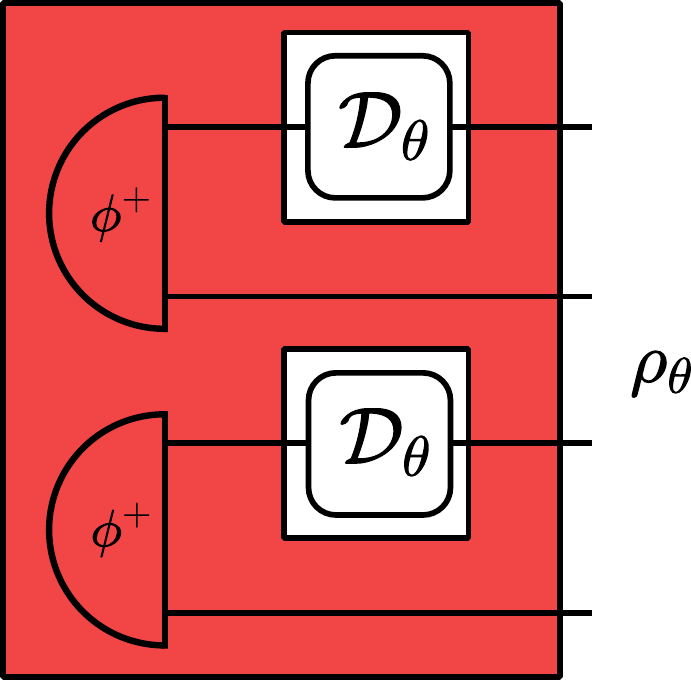}
     }
     \hfill
     \subfloat[\label{fig:Parallel}]{%
        \includegraphics[width=0.47\columnwidth]{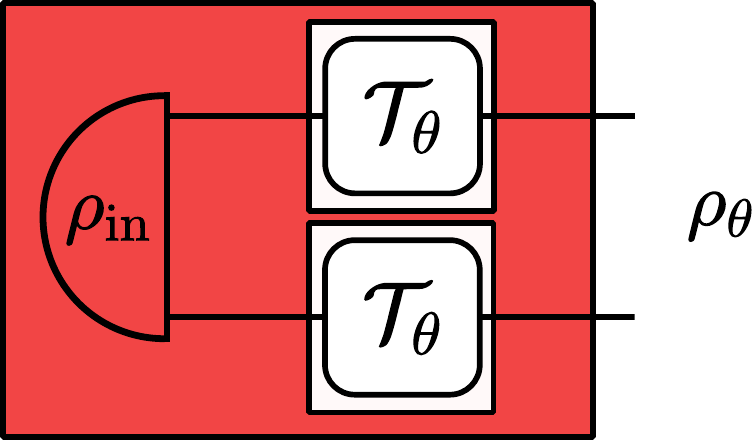}
     }
        \caption{(a): Optimal strategy for the parameter estimation of $N=2$ depolarizing channels as considered in~\cite{frey}. This so-called ``2-Choi circuit'' is made of a parallel use of the two depolarizing channels represented by the rounded square boxes, with maximally entangled input states $\ket{\phi^+}:=\frac{1}{\sqrt{d}}\kket{\id}=\frac{1}{\sqrt{d}}\sum_{i}\ket{i}\otimes\ket{i}$. (b): Optimal strategy for $N=2$ thermalizing channels as defined in~\cite{mukhopadhyay} (in the limit $t\to\infty$). This circuit simply consists of a parallel use of the two thermalizing channels represented by the rounded square boxes.}
        \label{fig:optimal_circuits}
\end{figure}

As shown in Appendix~\ref{app:Frey_Mukhopadhyay}, for each of the two tasks introduced in~\cite{frey,mukhopadhyay}, we could come up with an example of a process with definite causal order that outperforms the quantum switch, yielding a larger QFI. The two corresponding circuits are depicted in Fig.~\ref{fig:optimal_circuits}. Both are parallel circuits, which ensures they are compatible with a definite causal order. The circuit presented in Fig.~\ref{fig:2TQOD} consists of two independent uses of a sub-circuit, where the probe system sent through the depolarizing channel is initially prepared along with an auxiliary system, in a maximally entangled state $\ket{\phi^+}:=\frac{1}{\sqrt{d}}\kket{\id}=\frac{1}{\sqrt{d}}\sum_{i}\ket{i}\otimes\ket{i}$. 
We term this sub-circuit the ``Choi circuit'', since its output state is proportional to the Choi matrix of the channel $\mathcal{D}_\theta$.
The overall circuit is thus termed the 2-Choi circuit, because it consists of two independent uses of the Choi circuit. The circuit presented in Fig.~\ref{fig:Parallel} simply applies the two copies of the thermalizing channel in parallel, with no constraint on the input state $\rho_{\text{in}}$ of the probe system.

These two examples show that a fairer comparison should be made in order to claim an advantage of the quantum switch, or more generally of causally indefinite order processes, in the corresponding tasks. 
To claim an advantage of the quantum switch in such a scenario, one should instead make a comparison between its performance and the performance of the whole class of processes with a definite causal order. 
As proposed in~\cite{zhao}, one way to do this is to derive a bound on the QFI of any process with a definite causal order, and then show that the bound can be outperformed by the quantum switch. 
In general, however, analytically finding such a bound is a difficult task. 
Another approach, based on semidefinite programming (SDP) techniques (which are commonly used in the study of causally indefinite processes), is to optimize the QFI of the output state over classes of strategies with definite or indefinite causal orders. 
This was first proposed in the framework introduced in~\cite{Altherr} for strategies with a definite causal order, and then extended to some different classes of strategies with indefinite causal order in~\cite{liu}.

In what follows, we give a formal definition of the metrological task we consider and then show how strategies can be compared in this task using SDPs.


\section{Comparing the metrological performance of whole classes of processes}
\label{sec:SDPs}


As motivated above, a better understanding of the possible advantage of indefinite causal order processes in quantum metrology requires a broader approach that allows for comparison of classes of processes with definite causal order with classes of processes with indefinite causal order. This is the purpose of the framework introduced in~\cite{Altherr,liu}. We first review this work and then extend it, considering recently introduced classes of circuits, in order to better qualify the advantage of indefinite causal order processes in quantum metrology.


\subsection{The metrological task: optimizing the quantum Fisher information over a full class of strategies}
\label{sec:task}

As already introduced, given $N$ queries to a quantum channel $\mathcal{C}_\theta$ that depends on an unknown parameter $\theta$, we want to find the strategy that maximizes the QFI of its output state $\rho_\theta$, over a given class of strategies~$\textup{\textsf{\textbf{W}}}$. More formally, a strategy can be described as a process matrix $W\in \mathcal{L}(\mathcal{H}^{A^{IO}_\mathcal{N}F})$, that can be fed $N$ copies of $C_\theta$ (in the Choi representation)---i.e., with the ($d^{2N}\times d^{2N}$) Choi matrix $C_\theta^{\otimes N} \in \mathcal{L}(\mathcal{H}^{A^{IO}_\mathcal{N}})$---in order to produce an output state $\rho_\theta = C_\theta^{\otimes N} * W \in \mathcal{L}(\mathcal{H}^F)$. Mathematically, this boils down to determining
\begin{equation}
    J_\theta^{\textup{\textsf{\textbf{W}}}}(C_\theta^{\otimes N}) := \max_{W\in \textup{\textsf{\textbf{W}}}} J_\theta(C_\theta^{\otimes N} * W),
    \label{eq:max_QFI_class}
\end{equation}
where $J_\theta^{\textup{\textsf{\textbf{W}}}}(C_\theta^{\otimes N})$ is the optimal QFI over the class of strategies $\textup{\textsf{\textbf{W}}}$ using $N$ queries to $\mathcal{C}_\theta$, with respect to the parameter $\theta$. Following~\cite{Altherr,liu}, it is possible to rewrite $J_\theta^{\textup{\textsf{\textbf{W}}}}(C_\theta^{\otimes N})$ as an SDP for several relevant classes $\W$, allowing efficient numerical methods to be used to calculate  $J_\theta^{\W}(C_\theta^{\otimes N})$.


\subsection{Solving the task through SDP optimization}
\label{sec:SDP}

The framework introduced in~\cite{Altherr,liu} proposes a method to rewrite Eq.~\eqref{eq:max_QFI_class} as an SDP, which can then be solved numerically to obtain an optimal strategy for the metrological task defined above, over a given class of strategies with or without a definite causal order. In order to generalize this framework to new classes of strategies $\textup{\textsf{\textbf{W}}}$, we first give a sketch of the arguments which will next serve as a base for the extension to the new classes.
For further details, some clarifications, and proofs of the arguments, see Appendix~\ref{app:details_step_by_step}.

First, we use the convexity of the QFI to restrict the optimization over the strategies $W\in \textup{\textsf{\textbf{W}}}$ in Eq.~\eqref{eq:max_QFI_class} to pure (i.e., rank-one) process matrices, such that $W = \ketbra{w}{w}\in \textup{\textsf{\textbf{W}}}$; see Appendix~\ref{app_subsubsec:restrict_pure}. For all the classes considered in~\cite{Altherr,liu}, as well as for the QC-QCs that we shall introduce below, this step can be done without loss of generality since any process matrix in these classes can be purified within the same class, up to enlarging the future space $\mathcal{H}^F$ (which our definitions of the different classes allow for; see below). 
But, in general, this is not a trivial step, as it might be that the purifications of some process matrices in a given class of strategies belong to a larger class of strategies. 
This is in particular the case for the class of QC-CCs presented below.%
\footnote{For instance, the ``classical switch'' (the classical counterpart of the quantum switch) is a QC-CC whose purification is a QC-QC but no longer a QC-CC~\cite{wechs21}.}  
The approach used here then only allows one to obtain the optimal QFI within this larger purifying class---which thus only gives an upper bound for the QFI with respect to the class initially considered.

Restricting to pure process matrices allows one, in the expression of the QFI of the output state $\rho_\theta = C_\theta^{\otimes N} * W = C_\theta^{\otimes N} * \ketbra{w}{w}$ from Eq.~\eqref{eq:QFI_Fujiwara}, to turn the optimization over ensemble decompositions of size $q$ of $\rho_\theta$, $\{\ket{\psi_{\theta,j}}\}_{j=1}^q$, into an optimization over ensemble decompositions of the same size $q$, $\{\ket{C_{\theta,j}}\}_{j=1}^q$, of the channels' operator $C_\theta^{\otimes N}$ directly:
\begin{align}
    J_\theta(\rho_\theta)&=4 \min_{\{\ket{\psi_{\theta,j}}\}_{j=1}^q}\Tr\left(\sum_{j=1}^q\ketbra{\dot{\psi}_{\theta,j}}{\dot{\psi}_{\theta,j}}\right) \notag \\
    &= 4 \min_{\{\ket{C_{\theta,j}}\}_{j=1}^q}\Tr\left( \Big( \sum_{j=1}^q\ketbra{\dot{C}_{\theta,j}}{\dot{C}_{\theta,j}} \Big) * W \right).\label{eq:objective_SDP_h}
\end{align}
See Appendix~\ref{app_subsubsec:optim_decomp_C} for more details. Here we assume that the dependency of the channels on $\theta$ is smooth enough, so that similarly to $\min_{\{\ket{\psi_{\theta,j}}\}}$, the minimum $\min_{\{\ket{C_{\theta,j}}\}}$ is taken over continuously differentiable decompositions.

Then, any such ensemble decomposition of size $q$ of $C_\theta^{\otimes N}$ can be expressed in terms of a fixed ensemble decomposition of size $q$ of $C_\theta^{\otimes N}$, $\{\ket{C_{\theta,j}^0}\}_{j=1}^q$, and a (sufficiently smooth) $q\times q$ unitary $V_\theta$ that relates the two decompositions. This allows us to replace the optimization over $\{\ket{C_{\theta,j}}\}_{j=1}^q$ by an optimization over $V_\theta$; see Appendix~\ref{app_subsubsec:optim_V}. By realizing that the objective function of the optimization only depends on the $q\times q$ Hermitian matrix $h = i \dot{V}_\theta V_\theta^\dagger$,  the optimization over $V_\theta$ is, in turn, replaced by an optimization over such Hermitian matrices $h$, cf.\ Appendix~\ref{app_subsubsec:optim_h}.
One thereby obtains the expression
\begin{align}
    J_\theta(\rho_\theta)&=\min_{h \in \mathbb{H}_q} \Tr\left(\Omega_\theta(h) \, \tilde{W} \right), \label{eq:QFI_h}
\end{align}
where $\mathbb{H}_q$ denotes the set of $q \cross q$ Hermitian matrices, and $\tilde{W} = \Tr_F W$. Here we introduced the so-called ``performance operator''~\cite{Chiribella_2016,Altherr}
\begin{equation}
\Omega_\theta(h) = 4 \left( (\dot{\mathbf C}_\theta^0 - i \mathbf C_\theta^0 h ) (\dot{\mathbf C}_\theta^0 - i \mathbf C_\theta^0 h )^\dagger \right)^T, \label{eq:def_Omega}
\end{equation}
where ${}^T$ denotes transposition in the computational basis, and with $\mathbf{C}_{\theta}^0 = \left(\ket{C_{\theta,1}^0},\ldots,\ket{C_{\theta,q}^0}\right)$ being the $d^{2N}\times q$ matrices whose column vectors are the ($d^{2N}$-dimensional) elements of the fixed decomposition $\left\{\ket{C_{\theta,j}^0}\right\}_{j=1}^q$  of $C_\theta^{\otimes N}$ (and with $\dot{\mathbf C}_{\theta}^0$ its derivative with respect to $\theta$).
As one can observe, the QFI of $\rho_\theta$ is obtained by optimizing over $\tilde{W}$, the process matrix of the strategy with the global future traced out.
Thus, the action of $W$ on the global future $F$, and in particular the dimension of the Hilbert space $\mathcal{H}^F$, do not have to be specified at this level. 
One only needs to fix the dimension of $\mathcal{H}^F$ when one wants to purify $\tilde{W}$ to obtain the underlying strategy in full.

Then, injecting Eq.~\eqref{eq:QFI_h} into Eq.~\eqref{eq:max_QFI_class}, one obtains the expression of the optimal QFI over a given class of strategies $\textup{\textsf{\textbf{W}}}$:
\begin{equation}
     J_\theta^{\textup{\textsf{\textbf{W}}}}(C_\theta^{\otimes N}) = \max_{\tilde{W}\in \Tr_F \!\textup{\textsf{\textbf{W}}}} \, \min_{h\in \mathbb{H}_q} \,\Tr\left(\Omega_\theta(h)\,\tilde{W}\right),
     \label{eq:max_QFI_SDP}
\end{equation}
with $\Tr_F \textup{\textsf{\textbf{W}}} := \{ \Tr_F W, \, W \in \textup{\textsf{\textbf{W}}}\}$. Finally, because the objective function is convex in $h$ and linear (thus concave) in $\tilde{W}$, and assuming that $\textup{\textsf{\textbf{W}}}$ is compact (as will be the case for all classes under consideration here), one can exchange the minimization with the maximization using Fan's minimax theorem~\cite{fan53}, and obtain (see Appendix~\ref{app_subsubsec:swap_max_min})
\begin{equation}
     J_\theta^{\textup{\textsf{\textbf{W}}}}(C_\theta^{\otimes N}) = \min_{h\in \mathbb{H}_q}\,
     \max_{\tilde{W}\in \Tr_F \!\textup{\textsf{\textbf{W}}}}\,
     \Tr(\Omega_\theta(h)\,\tilde{W}).
     \label{eq:minmax_QFI_SDP}
\end{equation}
Since this equation is a min max problem, it is not straightforward in general to solve it numerically. 
As we shall see, the classes $\W$ that we will be interested in can all be characterized with semidefinite constraints, making the inner maximization problem an SDP. 
By first fixing $h$ and using duality to turn the maximization into a minimization, and after conveniently linearizing the (\emph{a priori} quadratic in $h$) constraints that involve $\Omega_\theta(h),$ one can then rewrite Eq.~\eqref{eq:minmax_QFI_SDP} itself as an SDP (Appendix~\ref{app_subsubsec:trade_for_dual}).
Solving this SDP gives both $J_\theta^{\textup{\textsf{\textbf{W}}}}(C_{\theta}^{\otimes N})$, the QFI optimized over the class $\textup{\textsf{\textbf{W}}}$, and $h^{(\text{opt})}$, the value of the optimal $h$ (Appendix~\ref{app_subsubsec:solve_SDP}). Fixing $h=h^{(\text{opt})}$ and imposing an extra condition on $\tilde{W}$ that we discuss in Appendix~\ref{app_subsubsec:reconstruct_Wopt} (see also Appendix~\ref{app:subsec:saddle_condition}), one can then also recover an optimal strategy $\tilde{W}^{(\text{opt})}$ from Eq.~\eqref{eq:max_QFI_SDP}. 
An optimal strategy $W^{(\text{opt})}$ can then finally be obtained by purifying $\tilde{W}^{(\text{opt})}$ using the global future $F$.
As mentioned previously, since not all QC-CCs are purifiable as QC-CCs this approach computes the optimal QFI for a larger class than \textup{\textsf{QC-CC}} and that includes all the purifications of QC-CCs. We denote the QFI of this larger class by $ J^{\textup{\textsf{QC-CC}}}_{\theta,\textup{purif}}(C_{\theta}^{\otimes N})$, and emphasize that it is an upper bound of $ J_\theta^{\textup{\textsf{QC-CC}}}(C_{\theta}^{\otimes N}) $.


\section{Classes of strategies}
\label{sec:classes}

\begin{figure}[htbp]
      \centering
          \includegraphics[width=0.8\columnwidth]{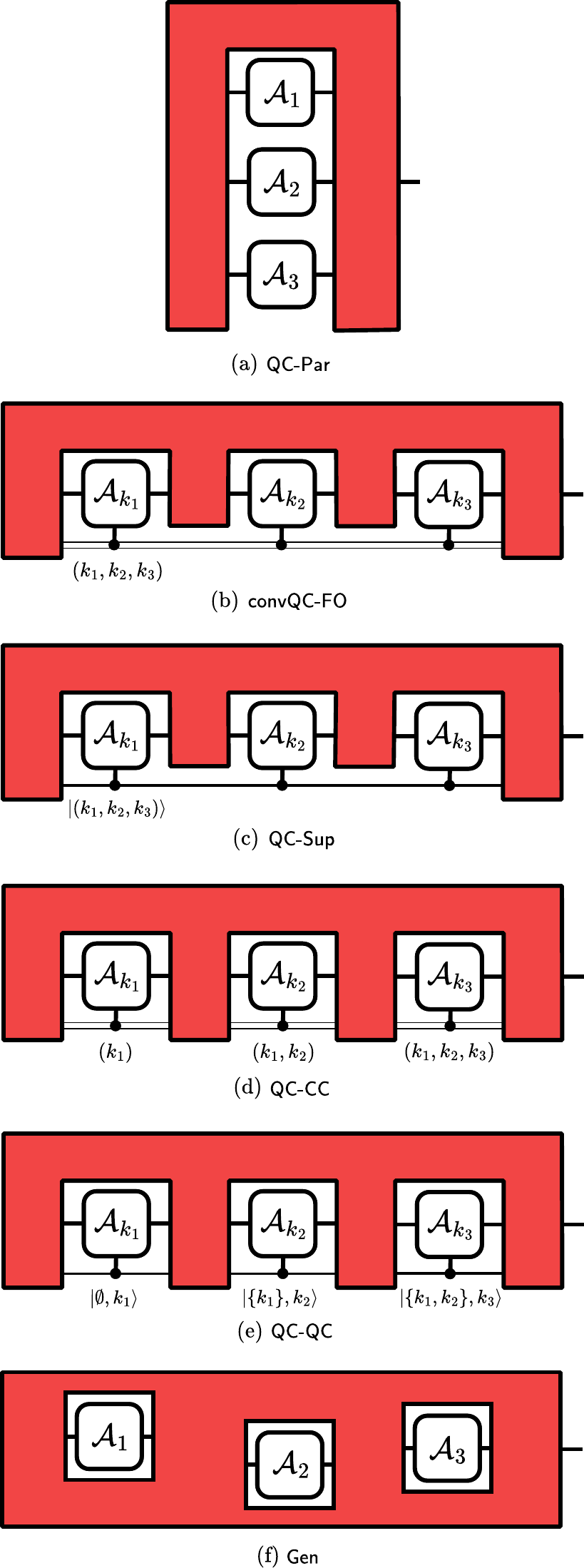}
          \caption{Abstract graphical representations of the different classes of strategies considered in this work, for $N=3$.
          (a) \textup{\textsf{QC-Par}}: parallel composition of the quantum channels. (b) \textup{\textsf{convQC-FO}}: convex combinations of strategies with fixed causal order. The order is (potentially probabilistically) fixed from the beginning of the process; it can be encoded physically in a classical system, represented by the double line. (c) \textup{\textsf{QC-Sup}}: coherent superposition of strategies with fixed causal order; the control system is quantum (single line) and fixed from the beginning of the process. (d) \textup{\textsf{QC-CC}}: the control system is classical and may evolve dynamically along the process. (e) \textup{\textsf{QC-QC}}: the control system is quantum and may evolve dynamically along the process. Note that the backbone of the ``combs'' from (a) to (e) may contain some quantum memory channels, which remain internal to the processes~\cite{wechs21}. (f) \textup{\textsf{Gen}}: representation of a general, arbitrary strategy.}
         \label{fig:classes_process}
 \end{figure}

With the metrological task defined and the methodology to solve it presented, we review in this section the different classes of strategies $\W$ that we are going to compare and their characterizations in terms of SDP constraints. 
We recall that throughout this work we are interested in the scenario where strategies are provided with $N$ copies of a quantum channel $\mathcal{C}_\theta$, as illustrated in Fig.~\ref{fig:strategy}. 
In more general scenarios, however, strategies can connect $N$ (potentially different) quantum operations $\mathcal{A}_k$. We first review several classes of strategies that were studied in \cite{liu}: \textup{\textsf{QC-Par}} (strategies that compose in parallel), \textup{\textsf{(conv)QC-FO}} (strategies with a fixed order, which are what are generally considered in metrology), \textup{\textsf{QC-Sup}} (superposition) and \textup{\textsf{Gen}} (arbitrary strategies or process matrices). We then consider two additional physically relevant classes that were introduced in \cite{wechs21}: \textup{\textsf{QC-CC}} (quantum circuits with classical control of causal order) and \textup{\textsf{QC-QC}} (quantum circuits with quantum control of causal order). We present in Fig.~\ref{fig:classes_process} an abstract representation of the different classes of strategies considered in this section.

As already noted, in the calculation of $J_\theta^{\W}(C_\theta^{\otimes N})$ one only needs characterizations of the sets $\Tr_F\W$.%
\footnote{Note, in particular, that~\cite{liu} made a simplifying assumption (see their Eq.~(9)) about the structure of $\Tr_F\W$ for the classes they considered, but which is not satisfied by some of the classes we introduce here. We hence rederive the SDP problems for $J_\theta^{\W}(C_\theta^{\otimes N})$ directly from the full characterizations of each $\W$ we give here; cf.\ Appendix~\ref{app_subsec:explicit_SDPs}.}
In the definitions of the classes $\W\subseteq \mathcal{L}(\mathcal{H}^{A^{IO}_{\mathcal{N}}F})$ that follow, one should thus understand the dimension of $\mathcal{H}^F$ to be free, in contrast to that of the $\mathcal{H}^{A^{IO}_k}$, which are fixed by the dimension of $C_\theta$.
That is, a process $W\in\W$ (for each $\W$ specified below) if, for some dimension of $\mathcal{H}^F$, $W$ satisfies the corresponding constraints we specify for $\W$.

Further details and, in particular, explicit formulations of the (primal and dual) SDP problems computing $J_\theta^{\W}(C_\theta^{\otimes N})$ for the different classes are given in Appendix~\ref{app_subsec:explicit_SDPs}.

\subsection{\text{\textup{\textsf{QC-Par}}}, \text{\textup{\textsf{(conv)QC-FO}}}, \text{\textup{\textsf{QC-Sup}}} and \text{\textup{\textsf{Gen}}}}

One of the simplest classes of quantum strategies is to compose the $N$ operations $\mathcal{A}_k$ (with $1\le k \le N$) in parallel, as depicted in Fig.~\ref{fig:classes_process}(a).
In such a quantum circuit with operations used in parallel (QC-Par), each $\mathcal{A}_k$ is applied to part of a multipartite state that is in general entangled across the $\mathcal{A}_k$ and, eventually, with an ancillary system. A quantum channel can then map the resulting output state to the global future $F$.
This class has been extensively studied and the corresponding class $\textsf{QC-Par}$ of process matrices can be characterized as~\cite{Chiribella2009}
\begin{align}
    \text{\textup{\textsf{QC-Par}}}=\left\{ \begin{array}{l}
        W \in \mathcal{L}(\mathcal{H}^{A_\mathcal{N}^{IO}F}), \ W \ge 0 \\[2mm]
        \text{s.t.} \left\{ \begin{array}{l}
            \Tr_F W = W_{(I)} \otimes \id^{A_{\cal N}^O}, \\[1mm]
            \Tr W_{(I)} = 1, \\[1mm]
            W_{(I)} \ge 0
        \end{array} \right.
    \end{array} \!\! \right\}, \label{eq:QC-Par}
\end{align}
where $W_{(I)} \in {\cal L}({\cal H}^{A_{\cal N}^I})$ is simply an $N$-partite density matrix.

Another extensively studied class of strategies is quantum circuits with a fixed order (QC-FO)~\cite{wechs21}, also known as ``quantum strategies''~\cite{Gutoski2006} or ``quantum combs''~\cite{Chiribella08}.
This class corresponds to ``standard'' quantum circuits, where the $\mathcal{A}_k$ are applied sequentially, potentially with intermediate quantum channels applied between the $\mathcal{A}_k$ jointly on the outcome of the previous operation and some quantum memory.
Such QC-FOs include parallel circuits as a special case, and their process matrices can be formally characterized as
\begin{align}
    & \text{\textup{\textsf{QC-FO}}} \notag \\
    & =\left\{ \!\begin{array}{l}
        W \in \mathcal{L}(\mathcal{H}^{A_\mathcal{N}^{IO}F}), \ W \ge 0 \\[2mm]
        \text{s.t.} \left\{ \begin{array}{l}
            \Tr_F W = W_{(N)} \otimes \mathbbm{1}^{A_N^O}, \\[1mm]
            \forall\, n =1,\ldots,N-1, \\
            \qquad \Tr_{A_{n+1}^I} W_{(n+1)} = W_{(n)} \otimes \mathbbm{1}^{A_n^O}, \\[1mm]
            \Tr W_{(1)} = 1, \\[1mm]
            \forall\, n =1,\ldots,N, \, W_{(n)} \ge 0
        \end{array} \right.
    \end{array} \!\!\!\right\}, \label{eq:QC-FO}
\end{align}
where $W_{(n)} \in {\cal L}({\cal H}^{A_{\{1,\ldots,n-1\}}^{IO}A_n^I})$, for $n = 1,\ldots, N$.
Strictly speaking, the definition of $\textup{\textsf{QC-FO}}$ as just stated only includes circuits with the specific fixed causal order $\mathcal{A}_1,\ldots,\mathcal{A}_n$. 
However, any other permutation of this order would still define a class of circuits with a fixed causal order. 
Thus, we define the classes $\textup{\textsf{QC-FO}}_\pi$, which are defined analogously to  $\textup{\textsf{QC-FO}}$, but which correspond to the causal orders $\mathcal{A}_{\pi(1)},\ldots,\mathcal{A}_{\pi(n)}$, with $\pi $ a permutation of $\mathcal{N}$.
Note that for each $\pi$, $\textup{\textsf{QC-FO}}_\pi$ is convex; however, their union is not. Therefore we also define $\textup{\textsf{convQC-FO}}$, the convex hull of all the quantum circuits with a fixed causal order, as 
\begin{align}
    \text{\textup{\textsf{convQC-FO}}} := & \text{conv}(\cup_\pi \textup{\textsf{QC-FO}}_\pi) \notag \\
    = & \left\{ \!\begin{array}{l}
        W \in \mathcal{L}(\mathcal{H}^{A_\mathcal{N}^{IO}F}), \ W \ge 0 \\[2mm]
        \text{s.t.} \left\{ \begin{array}{l}
            W = \sum_{\pi} q_\pi W_\pi, \\[1mm]
            \forall\, \pi, W_\pi \in \text{\textup{\textsf{QC-FO}}}_\pi, \\[1mm]
            \sum_{\pi} q_\pi = 1, \, \forall\, \pi, q_\pi \ge 0
        \end{array} \right.
    \end{array} \!\!\right\}, \label{eq:convQC-FO}
\end{align}
where the sums are over all possible permutations $\pi$ of ${\cal N}$.
By definition, this class only comprises processes with a definite causal order, since they are defined as classical mixtures of processes with a fixed causal order. One can alternatively describe processes in \textup{\textsf{convQC-FO}} as being classically controlled, using a control state that is fixed, with some probability, from the beginning of the process; see Fig.~\ref{fig:classes_process}(b) for an abstract representation.

In analogy with the quantum switch, one can construct processes that use superpositions of different fixed causal orders by using a quantum, rather than a classical, control system whose state is fixed from the beginning of the process, and eventually given to the global future $F$, to coherently control the order. 
Such strategies, which are graphically represented in Fig.~\ref{fig:classes_process}(c), correspond to the class of quantum circuits with causal superposition (QC-Sup) that was introduced in~\cite{liu}, and which can be characterized as
\begin{align}
    \text{\textup{\textsf{QC-Sup}}}=\left\{ \begin{array}{l}
        W \in \mathcal{L}(\mathcal{H}^{A_\mathcal{N}^{IO}F}), \ W \ge 0 \\[2mm]
        \text{s.t.} \left\{ \begin{array}{l}
            \Tr_F W = \sum_{\pi} q_\pi \tilde{W}_\pi, \\[1mm]
            \forall\, \pi, \tilde{W}_\pi \in \Tr_F(\text{\textup{\textsf{QC-FO}}}_\pi), \\[1mm]
            \sum_{\pi} q_\pi = 1, \, \forall\, \pi, q_\pi \ge 0
        \end{array} \right.
    \end{array} \!\!\right\}. \label{eq:QC-Sup}
\end{align}
Similarly to the characterization of $\text{\textup{\textsf{convQC-FO}}}$, this characterization involves a convex mixture of processes with fixed causal order. 
Crucially, however, the convex decompositions apply to $\Tr_F W$ here, while for $\text{\textup{\textsf{convQC-FO}}}$ they applied to $W$ directly. 
This explains the main difference between the two classes: $\text{\textup{\textsf{QC-Sup}}}$ allows for processes with causal superposition, and hence indefinite causal order, while $\text{\textup{\textsf{convQC-FO}}}$ does not. 

One can also consider strategies corresponding to arbitrary process matrices, which are the most general way to consistently combine the operations $\mathcal{A}_k$, as depicted abstractly in Fig.~\ref{fig:classes_process}(f) (cf.~also Fig.~\ref{fig:strategy}).
We will denote this class, which includes all strategies with indefinite causal order, as \text{\textup{\textsf{Gen}}}, and it can be characterized as~\cite{araujo_witnessing_2015}
\begin{align}
    & \text{\textup{\textsf{Gen}}} \notag \\
    & =\left\{ \!\begin{array}{l}
        W \in \mathcal{L}(\mathcal{H}^{A_\mathcal{N}^{IO}F}), \ W \ge 0 \\[2mm]
        \text{s.t.} \left\{\! \begin{array}{l}
            \forall\, \emptyset \subsetneq \mathcal{K} \subseteq\mathcal{N}, \\
            \quad {}_{\prod_{k\in\mathcal{K}}[1-A_k^{O}]}(\Tr_{A_{\mathcal{N}\setminus\mathcal{K}}^{IO}F}W)=0, \\[2mm]
            \Tr W = \prod_{k\in{\cal N}}d_{A_k^O} 
        \end{array} \right.
    \end{array} \!\!\!\right\}\!, \label{eq:ICO}
\end{align}
with the generic short-hand notation $d_{X}:= \text{dim}(\mathcal{H}^X)$ and the ``trace-out-and-replace'' notation defined as
\begin{equation}
    {}_XW:=(\Tr_X W)\otimes \frac{\mathbbm{1}^X}{d_X}, \, \, {}_{[1-X]}W:=W-{}_XW.
\end{equation}
(The second part of the previous definition can be used recursively, in a commutative manner: e.g., ${}_{[1-X][1-Y]}W={}_{[1-X]}({}_{[1-Y]}W)={}_{[1-Y]}({}_{[1-X]}W)=W - {}_XW - {}_YW - {}_{XY}W$.) 
Note that not all strategies in $\textup{\textsf{Gen}}$ are known to have a physical representation in terms of a generalized form of quantum circuit or otherwise.


\subsection{\text{\textup{\textsf{QC-CC}}} and \text{\textup{\textsf{QC-QC}}}}

The classes of strategies introduced above were all considered in~\cite{liu} and, with the exception of QC-Sup, were previously well studied.
Here we present two further classes that were recently introduced in~\cite{wechs21} and which are of particular physical interest.

The first of these are quantum circuits with classical control of causal order (QC-CC), which attempt to generalize $\textup{\textsf{convQC-FO}}$ as the most general strategies compatible with a well-defined causal order.
Contrary to QC-FOs, the causal order of a QC-CC is not necessarily fixed \emph{a priori}, but can be established on the fly: a classical control system is used at each step in the circuit to control which operation (amongst those not yet applied) to apply next; see Fig.~\ref{fig:classes_process}(d) for an abstract graphical representation. Such QC-CCs thus allow the causal order to be dynamically controlled.
QC-CCs can be characterized in terms of SDP constraints involving (positive semidefinite) matrices $W_{(k_1,\ldots,k_n)} \in \mathcal{L}(\mathcal{H}^{A_{\{k_1,\ldots,k_{n-1}\}}^{IO}A_{k_n}^I})$ and $W_{(k_1,\ldots,k_N,F)} \in \mathcal{L}(\mathcal{H}^{A_\mathcal{N}^{IO}F})$ as
\begin{align}
    & \text{\textup{\textsf{QC-CC}}} \notag \\
    & =\left\{ \!\begin{array}{l}
        W \in \mathcal{L}(\mathcal{H}^{A_\mathcal{N}^{IO}F}), \ W \ge 0 \\[2mm]
        \text{s.t.} \left\{ \begin{array}{l}
            W = \sum_{(k_1,\ldots,k_N)} W_{(k_1,\ldots,k_N,F)}, \\[1mm]
            \forall\, (k_1,\ldots,k_N), \\
            \quad \Tr_F W_{(k_1,\ldots,k_N,F)} = W_{(k_1,\ldots,k_N)} \otimes \mathbbm{1}^{A_{k_N}^O}, \\[1mm]
            \forall\, n = 1,\ldots,N-1, \forall\, (k_1,\ldots,k_n), \\
            \quad \sum_{k_{n+1}} \Tr_{A_{k_{n+1}^I}} W_{(k_1,\ldots,k_n,k_{n+1})} \\[-1mm]
            \hspace{30mm} = W_{(k_1,\ldots,k_n)} \otimes \mathbbm{1}^{A_{k_n}^O}, \\[1mm]
            \sum_{k_1} \Tr W_{(k_1)} = 1, \\[2mm]
            \forall\, n = 1,\ldots,N, \, \forall\,(k_1,\ldots,k_n), \\
            \quad W_{(k_1,\ldots,k_n)} \ge 0, W_{(k_1,\ldots,k_N,F)} \ge 0
        \end{array} \right.
    \end{array} \!\!\!\right\}, \label{eq:QC-CC}
\end{align}
where for all $n$-tuples $(k_1,\ldots,k_n)$, we implicitly assume that $k_i \neq k_j$ for $i\neq j$.
For $N \leq 3$, it was shown~\cite{wechs19} that $\textup{\textsf{QC-CC}}$ constitutes the class of all circuits with a definite causal order (so-called ``causally separable'' processes), while it is still an open question for $N>3$.

The second new class we consider are the quantum circuits with quantum control of causal order (QC-QC), which can be thought of as the
quantum counterpart of QC-CCs.
QC-QCs use a quantum control system to coherently control which operation is applied at each step, allowing for both coherent causal superposition and dynamical control of causal order, as illustrated in Fig.~\ref{fig:classes_process}(e).
QC-QCs may thus exhibit indefinite causal order, and are the most general class of processes known to have a clear physical interpretation~\cite{wechs21,purves21}, which gives them particular practical importance for understanding any potential advantages from indefinite causal order  in quantum metrology.
This class can be characterized in terms of (positive semidefinite) matrices $W_{(\mathcal{K}_{n-1},k_n)} \in \mathcal{L}(\mathcal{H}^{A_{\mathcal{K}_{n-1}}^{IO}A_{k_n}^I})$, for all strict subsets $\mathcal{K}_{n-1}$ of $\mathcal{N}$ and all $k_n \in \mathcal{N}\setminus \mathcal{K}_{n-1}$, as
\begin{align}
    & \text{\textup{\textsf{QC-QC}}} \notag \\
    & =\left\{ \!\begin{array}{l}
        W \in \mathcal{L}(\mathcal{H}^{A_\mathcal{N}^{IO}F}), \ W \ge 0 \\[2mm]
        \text{s.t.} \left\{ \begin{array}{l}
            \Tr_F W = \sum_{k_N \in \mathcal{N}}W_{(\mathcal{N}\setminus \{k_N\},k_N)}\otimes \mathbbm{1}^{A_{k_N}^O}, \\[1mm]
            \forall\, n = 1,\ldots,N-1, \forall\, \emptyset \subsetneq {\cal K}_n \subsetneq {\cal N}, \\
            \qquad \sum_{k_{n+1}\in \mathcal{N}\setminus\mathcal{K}_n} \Tr_{A_{k_{n+1}}^I}W_{(\mathcal{K}_n,k_{n+1})} \\[-1mm]
            \hspace{12mm} = \sum_{k_n \in \mathcal{K}_n} W_{(\mathcal{K}_n\setminus \{k_n\},k_n)}\otimes \mathbbm{1}^{A_{k_n}^O}, \\[1mm]
            \sum_{k_1 \in \mathcal{N}} \Tr W_{(\emptyset,k_1)} = 1, \\[2mm]
            \forall\, n = 1,\ldots,N, \, \forall\, {\cal K}_{n-1}\subsetneq{\cal N}, \\
            \qquad \forall\, k_n \in \mathcal{N}\setminus\mathcal{K}_{n-1}, W_{({\cal K}_{n-1},k_n)} \ge 0
        \end{array} \right.
    \end{array} \!\!\!\right\}, \label{eq:QC-QC}
\end{align}
where the notation ${\cal K}_n$ implicitly assumes that $|{\cal K}_n| = n$.


\subsection{Inclusion relations between the classes}

Before considering the power of these different classes in the metrological task, let us first discuss the relation between these classes themselves.

First, note that in the rather trivial case of $N=1$, all the classes introduced above coincide.

For $N = 2$, however, there are several non-trivial inclusion relations between the classes.
Since parallel circuits can be seen as particular instances of circuits with fixed orders, $\textup{\textsf{QC-Par}}$ is a subclass of $\textup{\textsf{QC-FO}}$, while $\textup{\textsf{QC-FO}}$ (for any order $\pi$) is evidently a subclass of $\textup{\textsf{convQC-FO}}$. It is easy to see that these inclusions are moreover strict.
For $N=2$, no dynamical control of the causal order can be considered since the choice of the first operation also forces the other operation to be second independently of what first operation is applied.
Hence $\textup{\textsf{convQC-FO}}$ and $\textup{\textsf{QC-CC}}$ coincide, as do likewise $\textup{\textsf{QC-Sup}}$ and $\textup{\textsf{QC-QC}}$.
The quantum switch, $W^\text{QS}$, is in $\textup{\textsf{QC-QC}}$ yet not in $\textup{\textsf{convQC-FO}}$ which shows this inclusion is, however, strict.
Finally, $\textup{\textsf{Gen}}$ includes, by construction, all other strategies and hence all these classes, and one can easily construct examples to show it strictly includes $\textup{\textsf{QC-QC}}$~\cite{oreshkov12,wechs21}.

In the approach we use to compute the optimal QFI we optimize over $\tilde{W} \in \Tr_F \textup{\textsf{\textbf{W}}}$.
When the global future is traced out, the inclusion relations between the classes are slightly altered. 
As it turns out, in this case $\Tr_F\textup{\textsf{convQC-FO}}$, $\Tr_F \textup{\textsf{QC-CC}}$, $\Tr_F\textup{\textsf{QC-Sup}}$ and $\Tr_F\textup{\textsf{QC-QC}}$ all coincide since, in addition to there being no dynamical order for $N=2$, the possibility of including circuits with causal superposition, which would distinguish $\textup{\textsf{QC-CC}}$ from $\textup{\textsf{QC-Sup}}$ and $\textup{\textsf{QC-QC}}$, vanishes when $F$ is traced out. 
The different inclusion relations for $N=2$, both without and, resp., when tracing over $F$ are 
\begin{widetext}
\begin{align}
&\textup{\textsf{QC-Par}} &&\!\!\!\!\subsetneq\!\!\!\!& &\textup{\textsf{QC-FO}} &&\!\!\!\!\subsetneq\!\!\!\!& &\textup{\textsf{convQC-FO}} &&\!\!\!\!=\!\!\!\!& &\textup{\textsf{\textbf{QC-CC}}} && \!\!\!\!\subsetneq\!\!\!\!\!& &\textup{\textsf{QC-Sup}} &&\!\!\!\!=\!\!\!\!& &\textup{\textsf{\textbf{QC-QC}}} &&\!\!\!\!\subsetneq\!\!\!\!& &\textup{\textsf{Gen}},\label{eq:hierarchy_class_N2} \\[2mm]
\Tr_F\,&\textup{\textsf{QC-Par}} &&\!\!\!\!\subsetneq\!\!\!\!& \Tr_F\,&\textup{\textsf{QC-FO}} &&\!\!\!\!\subsetneq\!\!\!\!& \Tr_F\,&\textup{\textsf{convQC-FO}} &&\!\!\!\!=\!\!\!\!& \bm{\bmTr\nolimits_{F}}\,&\textup{\textsf{\textbf{QC-CC}}} &&\!\!\!\!=\!\!\!\!& \Tr_F\,&\textup{\textsf{QC-Sup}} &&\!\!\!\!=\!\!\!\!& \bm{\bmTr\nolimits_{F}}\,&\textup{\textsf{\textbf{QC-QC}}} &&\!\!\!\!\subsetneq\!\!\!\!& \Tr_F\,&\textup{\textsf{Gen}},
\label{eq:hierarchy_class_N2_trF}
\end{align}
\begin{figure}[t]
    \centering
     \subfloat[\label{fig:strategies_N2_avecF}]{%
        \includegraphics[width=0.47\columnwidth]{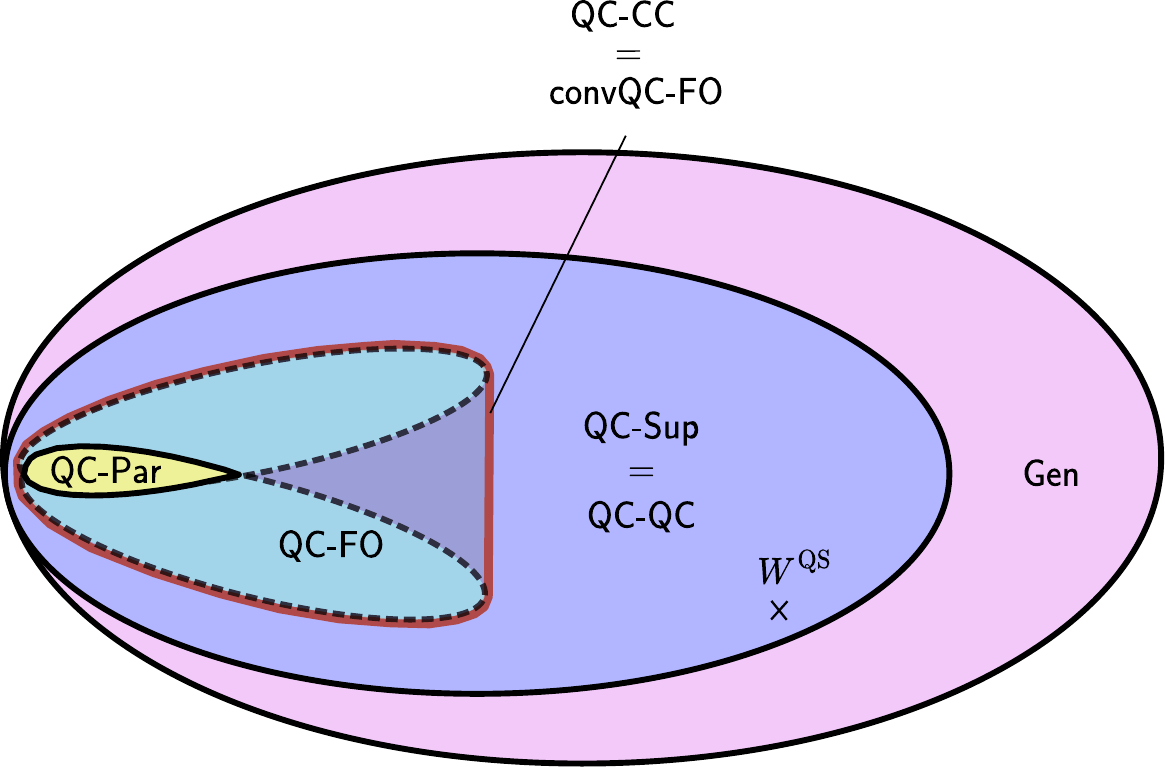}
     }
     \hfill
     \subfloat[\label{fig:strategies_N2_sansF}]{%
        \includegraphics[width=0.47\columnwidth]{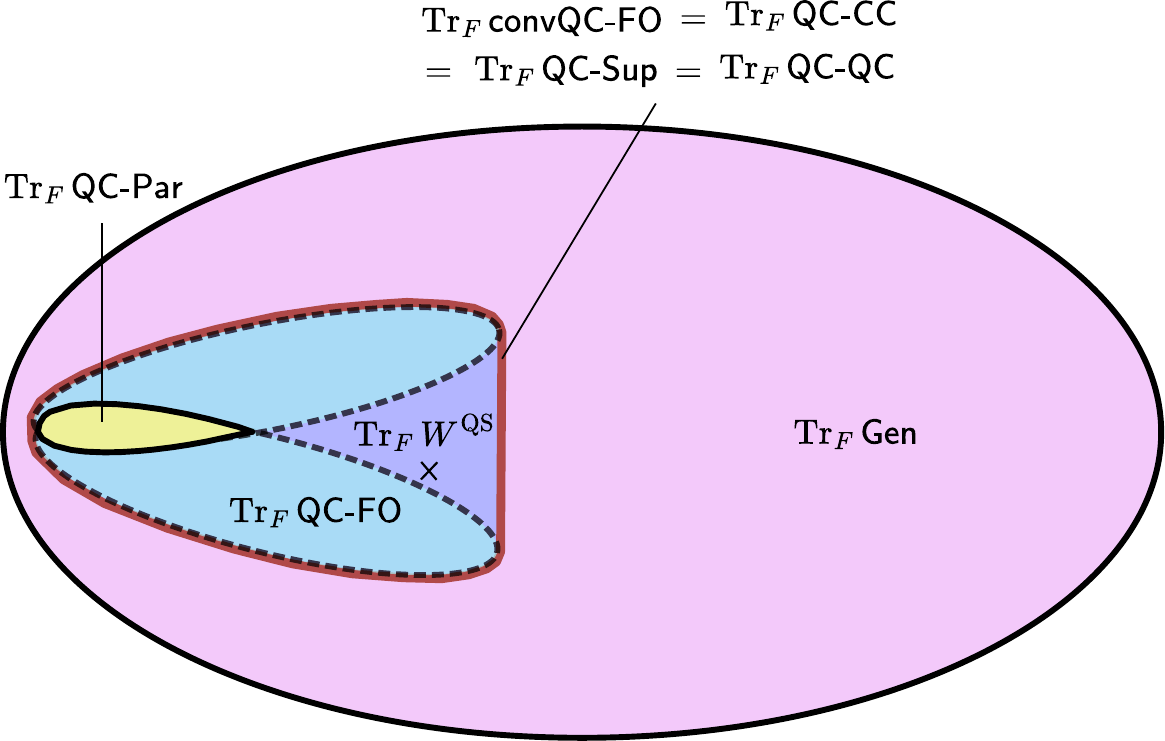}
     }
        \caption{Relations between the different classes of strategies with and without definite causal order for $N = 2$, in the cases where the global future $F$ is (a) not traced out; (b) traced out.}
        \label{fig:strategies_pur_N2_v2}
\end{figure}
\end{widetext}
where the new classes we study here that have not previously been considered in quantum metrology are indicated in bold. 
These are graphically represented in Figs.~\ref{fig:strategies_N2_avecF} and Fig.~\ref{fig:strategies_N2_sansF}, respectively.

For $N \geq 3$, the inclusion relations between $\textup{\textsf{QC-Par}}$, $\textup{\textsf{QC-FO}}$ and $\textup{\textsf{convQC-FO}}$ remain the same, and $\textup{\textsf{Gen}}$ likewise includes all the other classes. 
However, due to the possibility of dynamical control of causal order that arises starting with $N=3$, the equalities between $\textup{\textsf{convQC-FO}}$ and $\textup{\textsf{QC-CC}}$, as well as between $\textup{\textsf{QC-Sup}}$ and $\textup{\textsf{QC-QC}}$ break down.
$\textup{\textsf{QC-Sup}}$ becomes a strict subclass of $\textup{\textsf{QC-QC}}$, as shown by the so-called Grenoble process~\cite{wechs21} that gives an explicit example of a QC-QC that can readily be verified to not be in $\textup{\textsf{QC-Sup}}$.
Moreover, neither $\textup{\textsf{QC-CC}}$ nor $\textup{\textsf{QC-Sup}}$ is a subset of the other, since $\textup{\textsf{QC-CC}}$ contains circuits with dynamical control of causal order but without causal superposition (such as a ``classical switch'', where the first operation prepares a control system that determines the order between the remaining ones~\cite{chirib13,wechs21}), while $\textup{\textsf{QC-Sup}}$ contains circuits with causal superposition but without dynamical control of causal order (such as the generalization of the quantum switch to $N$ operations).

In the case where $F$ is traced out, $\Tr_F \textup{\textsf{convQC-FO}}$ and $\Tr_F \textup{\textsf{QC-Sup}}$ again coincide by construction, and are subclasses of $\Tr_F \textup{\textsf{QC-CC}}$, which allows for dynamical control of the causal order. $\Tr_F \textup{\textsf{QC-CC}}$ is a strict subclass of $\Tr_F \textup{\textsf{QC-QC}}$, as again shown by the Grenoble process~\cite{wechs21} which retains its properties even when $F$ is traced out. 

\begin{figure*}[ht]
     \centering
     \subfloat[\label{fig:strategies_N3_avecF}]{%
        \includegraphics[width=0.95\columnwidth]{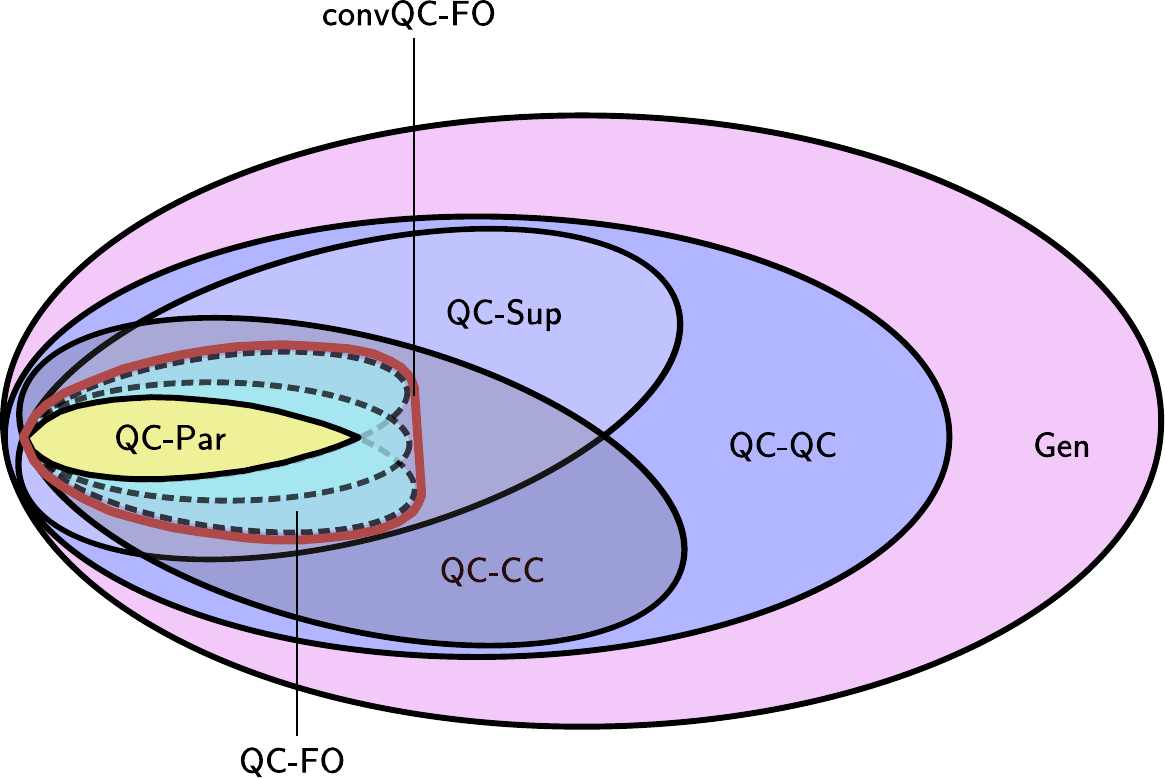}
     }
	 \hfill
     \subfloat[\label{fig:strategies_N3_sansF}]{%
        \includegraphics[width=0.95\columnwidth]{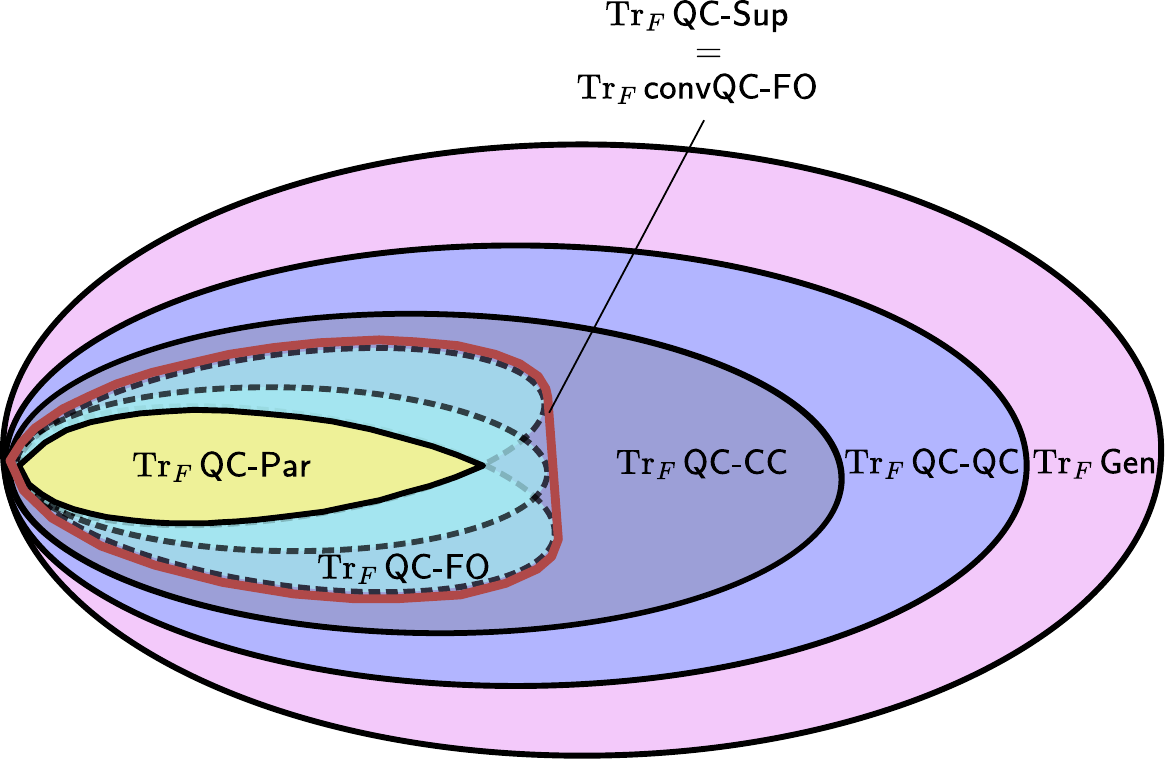}
     }
        \caption{Relations between the different classes of strategies with and without definite causal order for $N \ge 3$, in the cases where the global future $F$ is (a) not traced out; (b) traced out.}
        \label{fig:strategies_N3}
\end{figure*}

The different inclusion relations for $N\ge 3$, both without and, resp., when tracing over $F$ are thus now
\begin{widetext}
\begin{align}
&\textup{\textsf{QC-Par}} &&\!\!\!\!\subsetneq\!\!\!\!& &\textup{\textsf{QC-FO}} &&\!\!\!\!\subsetneq\!\!\!\!& &\textup{\textsf{convQC-FO}} &&\!\!\!\!\subsetneq\!\!\!\!& & \qquad\,\,\,\, \begin{matrix}
	\textup{\textsf{\textbf{QC-CC}}} \\
	\textup{\textsf{QC-Sup}}
\end{matrix}  &&\!\!\!\!\subsetneq\!\!\!\!& &\textup{\textsf{\textbf{QC-QC}}} &&\!\!\!\!\subsetneq\!\!\!\!& &\textup{\textsf{Gen}},\label{eq:hierarchy_class_N3} \\[2mm]
\Tr_F\,&\textup{\textsf{QC-Par}} &&\!\!\!\!\subsetneq\!\!\!\!& \Tr_F\,&\textup{\textsf{QC-FO}} &&\!\!\!\!\subsetneq\!\!\!\!& \Tr_F\,&\textup{\textsf{convQC-FO}} &&\!\!\!\!= \!\!\!\!& \Tr_F\,&\textup{\textsf{QC-Sup}} \ \subsetneq \ \bm{\bmTr\nolimits_{F}}\textup{\textsf{\textbf{QC-CC}}} &&\!\!\!\!\subsetneq\!\!\!\!& \bm{\bmTr\nolimits_{F}}\,&\textup{\textsf{\textbf{QC-QC}}} &&\!\!\!\!\subsetneq\!\!\!\!& \Tr_F\,&\textup{\textsf{Gen}},
\label{eq:hierarchy_class_N3_trF}
\end{align}
\end{widetext}
where $\textup{\textsf{QC-CC}}$ and $\textup{\textsf{QC-Sup}}$ are written on top of each other because they are incomparable. 
These relations are graphically represented in Figs.~\ref{fig:strategies_N3_avecF} and Fig.~\ref{fig:strategies_N3_sansF}, respectively.

We will now focus on different choices of external channel $\mathcal{C}_\theta$, in the cases $N=2,3$, in order to provide a quantitative comparison between the different classes previously introduced, and reassess the advantage of indefinite causal order for the quantum metrological task of channel parameter estimation.


\section{Results}
\label{sec:results}


A comparison of the classes \textup{\textsf{QC-Par}}, \textup{\textsf{QC-FO}}, \textup{\textsf{QC-Sup}} and \textup{\textsf{Gen}} through the metrological task presented above was already made in~\cite{liu} for different choices of external channels $\mathcal{C}_\theta$, both for $N=2$ and $N=3$. 
The classes \textup{\textsf{QC-CC}} and \textup{\textsf{QC-QC}}, however, are potentially more pertinent classes for understanding advantages arising from indefinite causal order.
Indeed, the former corresponds to the most general strategies with a well-defined causal order, while the latter describes the most general strategies with a known physical implementation in terms of quantum-controlled circuits.
Here, we study how these classes perform compared to those studied in~\cite{liu} in the metrological task.
First, we establish general hierarchies for the QFI values $J_\theta^{\W}(C_\theta^{\otimes N})$ corresponding to the different classes $\mathbf{\W}$ introduced above, as well as the QFI value $J^{\textup{\textsf{QC-CC}}}_{\theta,\textup{purif}}(C_\theta^{\otimes N})$ for the purification of QC-CCs (cf.\ Sec.~\ref{sec:SDP}). These hierarchies are based on the inclusion relations between the different classes presented above, which hold for any choice of channel $\mathcal{C}_\theta$.
Then, we give a quantitative comparison and provide numerical hierarchies for particular choices of channel $\mathcal{C}_\theta$, using the SDP approach.


\subsection{General hierarchies}

To start with, let us review the general hierarchies for $N=2$ and then for $N \ge 3$ that follow from the inclusion relations given by Eqs.~\eqref{eq:hierarchy_class_N2}--\eqref{eq:hierarchy_class_N3_trF}.

First, note that it is a general fact that (for any $N$), by the convexity of the QFI, $J_\theta^{\textup{\textsf{convQC-FO}}}(C_\theta^{\otimes N}) = J_\theta^{\textup{\textsf{QC-FO}}}(C_\theta^{\otimes N})$. Hence, we will not consider $J_\theta^{\textup{\textsf{convQC-FO}}}(C_\theta^{\otimes N})$ explicitly any further.
Secondly, as we discussed in Sec.~\ref{sec:SDP}, not all QC-CCs are purifiable as QC-CCs; instead, the set of purifications of QC-CCs gives a subset of QC-QCs.
We thus recall that, since the approach we use to calculate the QFI optimizes the QFI within the larger purifying class, it does not directly provide $J_\theta^{\textup{\textsf{QC-CC}}}(C_\theta^{\otimes N})$ but instead an upper bound on this quantity that we call $J^{\textup{\textsf{QC-CC}}}_{\theta,\textup{purif}}(C_\theta^{\otimes N})$.


\subsubsection{$N=2$ case}

Since for $N=2$, $\textup{\textsf{convQC-FO}} = \textup{\textsf{QC-CC}}$ (see Eq.~\eqref{eq:hierarchy_class_N2}), we have that $J_\theta^{\textup{\textsf{QC-CC}}}(C_\theta^{\otimes 2}) = J_\theta^\textup{\textsf{convQC-FO}}(C_\theta^{\otimes 2})  = J_\theta^{\textup{\textsf{QC-FO}}}(C_\theta^{\otimes 2})$.
Furthermore, as written in Eq.~\eqref{eq:hierarchy_class_N2_trF}, in the $N=2$ case $\Tr_F \textup{\textsf{QC-CC}}$ coincides with $\Tr_F \textup{\textsf{QC-Sup}}$ and $\Tr_F \textup{\textsf{QC-QC}}$, so it must be that $J^{\textup{\textsf{QC-CC}}}_{\theta,\textup{purif}}(C_\theta^{\otimes 2}) = J_\theta^{\textup{\textsf{QC-Sup}}}(C_\theta^{\otimes 2}) = J_\theta^{\textup{\textsf{QC-QC}}}(C_\theta^{\otimes 2})$.
Combined with the other inclusion relations presented in Eq.~\eqref{eq:hierarchy_class_N2} we thus obtain, for any quantum channel $\mathcal{C}_\theta$, the following hierarchy between the classes (where the QFI values for the new classes considered in this work are again highlighted in bold):
\begin{widetext}
    \begin{equation}
       J_\theta^{\textup{\textsf{QC-Par}}}(C_\theta^{\otimes 2}) \leq J_\theta^{\textup{\textsf{QC-FO}}}(C_\theta^{\otimes 2})=\bm{J_{\theta}^{\textup{\textsf{\textbf{QC-CC}}}}(C_\theta^{\otimes 2})} \leq \bm{J^{\textup{\textsf{\textbf{QC-CC}}}}_{\theta,\textup{\textbf{purif}}}(C_\theta^{\otimes 2})} = J_\theta^{\textup{\textsf{QC-Sup}}}(C_\theta^{\otimes 2}) = \bm{J_\theta^{\textup{\textsf{\textbf{QC-QC}}}}(C_\theta^{\otimes 2})} \leq J_\theta^{\textup{\textsf{Gen}}}(C_\theta^{\otimes 2}).
       \label{eq:hierarchy_N2_pur}
    \end{equation}
\end{widetext}

As we will see in the examples below in Sec.~\ref{subsubsec:examples1} (see also~\cite{liu}), the inequalities in this hierarchy are strict for at least some choice of external channels $\mathcal{C}_\theta$, and so Eq.~\eqref{eq:hierarchy_N2_pur} cannot be strengthened further in general.


\subsubsection{$N\ge3$  case}
\label{sec:N_3_general}

For $N\ge3$, as indicated in Eq.~\eqref{eq:hierarchy_class_N3_trF} we have $\Tr_F \textup{\textsf{QC-Sup}} \subsetneq \Tr_F \textup{\textsf{QC-CC}}$, which directly implies that $J_\theta^{\textup{\textsf{QC-Sup}}}(C_\theta^{\otimes N}) \leq  J^{\textup{\textsf{QC-CC}}}_{\theta,\textup{purif}}(C_\theta^{\otimes N})$. 
We likewise have trivially that $J_\theta^{\textup{\textsf{QC-CC}}}(C_\theta^{\otimes N}) \leq  J^{\textup{\textsf{QC-CC}}}_{\theta,\textup{purif}}(C_\theta^{\otimes N})$.
However, the relation between $J_\theta^{\textup{\textsf{QC-Sup}}}(C_\theta^{\otimes N})$ and $J_\theta^{\textup{\textsf{QC-CC}}}(C_\theta^{\otimes N})$ is \emph{a priori} unclear, and may in principle differ depending on $\mathcal{C}_\theta$.
Combing this with the inclusion relations presented in Eq.~\eqref{eq:hierarchy_class_N3}, we obtain, for any quantum channel $\mathcal{C}_\theta$, the following hierarchies between the QFIs:
\begin{widetext}
\begin{equation}
      J_\theta^{\textup{\textsf{QC-Par}}}(C_\theta^{\otimes N})
	  \leq 
	  J_\theta^{\textup{\textsf{QC-FO}}}(C_\theta^{\otimes N})
	  \leq
       \begin{matrix}

       \\[1mm]
J_\theta^{\textup{\textsf{QC-Sup}}}(C_\theta^{\otimes N})
 \\[2mm]	\underbrace{\bm{J_\theta^{\textup{\textsf{\textbf{QC-CC}}}}(C_\theta^{\otimes N})}}_{=?}
	\end{matrix}
	\leq 
	\bm{J^{\textup{\textsf{\textbf{QC-CC}}}}_{\theta,\textup{\textbf{purif}}}(C_\theta^{\otimes N}) }
	\leq 
	\bm{J_\theta^{\textup{\textsf{\textbf{QC-QC}}}}(C_\theta^{\otimes N})}
	\leq 
	J_\theta^{\textup{\textsf{Gen}}}(C_\theta^{\otimes N}).
       \label{eq:hierarchy_N3_dual}
    \end{equation}
\end{widetext}
The question mark here indicates that, as discussed above, the value of $J_\theta^{\textup{\textsf{QC-CC}}}(C_\theta^{\otimes N})$ can only be bounded but not evaluated explicitly.
The fact that $J_\theta^{\textup{\textsf{QC-CC}}}(C_\theta^{\otimes N})$ and $J_\theta^{\textup{\textsf{QC-Sup}}}(C_\theta^{\otimes N})$ can in general not be compared is illustrated here by writing them on top of each other.


\subsection{Examples}

We now look at some examples to illustrate the hierarchies previously introduced and to study the tightness of the inequalities they contain for different families of channels. 
For each example considered, one can then conclude whether one can obtain an advantage using indefinite causal order strategies over definite causal order strategies.


\subsubsection{Qubit depolarizing, Pauli, and thermalizing channels}\label{subsubsec:examples1}

Using the SDP approach for $N = 1,2,3$ (qubit) depolarizing channels $\mathcal{D}_\theta$ as defined by Eq.~\eqref{eq:depo_channel}, with Choi matrix $D_\theta$, we obtain the following equalities between the QFIs of the different classes, up to a numerical error of $10^{-6}$: 
    \begin{widetext}
    \begin{equation}
        J_\theta^{\textup{\textsf{QC-Par}}}(D_\theta^{\otimes N})=J_\theta^{\textup{\textsf{QC-FO}}}(D_\theta^{\otimes N})=\bm{J_\theta^{\textup{\textsf{\textbf{QC-CC}}}}(D_\theta^{\otimes N})} = J_\theta^{\textup{\textsf{QC-Sup}}}(D_\theta^{\otimes N}) = \bm{J_\theta^{\textup{\textsf{\textbf{QC-QC}}}}(D_\theta^{\otimes N})} = J_\theta^{\textup{\textsf{Gen}}}(D_\theta^{\otimes N})= \frac{3N}{(1-\theta)(1+3\theta)}.
        \label{eq:hierarchy_depol}
    \end{equation}
    \end{widetext}
	
The optimal strategy we found using the SDP approach corresponds to the $N$-Choi circuit, which directly generalizes the $2$-Choi circuit we described in Sec.~\ref{sec:critique}. 
Indeed, solving the SDP optimization problem for the largest class \textup{\textsf{Gen}} we found precisely the strategy $\Tilde{W}=\frac{1}{2^N}\id^{A_{\mathcal{N}}^{IO}}$, which one can purify into the $N$-Choi circuit (see Appendix~\ref{app:Frey}).
Since the $N$-Choi circuit is a parallel circuit, we conclude that there is no advantage of indefinite causal order strategies over definite causal order strategies for $N=1,2,3$ depolarizing channels $\mathcal{D}_\theta$, contrary to what was claimed in~\cite{frey}.
We conjecture that the above equalities hold for any $N$ and any probe-dimension $d$ (indeed, we verified this for $N=1,2$ also in the qutrit case $d=3$), and that the corresponding optimal strategy remains the $N$-Choi circuit.

 One can extend these results (in the qubit case) by considering more general Pauli channels $ \mathcal{P}_\vartheta$, defined as
  \begin{equation}
    \mathcal{P}_\vartheta(\rho) = \vartheta\rho + (1-\vartheta)(\alpha \sigma_x\rho \sigma_x + \beta \sigma_y\rho \sigma_y + \gamma \sigma_z\rho \sigma_z),
    \label{eq:pauli_channel}
    \end{equation}
     where $\sigma_x, \sigma_y$ and $\sigma_z$ are the Pauli matrices, and $\alpha,\beta,\gamma \in [0,1]$ such that $\alpha + \beta + \gamma = 1$. 
This type of channel describes many usual noise channels considered in quantum information theory, such as depolarizing, phase-flip and bit-flip channels. 
Here, we denoted the metrological parameter by $\vartheta$, rather than $\theta$, to give the Pauli channel expression a symmetric form. By considering the change of variable $\vartheta \to \frac{1+3\theta}{4}$, and the choice of parameters $\alpha=\beta=\gamma =1/3$, one recovers the depolarizing channel as defined by Eq.~\eqref{eq:depo_channel}.
     
The same equalities given in Eq.~\eqref{eq:hierarchy_depol} between the QFIs of all the different classes were again obtained for Pauli channels with $N=1,2,3$: $\forall\, (\alpha,\beta,\gamma), \, J_\vartheta^{\textup{\textsf{\textbf{W}}}}(P_\vartheta^{\otimes N}) = \frac{N}{\vartheta(1-\vartheta)}$, for all classes $\textup{\textsf{\textbf{W}}}$ considered in this work. 
As a sanity check one can verify that, with the change of variable $\vartheta \to \frac{1+3\theta}{4}$, we recover the optimal QFI of Eq.~\eqref{eq:hierarchy_depol}, and an optimal strategy is again the $N$-Choi circuit in that case (see again Appendix~\ref{app:Frey}). 
Note that this was already proven in Ref.~\cite{fujiwara03} for the class $\textup{\textsf{QC-Par}}$. 
We hence see that no advantage of indefinite causal order can be obtained in the metrological task for Pauli channels.
As for depolarizing channels, we conjecture that the above equalities holds for any $N$, and that the corresponding optimal strategy remains the $N$-Choi circuit.

The exact same study for the thermalizing channels defined by Eq.~\eqref{eq:therma_channel} leads again to an identical conclusion. 
For $N=1,2,3$ thermalizing channels $\mathcal{T}_\theta$, we numerically found the optimal QFI to be $N \frac{\epsilon^2 \,e^{-\epsilon/\theta}}{\theta^4\, (1+e^{-\epsilon/\theta})^2}$ for all the classes $\textup{\textsf{\textbf{W}}}$ considered in this work (see Appendix~\ref{app:Mukhopadhyay}).
We conclude that, contrary to the advantage claimed in~\cite{mukhopadhyay}, no advantage of indefinite causal order strategies can be found for $N=1,2,3$. 
The optimal circuit was found to be the parallel circuit generalizing that presented in Fig.~\ref{fig:Parallel} for $N=2$.
We also conjecture that these results hold for any value of $N$, and that the optimal strategy still consists of $N$ thermalizing channels used in parallel, with any input state.

Since no advantage could be obtained for depolarizing, Pauli, or thermalizing channels, we consider further examples of channels in the following subsections.


\subsubsection{Rotation composed with an amplitude damping channel}

Let us consider a $z$-rotation on the Bloch sphere of an angle $\theta$ that we would like to estimate, and which is described by the unitary $U_z(\theta) = e^{-i\theta\sigma_z/2}$.
As it turns out, this choice of channel does not display any gap between the QFI of the different classes. To see a gap, following Ref.~\cite{liu} we then compose the rotation with an amplitude damping channel $\mathcal{A}$, defined by its two Kraus operators:
\begin{align}
    E_0 &= \ketbra{0}{0}+\sqrt{1-u}\ketbra{1}{1}, \nonumber\\
     E_1 &= \sqrt{u}\ketbra{0}{1},
\end{align}
where $u$ is the decay parameter. The composition of the rotation and the amplitude damping channel then defines the channel we consider,
\begin{equation}
    \mathcal{A}_\theta = \mathcal{A} \circ \mathcal{U}_z(\theta), \label{eq:def_Atheta}
\end{equation}
with $\mathcal{U}_z(\theta):\rho \mapsto U_z(\theta) \rho U_z^\dagger(\theta)$, and we will write its corresponding Choi matrix as $A_\theta$. 
It is easy to see that when optimizing over any of the classes under consideration here, the QFI will not depend on the specific value of $\theta$,%
\footnote{Indeed if one changes $\theta \to \theta+\theta'$, one can write $\mathcal{A}_{\theta+\theta'} = \mathcal{A}_\theta \circ \mathcal{U}_z(\theta')$, and ``correct'' the first rotation (with angle $\theta'$) by incorporating its inverse in the process matrices that we optimize over.}
so that one can fix $\theta$ to any given value.

For $N=2$, solving numerically the SDP to calculate the QFI for the different classes we find, for a range of values of $u$, the following hierarchy (where equalities are obtained up to an  error of $10^{-6}$)
\begin{widetext}
 \begin{equation}
       J_{\theta}^{\textup{\textsf{QC-Par}}}(A_\theta^{\otimes 2}) < J_{\theta}^{\textup{\textsf{QC-FO}}}(A_\theta^{\otimes 2}) = \bm{J_{\theta}^{\textup{\textsf{\textbf{QC-CC}}}}(A_\theta^{\otimes 2})} < \bm{J^{\textup{\textsf{\textbf{QC-CC}}}}_{\theta,\textup{\textbf{purif}}}(A_\theta^{\otimes 2})} = J_{\theta}^{\textup{\textsf{QC-Sup}}}(A_\theta^{\otimes 2}) = \bm{J_{\theta}^{\textup{\textsf{\textbf{QC-QC}}}}(A_\theta^{\otimes 2})}=J_{\theta}^{\textup{\textsf{Gen}}}(A_\theta^{\otimes 2})
       \label{eq:hierarchy_N2_AD}
    \end{equation}
\end{widetext}
(except at $u=0,1$, when the hierarchy collapses completely).
The numerical values of the QFI are given in Appendix~\ref{app:hierarchy_AD_data}. 
In this case all the inequalities in Eq.~\eqref{eq:hierarchy_N2_pur} are found to be strict---i.e., we obtain strict separations between the corresponding classes---except for the last one---i.e., more general strategies in $\textup{\textsf{Gen}}$ do not outperform causal superpositions. Since \textup{\textsf{QC-CC}} constitutes the set of causally separable processes for $N=2$, Eq.~\eqref{eq:hierarchy_N2_AD} provides a numerical proof that, for the quantum channels $\mathcal{A}_\theta$, causally nonseparable processes outperform causally separable ones. 

For the $N=3$ scenario, by conducting similar numerical studies we found (again, except for $u=0,1$) the following hierarchy for these channels:
\begin{widetext}
\begin{equation}
       J_\theta^{\textup{\textsf{QC-Par}}}(A_\theta^{\otimes 3})<J_\theta^{\textup{\textsf{QC-FO}}}(A_\theta^{\otimes 3})\leq \underbrace{\bm{J_\theta^{\textup{\textsf{\textbf{QC-CC}}}}(A_\theta^{\otimes 3})}}_{=?} \leq J_\theta^{\textup{\textsf{QC-Sup}}}(A_\theta^{\otimes 3}) = \bm{J^{\textup{\textsf{\textbf{QC-CC}}}}_{\theta,\textup{\textbf{purif}}}(A_\theta^{\otimes 3})} < \bm{J_\theta^{\textup{\textsf{\textbf{QC-QC}}}}(A_\theta^{\otimes 3})}<J_\theta^{\textup{\textsf{Gen}}}(A_\theta^{\otimes 3}).
       \label{eq:hierarchy_N3_AD}
    \end{equation}
\end{widetext}
The numerical results are again provided in Appendix~\ref{app:hierarchy_AD_data}. 
In this case all the inequalities in Eq.~\eqref{eq:hierarchy_N3_dual} are found to be strict except between $J^{\textup{\textsf{QC-Sup}}}(A_\theta^{\otimes 3})$ and $J^{\textup{\textsf{QC-CC}}}_{\theta,\textup{purif}}(A_\theta^{\otimes 3})$, which were found to coincide here.

From Eq.~\eqref{eq:hierarchy_N3_AD}, one sees that QC-QC strategies strictly outperform both QC-CC and QC-Sup strategies. 
Although the optimal QC-QC strategy could be obtained from the primal SDP problem, 
its form is rather complicated so we omit it here.
Since in the $N=3$ case the $\textup{\textsf{QC-CC}}$ class once again coincides with the set of causally separable processes~\cite{wechs19}, Eq.~\eqref{eq:hierarchy_N3_AD} gives a numerical proof of the strict advantage of indefinite causal order strategies over definite causal order strategies in the metrological task we study. 
The fact that QC-QCs strictly outperform QC-Sups shows that the additional features enabled by the more general control structure of QC-QCs---which, compared to QC-Sups, do not ``just'' involve superpositions of fixed causal orders---are potential resources for this metrological task.
Eq.~\eqref{eq:hierarchy_N3_AD} also shows that QC-QC strategies are outperformed by more general causally indefinite strategies in that case. 
Yet, it remains unclear whether such strategies can be described as generalized quantum circuits, or are even consistent with the laws of physics.

As we already discussed earlier, the QFI for $\textup{\textsf{QC-CC}}$ could only be upper-bounded, but not evaluated explicitly. 
We found moreover that the optimal process matrix $\tilde{W}_\textup{\textsf{QC-CC}}^{(\text{opt})} \in \mathcal{L}(\mathcal{H}^{A_\mathcal{N}^{IO}})$, for a given value of $\theta$, obtained numerically by solving the primal SDP problem belonged to $\Tr_F \textup{\textsf{QC-Sup}} \subsetneq \Tr_F \textup{\textsf{QC-CC}}$. 
The purification $W_\textup{\textsf{QC-CC}}^{(\text{opt})} \in \mathcal{L}(\mathcal{H}^{A_\mathcal{N}^{IO}F})$ of this strategy is thus contained in $\textup{\textsf{QC-Sup}}$ and, in principal, one may have $W_{\textup{\textsf{QC-CC}}}^{(\text{opt})} \notin \textup{\textsf{QC-CC}}$ (cf.\ Fig.~\ref{fig:strategies_N3_avecF}).
This is, in fact, the case for the process we found numerically in this case. 
This implies that using purifications to extend strategies that are optimal over $\Tr_F \textup{\textsf{QC-CC}}$ does not provide much insight about the optimal QC-CC here. 
One may hope that alternative methods could be found to extend the optimized process over $\Tr_F \textup{\textsf{QC-CC}}$ on $F$, but we could not find such a procedure giving a process in $\textup{\textsf{QC-CC}}$ as desired. 
It thus remains unclear whether the upper bound $J^{\textup{\textsf{QC-CC}}}_{\theta,\textup{purif}}(A_\theta^{\otimes 3})$ is tight or not in this case.


\subsubsection{Rotation composed with random channels}

Finally, to get some insight on how typical the advantages of the various classes are, we also briefly comment on the hierarchies obtained using random channels defined in the same way as was done in Ref.~\cite{liu}. 
These channels are defined in analogy with the channel $\mathcal{A}_\theta$, replacing the amplitude damping channel by a random noise channel sampled from a set of CP and trace preserving maps.%
\footnote{We generated these random channels numerically using the function ``RandomSuperoperator'' from the \textsc{Matlab} library \textsc{Qetlab}~\cite{qetlab}. They were chosen to be qubit channels with 2 Kraus operators.} 
We again check for equalities up to a numerical precision of $10^{-6}$.

For the case of $N=2$, for 876 of the 1000 random channels sampled we obtained the hierarchy given by Eq.~\eqref{eq:hierarchy_N2_pur}, with all inequalities holding strictly. 
Most of  the remaining cases (120 out of 1000), corresponded to the hierarchy of Eq.~\eqref{eq:hierarchy_N2_AD}, where general strategies from \textup{\textsf{Gen}} provided no advantage over QC-QC and QC-Sup strategies. 
The other 4 cases differed from Eq.~\eqref{eq:hierarchy_N2_pur}~or~\eqref{eq:hierarchy_N2_AD} solely in that \textup{\textsf{QC-Sup}} provided no advantage over \textup{\textsf{QC-FO}}.

For $N=3$, we performed a similar test and found that, for 967 of the 1000 random channels we sampled, the hierarchy obtained was the same as the one already obtained for a rotation composed with an amplitude damping channel (i.e., $\mathcal{A}_\theta$ as defined in Eq.~\eqref{eq:def_Atheta}) and given by Eq.~\eqref{eq:hierarchy_N3_AD}. The 33 remaining cases corresponded to hierarchies different from Eq.~\eqref{eq:hierarchy_N3_AD} in the sense that either \textup{\textsf{QC-QC}} provided no advantage over \textup{\textsf{QC-Sup}} while \textup{\textsf{Gen}} still strictly outperformed \textup{\textsf{QC-QC}}, or both \textup{\textsf{QC-QC}} and \textup{\textsf{Gen}} performed no better than \textup{\textsf{QC-Sup}}. 
Overall, almost all of the random channels tested provided a strict advantage of QC-QC strategies over QC-Sup strategies and hence over any strategy with a definite causal order.


\section{Conclusions}
\label{sec:conclusion}


Causal indefiniteness has attracted significant recent attention as a novel resource in quantum information, often based on applications of the canonical quantum switch~\cite{chirib13}.
A number of recent works have, in particular, attempted to use the quantum switch to provide advantages in quantum metrology~\cite{mukhopadhyay,frey,zhao,Frey21,Chapeau-Blondeau,Chapeau-Blondeau21coherentcontrol,chapeau-blondeau22,chiribella22,kurdzialek,goldberg23,Chapeau_Blondeau_23} in scenarios where the goal is, broadly speaking, to
maximize the precision of estimation of an unknown parameter $\theta$ carried by a quantum channel $\mathcal{C}_\theta$, given $N$ queries to this quantum channel. 
In this paper we critically reassessed some of these claims. 
Taking the results of~\cite{frey,mukhopadhyay} for illustration, we argued that the advantages exhibited cannot be interpreted as being due to indefinite causal order, since the quantum switch was compared only to a single causally ordered protocol, rather than the entire class of causally definite strategies.
As we show, one can in fact obtain better performance for the channels they consider using a simple parallel probing strategy than with the quantum switch. 
The quantum switch is thus, contrary to the initial claims, not a resource for metrology for the depolarizing and thermalizing channels considered in~\cite{frey,mukhopadhyay}.
While we examined explicitly these two results, our general criticism applies also to many of the other claimed advantages~\cite{Frey21,Chapeau-Blondeau,Chapeau-Blondeau21coherentcontrol,chapeau-blondeau22,Chapeau_Blondeau_23}.
Similar criticisms have also been recently raised regarding purported advantages of the quantum switch in various thermodynamic tasks~\cite{capela23}.

To provide a more rigorous study of metrological advantages arising from causal indefiniteness, we considered the framework recently introduced in~\cite{Altherr,liu} that allows one to optimize the metrological performance over whole classes of strategies with different properties using SDPs.
We extended this framework to include the most general known classes of physically realizable circuits with definite (\textup{\textsf{QC-CC}}) and indefinite (\textup{\textsf{QC-QC}}) causal order~\cite{wechs21}.
These classes notably allow for dynamical control of the order in which the channels are probed, or queried, allowing this to be determined on the fly, either incoherently (for QC-CCs) or coherently in a quantum superposition (for QC-QCs).
We were thus able to study whether such dynamical control is useful in metrology, and to compare a wide range of physically pertinent classes of strategies: parallel (\textup{\textsf{QC-Par}}), fixed order (\textup{\textsf{QC-FO}}), and classically controlled (\textup{\textsf{QC-CC}}) strategies, all which are causally definite, and causal superpositions (\textup{\textsf{QC-Sup}}), quantum control (\textup{\textsf{QC-QC}}) and general strategies (\textup{\textsf{Gen}}), which also include causally indefinite processes.

We studied the performance of these strategies with two and three queries to the channels $\mathcal{C}_\theta$ for several families of channels: general Pauli channels, thermalizing channels, a rotation composed with an amplitude damping channel, and for a rotation composed with different random channels.
For Pauli and thermalizing channels, we found that indefinite causal order provides no advantage, while for the other families \textup{\textsf{QC-Sup}}, \textup{\textsf{QC-QC}} and \textup{\textsf{Gen}} all provided advantages over causally definite strategies.
With three queries, in particular, we saw a strict hierarchy, with QC-QCs providing an advantage over QC-Sups which, in turn, outperformed QC-CCs.
We thus extended the hierarchies obtained in~\cite{liu} and showed that causal indefiniteness even beyond QC-Sups can provide advantages in metrology over the most general causally definite strategies.

As we saw, causally indefinite strategies are useful when dealing with some quantum channels, but for other families they provided no advantage.
It would be interesting to understand further when advantages can be obtained, and what properties of the channels $\mathcal{C}_\theta$ are behind such differences.
All the channels we studied here were non-unitary, which may appear a key omission given the prevalence of unitary channels in quantum metrological problems.
The recent results of~\cite{abbott23} show, however, that QC-QCs cannot provide advantages over QC-FOs in any task (even beyond metrology) in which one queries a single unitary $N$ times.
In the type of metrological task we considered here they hence cannot provide an advantage. 
Nonetheless, in more complex settings where several unitary channels need to be probed (as considered for instance in~\cite{zhao}), advantages may still be possible.
It would nonetheless be interesting to also see whether the class \textup{\textsf{Gen}} of general strategies, or some pertinent subclass (e.g., that of purifiable processes~\cite{Araujo2017purification}) can lead to advantages for unitary channels using causally indefinite strategies, even if the physical interpretation of such strategies remains unclear.
More generally, it would be interesting to consider the utility of causally indefinite strategies beyond the simple metrological task we consider here, for example looking at multi-parameter estimation, where more than one parameter must be estimated, or multi-channel parameter estimation, where different channels that depend on the same parameter must be probed.

Further work also remains to improve the approach used to compute the optimal QFI for $\textup{\textsf{QC-CC}}$, which is not closed under purifications, either by finding an alternative approach that computes exactly the optimal QFI over the class, or by finding better upper bounds on it. 
Indeed, given that in the metrological task at hand one uses $N$ copies of identical channels and, in any given run of a QC-CC strategy, the channels are combined in some fixed order, one may hope to show that QC-CCs provide no advantages over QC-FOs.
Even if the approach here was sufficient for our purposes and indeed allowed us to show some advantages of $\textup{\textsf{QC-QC}}$ over $\textup{\textsf{QC-CC}}$ and $\textup{\textsf{QC-Sup}}$ in particular, this would help to clarify the hierarchy between $\textup{\textsf{QC-CC}}$ and $\textup{\textsf{QC-Sup}}$ when $N \geq 3$.

Finally, one may wonder about the physical relevance of the class $\textup{\textsf{QC-Sup}}$, both in and beyond quantum metrology.
Causally indefinite strategies in $\textup{\textsf{QC-Sup}}$ exploit only ``static'' causal superpositions as in the quantum switch (as opposed to dynamical control of causal order). 
A recent work found that, in the asymptotic regime, causal superpositions provide no advantages over fixed order strategies~\cite{kurdzialek}. 
It would hence be interesting to understand whether QC-QCs, which allow for dynamical control of the causal order, could provide asymptotic advantages in such a setting.
More generally, it seems that $\textup{\textsf{QC-Sup}}$ fails to fully capture all processes with ``static'' causal superpositions.
Indeed, already by modifying slightly the description of the quantum switch and considering that the final control system is given as an input to a third operation, rather than to the future space $\mathcal{H}^F$, one obtains a process that formally is not in \textup{\textsf{QC-Sup}} yet conceptually is identical to the quantum switch.
This problem of formalizing more rigorously the distinction between static and dynamical causal order is the subject of a future work in preparation~\cite{dynamical_orders_in_prep}. This, in particular, could help clarify the origin of the advantage observed here of $\textup{\textsf{QC-QC}}$ over $\textup{\textsf{QC-Sup}}$.


\begin{acknowledgments}
We thank Qiushi Liu and Yuxiang Yang for helpful discussions and comments on this work. 

This research was funded in part by l’Agence Nationale de la Recherche (ANR) projects ANR-15-IDEX-02 and ANR-22-CE47-0012, and the PEPR integrated project EPiQ ANR-22-PETQ-0007 as part of Plan France 2030. For the purpose of open access, the authors have applied a CC-BY public copyright license to any Author Accepted Manuscript (AAM) version arising from this submission.
\end{acknowledgments}


\appendix


\section{Processes with definite causal order outperforming the quantum switch}
\label{app:Frey_Mukhopadhyay}


\subsection{Depolarizing channels}
\label{app:Frey}

Ref.~\cite{frey} considered depolarizing channels $\mathcal{D}_\theta$ of the form of Eq.~\eqref{eq:depo_channel} and compared the performance of the quantum switch, versus a simple use of the channels in sequence, in estimating the parameter $\theta$.

For $N=2$ uses of the channels, the quantum switch is described by the process matrix $W^{\text{QS}}$ of Eq.~\eqref{eq:PM_QS}, while the process that prepares some (pure) input state $\ket{\psi_{\text{in}}}$ and directly applies the two copies of $\mathcal{D}_\theta$ one after the other is described as the process matrix $W^{\text{Seq}} = \ketbra{\psi_{\text{in}}}{\psi_{\text{in}}}^{A_1^I}~\otimes~\kketbra{\id}{\id}^{A_1^OA_2^I}~\otimes~\kketbra{\id}{\id}^{A_2^OF}$. 
From these process matrices one can explicitly write the output state of the circuit in each case, obtained as $D_\theta^{\otimes 2}*W^{\text{QS}}$ and $D_\theta^{\otimes 2}*W^{\text{Seq}}$, respectively. 
As these have relatively simple expressions,%
\footnote{Specifically, one finds $D_\theta^{\otimes 2}*W^{\text{QS}} = \theta^2 \, \ketbra{+}{+}^c\otimes\ketbra{\psi}{\psi}^t + 2\theta(1-\theta)\, \ketbra{+}{+}^c\otimes\frac{\id^t}{d} + (1-\theta)^2 \, \frac{\id^c}{2}\otimes\frac{\id^t}{d} + (1-\theta)^2 \, \frac{\ketbra{0}{1}^c+\ketbra{1}{0}^c}{2d^2}\otimes\ketbra{\psi}{\psi}^t$ and $D_\theta^{\otimes 2}*W^{\text{Seq}} = \theta^2 \, \ketbra{\psi}{\psi} + (1-\theta^2) \, \frac{\id}{d}$ (the latter could of course have been obtained more directly).}
for which an eigendecomposition can easily be found, one can calculate the QFI directly using Eq.~\eqref{eq:QFI_simple}. 
For the case of qubit ($d=2$) depolarizing channels, we thus find%
\footnote{One could similarly obtain the (quite cumbersome) expression of the QFI for any dimension $d$, and recover the result given in~\cite{frey}. The facts that the QFI does not depend on the (pure) input states and that $J_\theta(D_\theta^{\otimes 2}*W^{\text{QS}}) > J_\theta(D_\theta^{\otimes 2}*W^{\text{Seq}})$ hold for any dimension.}

\begin{align}
    &J_\theta(D_\theta^{\otimes 2}*W^{\text{QS}}) = \frac{8(1 + 3 \theta^2)}{(1-\theta)(1+3\theta)(3+2\theta+3\theta^2)}, \\
    &J_\theta(D_\theta^{\otimes 2}*W^{\text{Seq}})  
    = \frac{4\theta^2}{1-\theta^4},
\end{align}
for any input state $\ket{\psi_{\text{in}}}$ in both circuits, in agreement with~\cite{frey}. 
One can then show that $J_\theta(D_\theta^{\otimes 2}*W^{\text{QS}}) > J_\theta(D_\theta^{\otimes 2}*W^{\text{Seq}})$, $\forall\, \theta \in [0,1)$, from which Ref.~\cite{frey} concluded that the quantum switch provides a metrological advantage with respect to the sequential strategy.

\begin{figure}[htbp]
     \centering
         \includegraphics[width=0.5\textwidth]{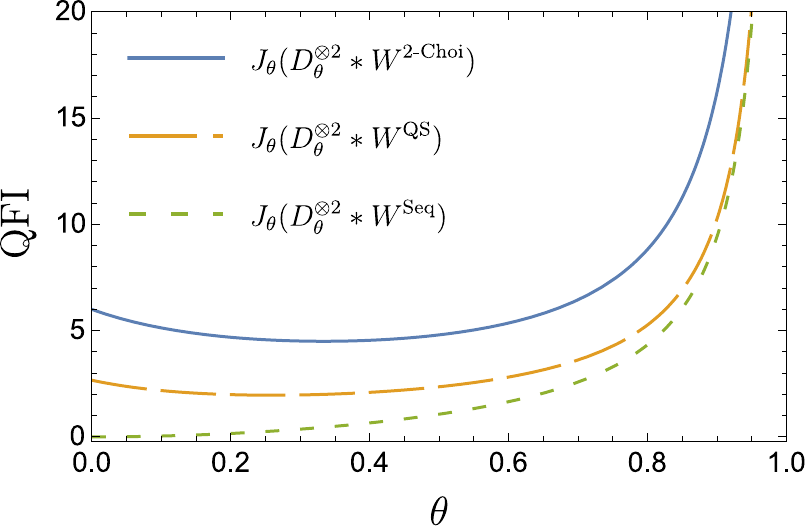}
         \caption{QFI of the output state of the quantum switch, of the sequential and of the 2-Choi strategies in the case of $N=2$ qubit ($d=2$) depolarizing channels, as defined in Eq.~\eqref{eq:depo_channel}.
         }
        \label{fig:QFI_Frey}
\end{figure}

Consider now the ``2-Choi circuit'' of Fig.~\ref{fig:2TQOD}.%
\footnote{For completeness, the process matrix of the $N$-Choi circuit can be written as $W^{N\text{-Choi}} = \bigotimes_{k=1}^N \ketbra{\phi^+}{\phi^+}^{A_k^IF_k}\otimes\ketbra{\id}{\id}^{A_k^OF_k'}$ (with a global future space $\mathcal{H}^F = \bigotimes_{k=1}^N \mathcal{H}^{F_kF_k'}$).}
Since it consists in two independent uses of the Choi circuit, whose output state is simply (up to a proportionality factor of $\frac{1}{d}$) the Choi matrix of the probed channel, by the additivity of the QFI one has $ J_\theta(D_\theta^{\otimes 2}*W^{\text{2-Choi}}) =  2J_\theta(\frac{1}{d}D_\theta) =  2J_\theta(\theta\ketbra{\phi^+}{\phi^+} + (1-\theta) \frac{\id}{d^2})$. Using Eq.~\eqref{eq:QFI_simple}, for the case of qubits ($d=2$) we find
\begin{equation}
    J_\theta(D_\theta^{\otimes 2}*W^{\text{2-Choi}}) = \frac{6}{(1-\theta)(1+3\theta)}.
\end{equation}
It is then straightforward to show that $J_\theta(D_\theta^{\otimes 2}*W^{\text{2-Choi}}) > J_\theta(D_\theta^{\otimes 2}*W^{\text{QS}})$, $\forall\, \theta \in [0,1)$, as can also be seen in Fig.~\ref{fig:QFI_Frey}. 
Hence, one cannot claim that the quantum switch has any advantage over general definite causal order strategies in this case.

Note that for $N$ qubit depolarizing channels, using again the additivity of the QFI, we readily obtain%
\footnote{More generally, for some arbitrary dimension $d$ one has $J_\theta(D_\theta^{\otimes N}*W^{\text{N-Choi}}) = \frac{(d^2-1)N}{(1-\theta)(1+(d^2-1)\theta)}$.
Note that Ref.~\cite{frey} also considered a ``quantum $N$-switch'' that superposes two reverse orders for $N$ operations. The value of the QFI obtained for that $N$-switch is again smaller than that of the $N$-Choi circuit given just above. (The analytical comparison being quite tedious, we verified this numerically for all values of $N,d \leq 100$; the observed behavior convinces us that this remains true for all values of $N,d$.)}
\begin{equation}
    J_\theta(D_\theta^{\otimes N}*W^{N\text{-Choi}}) = \frac{3N}{(1-\theta)(1+3\theta)}, \label{eq:J_N_depol}
\end{equation}
which is precisely the optimal QFI found numerically in Eq.~\eqref{eq:hierarchy_depol} for $N=1,2,3$ qubit depolarizing channels.

\subsection{Thermalizing channels}
\label{app:Mukhopadhyay}

Rather than depolarizing channels, Ref.~\cite{mukhopadhyay} considered the case with $N=2$ copies of a qubit thermalizing channel $\mathcal{T}_\theta$, as defined from the Kraus operators of Eq.~\eqref{eq:therma_channel} in the limit of infinite interaction time---in which case the channel reduces to the constant channel $\rho \mapsto \uptau_\theta:=p_\theta \ketbra{0}{0} + (1-p_\theta) \ketbra{1}{1}$, with $p_\theta = 1/(1+e^{-\epsilon/\theta})$.

When plugging the two copies of $\mathcal{T}_\theta$ into quantum switch, the output state can be computed from the process matrix $W^{\text{QS}}$ given in Eq.~\eqref{eq:PM_QS}.%
\footnote{Specifically, one finds $T_\theta^{\otimes 2}*W^{\text{QS}} = \frac{\id^c}{2}\otimes\uptau_\theta + \frac{\ketbra{0}{1}^c+\ketbra{1}{0}^c}{2}\otimes\uptau_\theta\ketbra{\psi}{\psi}\uptau_\theta$.}
It is found to depend on the initial state $\ket{\psi_{\text{in}}}$ of the target system; as in~\cite{mukhopadhyay}, we shall take it to be $\ket{\psi_{\text{in}}} = \ket{0}$.
As for the output of the basic sequential composition of the two copies, it is trivially equal to the constant output state $\uptau_\theta$ of the channel itself. From an eigendecomposition of these two output states, using Eq.~\eqref{eq:QFI_simple} with $\partial_\theta \rho_\theta = (\partial_{p_\theta} \rho_\theta)(\partial_\theta p_\theta)$, we find the corresponding QFIs to be
\begin{align}
J_\theta(T_\theta^{\otimes 2}*W^{\text{QS}}) &= \left(\frac{1}{p_\theta(1-p_\theta^2)}+\frac{1}{1-p_\theta}\right) \left(\partial_\theta p_\theta\right)^2, \\
J_\theta(T_\theta^{\otimes 2}*W^{\text{Seq}})  &= J_\theta(\uptau_\theta) = \left(\frac{1}{p_\theta}+\frac{1}{1-p_\theta}\right) \left(\partial_\theta p_\theta\right)^2,
\end{align}
which agrees with the result obtained in~\cite{mukhopadhyay}, and from which one clearly sees that $J_\theta(T_\theta^{\otimes 2}*W^{\text{QS}}) > J_\theta(T_\theta^{\otimes 2}*W^{\text{Seq}})$ for all $\theta>0$, thus suggesting some advantage for the quantum switch compared to the sequential strategy.

 \begin{figure}[t]
      \centering
          \includegraphics[width=0.5\textwidth]{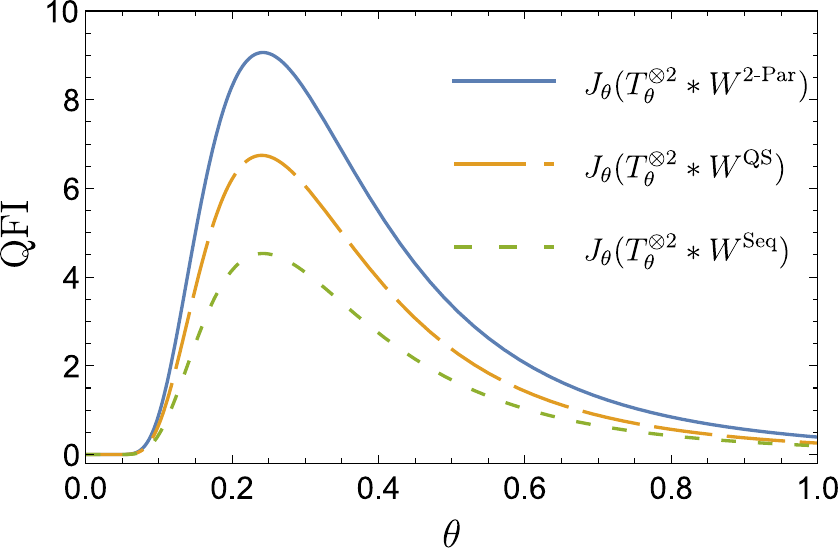}
          \caption{QFI of the output state of the quantum switch, of the sequential and of the parallel strategies in the case of $N=2$ thermalizing channels, as defined through Eq.~\eqref{eq:therma_channel}, in the limit $t\to\infty$. }
         \label{fig:QFI_Mukhoadhyay.pdf}
 \end{figure}

However, since one considers here two copies of the constant channel $\mathcal{T}_\theta$, it is more efficient to use them in parallel than in series, as in the process depicted in Fig.~\ref{fig:Parallel} (with its process matrix denoted $W^{2\text{-Par}}$), so as to have access to two copies of the output state, $\uptau_\theta^{\otimes 2}$ rather than just $\uptau_\theta$. By the additivity of the QFI, one gets
\begin{align}
J_\theta(T_\theta^{\otimes 2}*W^{2\text{-Par}})  &= J_\theta(\uptau_\theta^{\otimes 2}) =2 J_\theta(\uptau_\theta) \notag \\
& = 2\left(\frac{1}{p_\theta}+\frac{1}{1-p_\theta}\right) \left(\partial_\theta p_\theta\right)^2.
\end{align}
One can easily verify that $J_\theta(T_\theta^{\otimes 2}*W^{2\text{-Par}}) > J_\theta(T_\theta^{\otimes 2}*W^{\text{QS}})$, as also seen on Fig.~\ref{fig:QFI_Mukhoadhyay.pdf}. Thus, the parallel strategy outperforms the quantum switch for the present task, and no advantage of the quantum switch versus the general class of processes with definite causal order can be claimed here.

For $N$ thermalizing channels, the parallel strategy (with process matrix $W^{N\text{-Par}}$) gives
\begin{align}
J_\theta(T_\theta^{\otimes N}*W^{N\text{-Par}})  &= N\left(\frac{1}{p_\theta}+\frac{1}{1-p_\theta}\right) \left(\partial_\theta p_\theta\right)^2 \notag \\
& = N \frac{\epsilon^2 \,e^{-\epsilon/\theta}}{\theta^4\, (1+e^{-\epsilon/\theta})^2}.
\end{align}
This expression is precisely the optimal QFI for all the classes considered in this work, found numerically for $N=1,2,3$ thermalizing channels; see Sec.~\ref{subsubsec:examples1}.
(To obtain the numerical results we fixed $\epsilon = 1$; the result for generic values of $\epsilon$ is obtained by rescaling $\theta \to \theta/\epsilon$.)


\section{Computing the QFI and reconstructing the optimal process through SDP optimization}
\label{app:details_approach}


In this appendix we clarify further and complete the arguments for the approach we use---which was introduced in Refs.~\cite{Altherr,liu}, and which we only sketched in Sec.~\ref{sec:SDP}---to compute the QFI and reconstruct the optimal process in the metrological task at hand.


\subsection{Step-by-step description of the approach}
\label{app:details_step_by_step}

Recall that our aim is to optimize the QFI of the state that is output by the composition of the $N$ $\theta$-dependent channels $\cal{C}_\theta$ (represented by their Choi matrix $C_\theta$) with a process $W$ from a given (compact) class $\textsf{\textbf{W}}$, as given in Eq.~\eqref{eq:max_QFI_class} of the main text. That is,
\begin{equation}
    J_\theta^{\textup{\textsf{\textbf{W}}}}(C_\theta^{\otimes N}) = \max_{W\in \textup{\textsf{\textbf{W}}}} J_\theta(C_\theta^{\otimes N} * W), \label{eq:max_QFI_class_app}
\end{equation}
where the QFI $J_\theta$ of the output state $\rho_\theta = C_\theta^{\otimes N} * W$ can be computed, using Eq.~\eqref{eq:QFI_Fujiwara}, as
\begin{equation}
J_\theta(\rho_\theta)=4 \min_{\{\ket{\psi_{\theta,j}}\}_{j=1}^q}\Tr\left(\sum_{j=1}^q\ketbra{\dot{\psi}_{\theta,j}}{\dot{\psi}_{\theta,j}}\right),
\label{eq:QFI_Fujiwara_app}
\end{equation}
with the minimum being taken over all ensemble decompositions $\{\ket{\psi_{\theta,j}}\}_{j=1}^q$ of $\rho_\theta$ (that are assumed to be continuously differentiable, with $\ket{\dot{\psi}_{\theta,j}} = \partial_\theta\ket{\psi_{\theta,j}}$), of any given size $q \ge \max_\theta\rank \rho_\theta$.


\subsubsection{Restricting to pure processes}
\label{app_subsubsec:restrict_pure}

The first step in the approach is to restrict oneself to ``pure'' processes\footnote{Here by ``pure'' processes we simply refer to rank-1 processes, rather than to the more restrictive notion proposed in Ref.~\cite{Araujo2017purification}.} of the form $W = \ketbra{w}{w}$.

For many classes $\textsf{\textbf{W}}$ of interest, restricting to pure processes does not induce any loss of generality. 
This is the case when (apart from the requirement that $W\ge 0$) the conditions for a given process matrix $W$ to belong to a class $\textsf{\textbf{W}}$ only impose constraints on its partial trace $\Tr_F W$ directly, rather than constraints on the full $W$.
This is indeed the case for all the classes considered previously in~\cite{Altherr,liu}, as well as for the $\textsf{QC-QC}$ class that we also consider in this work; see the characterizations of Eqs.~\eqref{eq:QC-Par}, \eqref{eq:QC-FO}, \eqref{eq:QC-Sup}, \eqref{eq:ICO} and \eqref{eq:QC-QC}.
To see that one can indeed safely restrict oneself to pure processes for such classes, note that any ``mixed'' process matrix $W \in {\cal L}({\cal H}^{A_{\cal N}^{IO}F})$ in such a class $\textsf{\textbf{W}}$ can be ``purified'' into $W' = \ketbra{w'}{w'} \in {\cal L}({\cal H}^{A_{\cal N}^{IO}FF'})$ by introducing some auxiliary system $F'$, such that $\Tr_{F'} W' = W$. Then considering the global future system of $W'$ to be $FF'$, $\Tr_{FF'} W' = \Tr_F W$ also satisfies the required constraints for $W'$ to be in $\textsf{\textbf{W}}$ (recall that our definitions of the classes $\textsf{\textbf{W}}$ leaves open the specification of the future space).
Noting, further, that the QFI of the output state can only decrease by partially tracing out $F'$, i.e., $J_\theta(C_\theta^{\otimes N} * \ketbra{w'}{w'}) \ge J_\theta(C_\theta^{\otimes N} * W)$, it follows that it is indeed sufficient to consider pure processes to obtain the maximum in Eq.~\eqref{eq:max_QFI_class_app}.

For some other classes however (such as $\textsf{QC-CC}$ class that we also consider here), the purification argument does not work so straightforwardly: purifying a process matrix may in general take us outside of the class being considered into a larger one.
We still consider such purifications here, but the approach we use then only allows us in this case to obtain the optimal QFI in this larger class---which thus only provides an \emph{upper bound} (that we denote by $J_{\theta,\textup{purif}}^{\textup{\textsf{\textbf{W}}}}(C_\theta^{\otimes N})$) on the QFI of the class we initially consider.


\subsubsection{Trading the optimization over decompositions $\{\ket{\psi_{\theta,j}}\}_j$ of $\rho_\theta$ for an optimization over decompositions $\{\ket{C_{\theta,j}}\}_j$ of $C_\theta^{\otimes N}$}
\label{app_subsubsec:optim_decomp_C}

Once we restrict ourselves to pure processes $W = \ketbra{w}{w}$, it can be seen that for any fixed $q \ge \max_\theta\rank C_\theta^{\otimes N}$, the ensemble decompositions of size $q$ of $C_\theta^{\otimes N}$ and the ensemble decompositions of size $q$ of $\rho_\theta$, with $\rho_\theta = C_\theta^{\otimes N} * \ketbra{w}{w}$, are in one-to-one correspondence through the (``pure''\footnote{The link product, which was introduced in the ``mixed case'' (for matrices) in Eq.~\eqref{eq:def_link_prod} of the main text, can also be defined in the ``pure case'' as follows~\cite{wechs21}: for two vectors $\ket{a}\in \mathcal{H}^{XY}$ and $\ket{b}\in \mathcal{H}^{YZ}$ that are connected through a common system $Y$,
\begin{align}
    \ket{a} * \ket{b} := & \big( \id^{XZ} \otimes \bbra{\id}^{YY} \big) \big( \ket{a}^{XY} \otimes \ket{b}^{YZ} \big) \notag \\
    = & (\ket{a}^{T_Y}\otimes\id^Z)(\id^X\otimes\ket{b}) \quad \in \mathcal{H}^{XZ}.
\end{align}}) link product with $\ket{w}$---in the sense that for any ensemble decomposition $\{\ket{C_{\theta,j}}\}_{j=1}^q$ of $C_\theta^{\otimes N}$ of size $q$, there exists an ensemble decomposition $\{\ket{\psi_{\theta,j}}\}_{j=1}^q$ of $\rho_\theta$ of the same size $q$, such that $\ket{\psi_{\theta,j}} = \ket{C_{\theta,j}} * \ket{w} \ \forall\, j$, and vice versa.

The forward statement is trivial (using $\rho_\theta = C_\theta^{\otimes N} * W = \sum_{j=1}^q \ketbra{C_{\theta,j}}{C_{\theta,j}} * \ketbra{w}{w}$), and is obtained precisely by taking $\ket{\psi_{\theta,j}} = \ket{C_{\theta,j}} * \ket{w}$.

The reverse direction, on the other hand, is not so straightforward, because the link product is not trivial to invert~\cite{wechs21}. To see that it also holds, note first that since we consider $q \ge \max_\theta\rank C_\theta^{\otimes N}$, then there exists an ensemble decomposition $\{\ket{C_{\theta,j}^0}\}_{j=1}^q$ of $C_\theta^{\otimes N}$, of size $q$, which also gives (as above) an ensemble decomposition $\{\ket{\psi_{\theta,j}^0}\}_{j=1}^q$ of $\rho_\theta$ with $\ket{\psi_{\theta,j}^0} = \ket{C_{\theta,j}^0} * \ket{w}$. Consider then any other ensemble decomposition $\{\ket{\psi_{\theta,j}}\}_{j=1}^q$ of $\rho_\theta$, of size $q$. As $\{\ket{\psi_{\theta,j}}\}_{j=1}^q$ and $\{\ket{\psi_{\theta,j}^0}\}_{j=1}^q$ are ensemble decompositions of the same state and of the same size $q$, it follows from the HJW theorem~\cite{Hughston} (see also the next subsection for a similar argument) that the $\ket{\psi_{\theta,j}}$'s can be obtained from the $\ket{\psi_{\theta,j}^0}$'s as $\ket{\psi_{\theta,j}} = \sum_{k=1}^q U_{\theta,k,j} \ket{\psi_{\theta,k}^0}$, where the $U_{\theta,k,j}$'s are the coefficients of a $q\times q$ unitary. We then have $\ket{\psi_{\theta,j}} = \sum_{k=1}^q U_{\theta,k,j} \ket{C_{\theta,k}^0} * \ket{w} = \ket{C_{\theta,j}} * \ket{w}$ with $\ket{C_{\theta,j}} = \sum_{k=1}^q U_{\theta,k,j} \ket{C_{\theta,k}^0}$. It is then easy to verify that $\{\ket{C_{\theta,j}}\}_{j=1}^q$ thus defined indeed provides an ensemble decomposition of size $q$ of $C_\theta^{\otimes N}$.

There is a delicate issue here relating to the differentiability of the decompositions~\cite{fujiwara08}. We assume in this work that the channels ${\mathcal C}_\theta$ have a smooth enough dependency on $\theta$, so that starting from an ensemble decomposition of $\rho_\theta$ in which the $\ket{\psi_{\theta,j}}$'s are continuously differentiable, the decomposition of $C_\theta^{\otimes N}$ constructed above can also be taken such that the $\ket{C_{\theta,j}}$'s are continuously differentiable (we colloquially just write ``smooth'').%
\footnote{Admittedly, the exact requirements that this imposes on the channels ${\mathcal C}_\theta$ should still be clarified (as for instance in Ref.~\cite{fujiwara08}, where the families of quantum channels $\{{\mathcal C}_\theta\}_\theta$ are required to have continuously differentiable decompositions and to be ``piecewise regular''). This is however rather involved, and goes beyond the scope of our work.}
As the correspondence between the two decompositions is through $\ket{\psi_{\theta,j}} = \ket{C_{\theta,j}} * \ket{w}$, we can then write $\ket{\dot{\psi}_{\theta,j}} = \ket{\dot{C}_{\theta,j}} * \ket{w}$, and we can turn the optimization over $\{\ket{\psi_{\theta,j}}\}_{j=1}^q$ in Eq.~\eqref{eq:QFI_Fujiwara_app} into an optimization over the decompositions $\{\ket{C_{\theta,j}}\}_{j=1}^q$ of the channels' operator $C_\theta^{\otimes N}$ directly, so as to obtain
\begin{align}
J_\theta(\rho_\theta)&=4 \min_{\{\ket{C_{\theta,j}}\}_{j=1}^q}\Tr\Bigg(\Bigg(\sum_{j=1}^q\ketbra{\dot{C}_{\theta,j}}{\dot{C}_{\theta,j}}\Bigg) * W\Bigg) \notag \\
& = 4 \min_{\{\ket{C_{\theta,j}}\}_{j=1}^q}\Tr\Bigg(\!\Bigg(\sum_{j=1}^q\ketbra{\dot{C}_{\theta,j}}{\dot{C}_{\theta,j}}\Bigg)^{\!T} \tilde{W} \Bigg), \label{eq:min_decomp_Ctheta_j}
\end{align}
where in the second line we evaluated the link product and defined $\tilde{W} := \Tr_F W$.
Note at this point that the QFI of $\rho_\theta$ only depends on $\tilde{W}$, after tracing out the future space of the process $W$: indeed to actually obtain $\rho_\theta$ one should re-introduce that future space by purifying $\tilde{W}$ (see Sec.~\ref{app_subsubsec:restrict_pure} above), but all purifications are equivalent in the sense that they give the same QFI.


\subsubsection{Trading the optimization over $\{\ket{C_{\theta,j}}\}_j$ for an optimization over a unitary $V_\theta$}
\label{app_subsubsec:optim_V}

The next step is to write any ensemble decomposition $\{\ket{C_{\theta,j}}\}_{j=1}^q$ in terms of a fixed, ``reference'' decomposition $\{\ket{C_{\theta,j}^0}\}_{j=1}^q$ and a unitary matrix $V_\theta$.

Consider any fixed (smooth) ensemble decomposition $\{\ket{C_{\theta,j}^0}\}_{j=1}^q$ of $C_\theta^{\otimes N}$ (of size $q \ge \max_\theta\rank C_\theta^{\otimes N}$), and some other (smooth) decomposition $\{\ket{C_{\theta,j}}\}_{j=1}^q$ that we want to optimize. For convenience, let us denote by $\mathbf C_\theta^0 = \big(\ket{C_{\theta,1}^0},\ldots,\ket{C_{\theta,q}^0}\big)$ and $\mathbf C_\theta = \big(\ket{C_{\theta,1}},\ldots,\ket{C_{\theta,q}}\big)$ the $d_{A_{\cal N}^{IO}} \times q$ matrices whose column vectors are the elements of the two decompositions $\{\ket{C_{\theta,j}^0}\}_{j=1}^q$ and $\{\ket{C_{\theta,j}}\}_{j=1}^q$, respectively.
Since these are decompositions of the same operator and of the same size, it again follows from the HJW theorem\footnote{See Footnote~2 in the Supplemental Material of Ref.~\cite{Altherr} for a more detailed argument.}~\cite{Hughston} that the matrices $\mathbf C_\theta^0$ and $\mathbf C_\theta$ are related through $\mathbf C_\theta = \mathbf C_\theta^0 V_\theta$, for some $q\times q$ unitary $V_\theta$. We again assume that for two smooth decompositions $\{\ket{C_{\theta,j}^0}\}_{j=1}^q$ and $\{\ket{C_{\theta,j}}\}_{j=1}^q$ of $C_\theta^{\otimes N}$, the unitary $V_\theta$ that relates them can be taken to be smooth.

Conversely, given the fixed $\mathbf C_\theta^0$ and any (smooth) $q\times q$ unitary $V_\theta$, it is easily verified that the column vectors of $\mathbf C_\theta = \mathbf C_\theta^0 V_\theta$ provide an alternative (smooth) ensemble decomposition of $C_\theta^{\otimes N}$.

Hence, once a reference decomposition $\{\ket{C_{\theta,j}^0}\}_{j=1}^q$ is fixed, the optimization over $\{\ket{C_{\theta,j}}\}_{j=1}^q$ in Eq.~\eqref{eq:min_decomp_Ctheta_j} can be replaced by an optimization over such ($\theta$-dependent, smooth) $q\times q$ unitaries $V_\theta$.
Noting that $\sum_{j=1}^q \ketbra{\dot C_{\theta,j}}{\dot C_{\theta,j}} = \dot{\mathbf C}_\theta \dot{\mathbf C}_\theta^\dagger$, with $\dot{\mathbf C}_\theta = \dot{\mathbf C}_\theta^0 V_\theta + \mathbf C_\theta^0 \dot{V}_\theta = (\dot{\mathbf C}_\theta^0 + \mathbf C_\theta^0 \dot{V}_\theta V_\theta^\dagger ) V_\theta$, we obtain
\begin{equation}
J_\theta(\rho_\theta)=4 \min_{V_\theta}\Tr\!\left(\!\! \Big(\! (\dot{\mathbf C}_\theta^0 \!+\! \mathbf C_\theta^0 \dot{V}_\theta V_\theta^\dagger ) (\dot{\mathbf C}_\theta^0 \!+\! \mathbf C_\theta^0 \dot{V}_\theta V_\theta^\dagger )^\dagger \!\Big)^{\!T} \tilde{W}\!\right)\!. \label{eq:min_Vtheta}
\end{equation}


\subsubsection{Trading the optimization over $V_\theta$ for an optimization over a Hermitian matrix $h$}
\label{app_subsubsec:optim_h}

Looking more closely at Eq.~\eqref{eq:min_Vtheta}, one then sees that the objective function only depends on the $q\times q$ Hermitian matrix $h_\theta = i \dot{V}_\theta V_\theta^\dagger$.%
\footnote{The fact that $h_\theta$ is Hermitian follows directly from $\partial_\theta(V_\theta V_\theta^\dagger) = 0$.}

Conversely, from any (in general, $\theta$-dependent and smooth) $q\times q$ Hermitian matrix $h_\theta$, one can define a (smooth) $q\times q$ unitary $V_\theta$ that satisfies $h_\theta = i \dot{V}_\theta V_\theta^\dagger$, by taking $V_\theta = e^{-i H_\theta}$, where $H_\theta$ is an antiderivative (with respect to $\theta$) of $h_\theta$.

Hence, the optimization over $V_\theta$ in Eq.~\eqref{eq:min_Vtheta} is equivalent to an optimization over $q\times q$ Hermitian matrices $h_\theta$. Since the dependency of $h_\theta$ on $\theta$ no longer enters the objective function---e.g., the derivative of $h_\theta$ with respect to $\theta$ does not appear---from now on (and as we did in the main text) we simply drop the $\theta$ subscript in $h_\theta$.
Introducing the so-called ``performance operator''~\cite{Chiribella_2016,Altherr}
\begin{equation}
\Omega_\theta(h) = 4 \left( (\dot{\mathbf C}_\theta^0 - i \mathbf C_\theta^0 h ) (\dot{\mathbf C}_\theta^0 - i \mathbf C_\theta^0 h )^\dagger \right)^T, \label{eq:def_Omega_app}
\end{equation}
we now obtain
\begin{equation}
J_\theta(\rho_\theta)= \min_{h\in\mathbb{H}_q}\Tr\left( \Omega_\theta(h) \, \tilde{W}\right). \label{eq:min_htheta}
\end{equation}


\subsubsection{Swapping $\max_{\tilde W}$ and $\min_h$}
\label{app_subsubsec:swap_max_min}

Injecting now Eq.~\eqref{eq:min_htheta} into Eq.~\eqref{eq:max_QFI_class_app}, the QFI $J_\theta^{\textup{\textsf{\textbf{W}}}}(C_\theta^{\otimes N})$ is obtained as the result of a max-min optimization problem---with the maximization being now over $\tilde{W} = \Tr_F W \in \Tr_F\textsf{\textbf{W}}$. As the objective function $f(h,\tilde{W}) = \Tr\big( \Omega_\theta(h) \, \tilde{W}\big)$ is convex in $h$ and concave (even linear) in $\tilde{W}$, and as \textsf{\textbf{W}} (and hence $\Tr_F\textsf{\textbf{W}}$) is assumed to be compact, then by Fan's minimax theorem~\cite{fan53} one can swap $\max_{\tilde W}$ and $\min_h$, and write
\begin{align}
    J_\theta^{\textup{\textsf{\textbf{W}}}}(C_\theta^{\otimes N}) & = \max_{\tilde{W}\in \Tr_F\textup{\textsf{\textbf{W}}}} \ \min_{h\in\mathbb{H}_q} \ \Tr\left( \Omega_\theta(h) \, \tilde{W}\right) \notag \\
     & = \min_{h\in\mathbb{H}_q} \ \max_{\tilde{W}\in \Tr_F\textup{\textsf{\textbf{W}}}} \ \Tr\left( \Omega_\theta(h) \, \tilde{W}\right) .
 \label{eq_app:swap_max_min}
\end{align}


\subsubsection{Trading the inner ($\max_{\tilde W}$) SDP problem for its dual}
\label{app_subsubsec:trade_for_dual}

At this point, the inner ($\max_{\tilde W}$) optimization problem takes the standard form of a SDP problem. One can then consider the dual of this (primal) problem, which involves a minimization problem that we write for now in the generic form $\min_{{\cal S} \text{ s.t. \ldots}} S_\emptyset$: the precise form for the constraint ``${\cal S} \text{ s.t. \ldots}$'' will be specified, for each of the classes $\textup{\textsf{\textbf{W}}}$ under consideration, in Sec.~\ref{app_subsec:explicit_SDPs} below.

The structure of the problem considered here ensures that strong duality holds (see below), so that the optimal values of the primal and dual SDP problems match. We can then trade the former for the latter in Eq.~\eqref{eq_app:swap_max_min}, and obtain
\begin{equation}
    J_\theta^{\textup{\textsf{\textbf{W}}}}(C_\theta^{\otimes N})  = \min_{h\in\mathbb{H}_q} \ \min_{{\cal S} \text{ s.t. \ldots}} \ S_\emptyset .
 \label{eq_app:min_min}
\end{equation}


\subsubsection{Solving the global SDP problem}
\label{app_subsubsec:solve_SDP}

It remains now to solve Eq.~\eqref{eq_app:min_min}. The constraints we generically wrote for now as ``${\cal S} \text{ s.t. \ldots}$'' are linear in $\Omega_\theta(h)$, hence quadratic in $h$ (see Eq.~\eqref{eq:def_Omega_app}). One can however use a standard trick in convex optimization, which is to use Schur complements to linearize the constraints in $h$ (see again Sec.~\ref{app_subsec:explicit_SDPs} below).
This allows us to write Eq.~\eqref{eq_app:min_min} as an SDP problem, and solve it using standard tools.

We thus finally obtain the optimal QFI $J_\theta^{\textup{\textsf{\textbf{W}}}}(C_\theta^{\otimes N})$---or its upper bound $J_{\theta,\textup{purif}}^{\textup{\textsf{\textbf{W}}}}(C_\theta^{\otimes N})$, see Sec.~\ref{app_subsubsec:restrict_pure} above---for the class $\textup{\textsf{\textbf{W}}}$ under consideration.


\subsubsection{Reconstructing the optimal process}
\label{app_subsubsec:reconstruct_Wopt}

One may not only be interested in estimating the optimal QFI $J_{\theta}^{\textup{\textsf{\textbf{W}}}}(C_\theta^{\otimes N})$ (or its upper bound $J_{\theta,\textup{purif}}^{\textup{\textsf{\textbf{W}}}}(C_\theta^{\otimes N})$), but also in recovering the optimal process $\tilde{W}^\text{(opt)}\in \Tr_F\textup{\textsf{\textbf{W}}}$ that reaches it.

Our approach above provides us with the optimal value $h^\text{(opt)}$ of $h$ for the minimization problem of Eq.~\eqref{eq_app:min_min}, or equivalently, for the min-max problem of Eq.~\eqref{eq_app:swap_max_min}. What remains is to find the optimal value $\tilde{W}^\text{(opt)}$ of $\tilde{W}$.
It is however not enough to just fix $h=h^\text{(opt)}$ and minimize the objective function over $\tilde{W}$, as for the fixed value $h^\text{(opt)}$, there might be different values of $\tilde{W}$ giving the same value of the objective function---which may thus all be solutions to the min-max problem of Eq.~\eqref{eq_app:swap_max_min}, but not necessarily of the max-min problem, which is what we need in order to be assured that $\tilde{W}$ corresponds to a process achieving this QFI via the link of Eq.~\eqref{eq:min_htheta}. 

To make sure that one also gets a solution of the max-min problem, one must further impose that the derivative over $h$, after fixing $\tilde{W}$, vanishes. That is, one needs to further impose that
\begin{align}
    \partial_h \Tr(\Omega_\theta(h) \, \tilde{W})\big|_{h=h^\text{(opt)}}= 0 , \label{eq:saddle_derivative}
\end{align}
which can be translated%
\footnote{Here $\partial_h f(h) = 0$ means that the derivative of the function $f(h)$ with respect to all scalar variables that are used to define $h$ must be null.
Denoting by $h_{k,l}^\textbf{r}$ and $h_{k,l}^\textbf{i}$ the real and imaginary parts of the matrix elements $h_{k,l}$, respectively, this amounts to imposing that $\partial_{h_{k,l}^\textbf{r}} f(h) = \partial_{h_{k,l}^\textbf{i}} f(h) = 0$ for all $k,l$.
\\
For a Hermitian matrix $h$, all these partial derivatives can be encoded into a single matrix---which we take to formalize our notation $\partial_h f(h)$---with diagonal elements $[\partial_h f(h)]_{k,k} = \partial_{h_{k,k}^\textbf{r}} f(h)$ and non-diagonal elements $[\partial_h f(h)]_{k,l} = \frac12 \big(\partial_{h_{k,l}^\textbf{r}} f(h) + i\, \partial_{h_{k,l}^\textbf{i}} f(h)\big)$. (Indeed, using $h_{k,l}^\textbf{r} = h_{l,k}^\textbf{r}$ and $h_{k,l}^\textbf{i} = -h_{l,k}^\textbf{i}$ one gets $\partial_{h_{k,l}^\textbf{r}} f(h) = [\partial_h f(h)]_{k,l} + [\partial_h f(h)]_{l,k}$ and $\partial_{h_{k,l}^\textbf{i}} f(h) = -i\big([\partial_h f(h)]_{k,l} - [\partial_h f(h)]_{l,k}\big)$ for $k\neq l$.)
\\
With this definition of $\partial_h f(h)$, one gets in particular the nice properties that $\partial_h \Tr(Ah) = A$, $\partial_h \Tr(Bh^2) = Bh+hB$, and hence \mbox{$\partial_h\! \Tr[(F_0{+}F_1h)(F_0{+}F_1h)^\dagger G] = F_1^\dagger G(F_0{+}F_1h) \!+\! (F_0{+}F_1h)^\dagger G F_1$} for any matrices $A$, $B$, $F_0$, $F_1$ and $G$ that are independent from $h$. Writing $\Tr(\Omega_\theta(h) \, \tilde{W})$ in the latter form (through Eq.~\eqref{eq:def_Omega_app}), we indeed find that Eq.~\eqref{eq:saddle_derivative} translates into the condition of Eq.~\eqref{eq:saddle_derivative_v2}---which is also equivalent to that introduced in Algorithm~1 of Ref.~\cite{liu}.}
 here into the requirement that
 \begin{align}
    {\mathbf C_\theta^0}{}^\dagger \tilde{W}^T \big( \dot{\mathbf C}_\theta^0 - i \mathbf C_\theta^0 h^\text{(opt)} \big) \text{ is Hermitian} . \label{eq:saddle_derivative_v2}
\end{align}
We discuss and illustrate further the necessity for imposing Eq.~\eqref{eq:saddle_derivative} in Sec.~\ref{app:subsec:saddle_condition} below.

Hence, the optimal $\tilde{W}^\text{(opt)}$ can be obtained as the solution to the following SDP problem:
\begin{align}
    \begin{array}{rl}
        \tilde{W}^\text{(opt)} = & \arg \max_{\tilde{W}} \ \ \Tr\left( \Omega_\theta(h^\text{(opt)}) \, \tilde{W}\right) \\[2mm]
        & \ \text{s.t.} \left\{\begin{array}{l}
            \tilde{W}\in \Tr_F\textup{\textsf{\textbf{W}}}, \\[1mm]
            {\mathbf C_\theta^0}{}^\dagger \tilde{W}^T \big( \dot{\mathbf C}_\theta^0 - i \mathbf C_\theta^0 h^\text{(opt)} \big) \in \mathbb{H}_q.
        \end{array} \right.
    \end{array} \label{eq:tW_argmax}
\end{align}
From this, one can then take any purification of $\tilde{W}^\text{(opt)}$ to define the pure process $W^\text{(opt)} = \ketbra{w}{w}$ we started from, and which generates the optimal output state $\rho_\theta^\text{(opt)} = C_\theta^{\otimes N} * W^\text{(opt)}$.
Recall that when considering $\textup{\textsf{\textbf{W}}}$ to be \textup{\textsf{QC-Par}}, \textup{\textsf{QC-FO}}, \textup{\textsf{QC-Sup}}, \textup{\textsf{QC-QC}} or \textup{\textsf{Gen}}, it is guaranteed that $W^\text{(opt)}$ is in the desired class; for \textup{\textsf{QC-CC}} however, $\tilde{W}\in \Tr_F\textup{\textsf{QC-CC}}$ does in general not imply that its purification $W^\text{(opt)}$ is in \textup{\textsf{QC-CC}}, and we may indeed only get an upper bound on $J_\theta^{\textup{\textsf{QC-CC}}}(C_\theta^{\otimes N})$ (see Sec.~\ref{app_subsubsec:restrict_pure}).


\subsection{Explicit SDP formulations for the classes under consideration}
\label{app_subsec:explicit_SDPs}

In this section we shall provide the explicit formulations for the different SDP problems one encounters in the approach described above, for the different classes of processes that we consider in this work.

SDP problems are commonly encountered in the field of quantum information. These can be written in different ways; the standard form that we will use here consists in writing them as the following pair of ``primal'' and ``dual'' problems~\cite{watrous}:
\begin{equation}
    \begin{array}{rl}
        \max_{\cal W} & \Tr[{\cal A}\,{\cal W}] \\[2mm]
        \text{s.t.} & \Phi({\cal W}) = {\cal B}, \ {\cal W} \ge 0
    \end{array} \label{eq:primal_general_form}
\end{equation}
and
\begin{equation}
    \begin{array}{rl}
        \min_{\cal S} & \Tr[{\cal S}\,{\cal B}] \\[2mm]
        \text{s.t.} & \Phi^\dagger({\cal S}) \ge {\cal A},
    \end{array} \label{eq:dual_general_form}
\end{equation}
where ${\cal A}$ and ${\cal W}$ are Hermitian operators in some space ${\cal L}({\cal H}^\textsf{W})$, ${\cal B}$ and ${\cal S}$ are Hermitian operators in some space ${\cal L}({\cal H}^\textsf{S})$, $\Phi$ is a Hermiticity-preserving linear map from ${\cal L}({\cal H}^\textsf{W})$ to ${\cal L}({\cal H}^\textsf{S})$, and $\Phi^\dagger$ is its adjoint map (such that $\Tr[\Phi^\dagger({\cal S})\,{\cal W}] = \Tr[{\cal S}\,\Phi({\cal W})]$ for all ${\cal W}\in{\cal L}({\cal H}^\textsf{W})$, ${\cal S}\in{\cal L}({\cal H}^\textsf{S})$).

Weak duality ensures that the maximal value of the primal problem~\eqref{eq:primal_general_form} is no larger than the minimal value of the dual problem~\eqref{eq:dual_general_form}. In all situations we consider here, strong duality also holds, meaning that the two optimal values actually coincide. This indeed follows from verifying the so-called Slater's condition: specifically, this is ensured here by the fact that there always exists ${\cal W} > 0$ such that $\Phi({\cal W}) = {\cal B}$ and a Hermitian ${\cal S}$ such that $\Phi^\dagger({\cal S}) \ge {\cal A}$~\cite{watrous}.

The optimization problems we shall encounter will not readily be of the form above, as they will typically involve some constraints that are quadratic in $h$.
One can however resort to a standard trick to linearize these. Namely, for any $s\times s$ Hermitian matrix ${\mathbf S}$, and for any $s\times r$ matrix ${\mathbf C}$, let us define the Hermitian matrix ${\mathbf A}_\text{Schur}({\mathbf S},{\mathbf C})$ as a block matrix
\begin{equation}
    {\mathbf A}_\text{Schur}({\mathbf S},{\mathbf C}) := \left(\begin{array}{cc}
        \id & {\mathbf C}^\dagger \\[1mm]
        {\mathbf C} & {\mathbf S}
    \end{array} \right) \label{eq:ASchur}
\end{equation}
(with $\id$ being here the $r\times r$ identity matrix, so as to match the dimensions).
It then follows from the Schur Complement Lemma (see e.g. Theorem~1.12 in Ref.~\cite{Horn2005}), that
\begin{equation}
    {\mathbf A}_\text{Schur}({\mathbf S},{\mathbf C}) \ge 0 \quad \text{if and only if} \quad {\mathbf S} \ge {\mathbf C}{\mathbf C}^\dagger. \label{eq:ASchur_lemma}
\end{equation}
We will use this ``Schur complement trick'' to linearize constraints of the form $\tilde{\mathbf S} \otimes \id^{A_{k_N}^I} \ge \Tr_{A_{k_N}^O} \Omega_\theta(h)$, for some $k_N$. To write the right hand side in the form ${\mathbf C}{\mathbf C}^\dagger$, we will take
\begin{align}
    {\mathbf C} & = {\mathbf C}_{k_N}(h) := 2\Big(\bra{j}^{A_{k_N}^O} \big({\dot{\mathbf C}_\theta^0} - i {\mathbf C_\theta^0} h \big)^{\!*} \Big)_{j=1}^{d_{A_{k_N}^O}} \notag \\
    & = 2\Bigg(\!\bra{1}^{A_{k_N}^O} \!\big({\dot{\mathbf C}_\theta^0} - i {\mathbf C_\theta^0} h \big)^{\!*} \ \ \cdots \ \ \bra{d_{A_{k_N}^O}\!}^{A_{k_N}^O} \!\big({\dot{\mathbf C}_\theta^0} - i {\mathbf C_\theta^0} h \big)^{\!*} \!\Bigg) \label{eq:CkNh}
\end{align}
(where ${}^*$ denotes here the complex conjugate).
That is, ${\mathbf C}_{k_N}(h)$ a block matrix composed by the blocks $\big(\id^{A_{{\cal N}\backslash k_N}^{IO}A_{k_N}^I}\otimes\bra{j}^{A_{k_N}^O}\big) \big({\dot{\mathbf C}_\theta^0} - i {\mathbf C_\theta^0} h \big)^{\!*}$, where $\{\ket{j}^{A_{k_N}^O}\}_{j=1}^{d_{A_{k_N}^O}}$ denotes an orthonormal basis of ${\cal H}^{A_{k_N}^O}$ (with the $d_{A_{k_N}^O}$ blocks being put on a line, so as to turn the $d_{A_{\cal N}^{IO}}\times q$ matrix $\big({\dot{\mathbf C}_\theta^0} - i {\mathbf C_\theta^0} h \big)^{\!*}$ into a $\big(\frac{d_{A_{\cal N}^{IO}}}{d_{A_{k_N}^O}}\big)\times (d_{A_{k_N}^O} q)$ matrix).
With this, we define
\begin{align}
    {\mathbf A}_{\text{Schur},k_N}(\tilde{\mathbf S},h) := {\mathbf A}_\text{Schur}(\tilde{\mathbf S} \otimes \id^{A_{k_N}^I},{\mathbf C}_{k_N}(h)), \label{eq:def_ASchur_kN}
\end{align}
so that
\begin{align}
    & \tilde{\mathbf S} \otimes \id^{A_{k_N}^I} \ge \Tr_{A_{k_N}^O} \Omega_\theta(h) \notag \\
    & \hspace{10mm} \text{if and only if} \quad {\mathbf A}_{\text{Schur},k_N}({\mathbf S},h) \ge 0. \label{eq:ASchur_lemma_2}
\end{align}
Note that ${\mathbf A}_{\text{Schur},k_N}({\mathbf S},h)$ is linear in $h$, as desired.

Referring to these general formulations, we will now recall how to derive and write the SDP problems for the \textup{\textsf{QC-FO}} class, which were already given in Refs.~\cite{Altherr,liu}. We will then present the case of the \textup{\textsf{QC-CC}} and \textup{\textsf{QC-QC}} classes, which follow quite similar lines. For completeness and ease of reference, we will also give explicitly the SDP formulations for the \textup{\textsf{QC-Par}}, \textup{\textsf{QC-Sup}} and \textup{\textsf{Gen}} classes (whose derivation can also be found in Ref.~\cite{liu}).%
\footnote{As mentioned in the main text, the authors of Ref.~\cite{liu} made a simplifying assumption on the structure of the classes \textup{\textsf{\textbf{W}}} under consideration (or rather, on $\Tr_F\text{\textup{\textsf{\textbf{W}}}}$; see their Eq.~(9)), which led them to unify the derivation of the SDP problems for the different classes in a different way from ours. Our \textup{\textsf{QC-CC}} and \textup{\textsf{QC-QC}} classes do however not satisfy this simplifying assumption, so we could not readily use their derivation for our purposes. \\
Note also that the forms of the dual SDPs given in Refs.~\cite{Altherr,liu} are slightly different---but clearly equivalent---from the ones we give below: they can be obtained from ours by effectively ``renormalizing'' the dual variables in ${\cal S}$ (typically, taking $S_\emptyset = \lambda$ and replacing all other $S_{(\cdot)}$'s by $\lambda S_{(\cdot)}'$, for some new dual variables $\lambda$ and $S_{(\cdot)}'$; or in the \textup{\textsf{Gen}} case, replacing $S$ by $\lambda S'$ with $\Tr[S']=d_{A_{\cal N}^I}$). The way one uses the Schur complement trick to linearize the last constraint is then also slightly different (as one should then use Schur complements of matrices of the form $\left(\begin{smallmatrix}
    \lambda \id & {\mathbf C}^\dagger \\[1mm]
    {\mathbf C} & {\mathbf S}'
\end{smallmatrix}\right)$, rather than as in Eq.~\eqref{eq:ASchur}). \label{ftn:diff_SDPs}}


\subsubsection{SDP problems for the \textup{\textsf{QC-FO}} class}

Using Eq.~\eqref{eq:QC-FO}, the primal problem
\begin{align}
    & \max_{\tilde{W}\in \Tr_F\textup{\textsf{QC-FO}}} \ \Tr\left( \Omega_\theta(h) \, \tilde{W}\right) \notag \\[2mm]
    & = \max \ \Tr\left( \Big( \Tr_{A_N^O} \Omega_\theta(h) \Big) W_{(N)} \right) \notag \\
    & \quad \text{s.t.} \left\{ \begin{array}{l}
        \Tr W_{(1)} = 1, \\[1mm]
        \forall\, n = 1, \ldots, N{-}1, \ \Tr_{A_{n+1}^I} \! W_{(n+1)} = W_{(n)} \otimes \id^{A_n^O}, \\[1mm]
        \forall\, n = 1, \ldots, N, \ W_{(n)} \ge 0.
    \end{array} \right. \label{eq:primal_QCFO}
\end{align}
involved in Eq.~\eqref{eq_app:swap_max_min} can be written in the standard form of Eq.~\eqref{eq:primal_general_form}, with%
\footnote{Other (ultimately equivalent) choices are possible. E.g., given the redundancies in the way we wrote the constraints for $\tilde{W}\in \Tr_F\textup{\textsf{QC-FO}}$ (namely, that all $W_{(n)}$'s are uniquely determined by $W_{(N)}$, and that their positive semidefiniteness simply follows from that of $W_{(N)}$), one could just take ${\cal W} = W_{(N)}$, ${\cal A} = \Tr_{A_N^O}\Omega_\theta(h)$, with the same $\Phi$ and ${\cal B}$ as in Eq.~\eqref{eq:defs_SDP_QCFO}, after replacing $W_{(n)}$ in $\Phi$ by $\frac{1}{d_n^Od_{n+1}^O...d_{N-1}^O}\Tr_{A_n^O A_{\{n+1,\ldots,{N-1}\}}^{IO}A_N^I}W_{(N)}$.
The choice of ${\cal W}$ and ${\cal A}$ in Eq.~\eqref{eq:defs_SDP_QCFO}, however, allows us to follow nicely parallel derivations for the \textup{\textsf{QC-CC}} and \textup{\textsf{QC-QC}} classes; see Sec.~\ref{app:subsubsec_QCCC_QCQC} below.}
\begin{align}
    {\cal W} & = \bigoplus_{n=1}^N W_{(n)}, \notag \\
    {\cal A} & = \left(\bigoplus_{n=1}^{N-1} \mathbb{0}^{A_{\{1,\ldots,n-1\}}^{IO}A_n^I} \right) \oplus \Tr_{A_N^O}\Omega_\theta(h), \notag \\
    \Phi({\cal W}) & = \bigoplus_{n=0}^{N-1} \Phi_{(n)}({\cal W}) \notag \\
    \text{with } & \left\{\begin{array}{l}
        \Phi_{(0)}({\cal W}) = \Tr W_{(1)}, \\[2mm]
        \forall\, n =1,\ldots,N-1, \\
        \quad \Phi_{(n)}({\cal W}) = \Tr_{A_{n+1}^I} W_{(n+1)} - W_{(n)} \otimes \mathbbm{1}^{A_{n}^O},
    \end{array}\right. \notag \\
    {\cal B} & = 1 \oplus \bigoplus_{n=1}^{N-1} \mathbb{0}^{A_{\{1,\ldots,n\}}^{IO}}. \label{eq:defs_SDP_QCFO}
\end{align}
where $\mathbb{0}$ denotes a null matrix (in the space indicated by the superscript).

Introducing the dual variable
\begin{align}
    {\cal S} & = \bigoplus_{n=0}^{N-1} S_n \quad \in \bigoplus_{n=0}^{N-1} {\cal L}({\cal H}^{A_{\{1,\ldots,n\}}^{IO}})
\end{align}
(with ${\cal L}({\cal H}^{A_\emptyset^{IO}}) = \mathbb{R}$ for $n=0$), the adjoint of the linear map $\Phi$ is easily found to be%
\footnote{From the structure of ${\cal W}$, one can write $\Tr[\Phi^\dagger({\cal S})\,{\cal W}] = \sum_{n=1}^N \Tr[(\Phi^\dagger)_{(n)}({\cal S})\,W_{(n)}]$ and $\Phi^\dagger({\cal S}) = \bigoplus_{n=1}^N \, (\Phi^\dagger)_{(n)}({\cal S})$. Furthermore, the adjoint map satisfies $\Tr[\Phi^\dagger({\cal S})\,{\cal W}] = \Tr[{\cal S}\,\Phi({\cal W})] = \sum_{n=0}^{N-1} \Tr[S_n\,\Phi_{(n)}({\cal W})] = S_0 \Tr W_{(1)} + \sum_{n=1}^{N-1} \Tr[S_n(\Tr_{A_{n+1}^I} W_{(n+1)})] - \sum_{n=1}^{N-1} \Tr[S_n(W_{(n)} \otimes \mathbbm{1}^{A_{n}^O})] = S_0 \Tr W_{(1)} + \sum_{n=2}^N \Tr[(S_{n-1}\otimes\id^{A_n^I}) W_{(n)}] - \sum_{n=1}^{N-1} \Tr[(\Tr_{A_n^O} S_n)W_{(n)} ] = \sum_{n=1}^{N-1} \Tr[(S_{n-1}\otimes\id^{A_n^I}-\Tr_{A_n^O} S_n)W_{(n)} ] + \Tr[(S_{N-1}\otimes\id^{A_N^I}) W_{(N)}] = \sum_{n=1}^N \Tr[(\Phi^\dagger)_{(n)}({\cal S})\,W_{(n)}]$; as this must hold for all $W_{(n)}$'s, one can directly identify the terms inside the traces, for each term in the sums.}
\begin{align}
    \Phi^\dagger({\cal S}) & = \bigoplus_{n=1}^N \, (\Phi^\dagger)_{(n)}({\cal S}) \notag \\
    \text{with } & \left\{\begin{array}{l}
        \forall\, n =1,\ldots,N-1, \\
        \quad (\Phi^\dagger)_{(n)}({\cal S}) = S_{n-1}\otimes\id^{A_n^I} - \Tr_{A_n^O} S_n, \\[2mm]
        (\Phi^\dagger)_{(N)}({\cal S}) = S_{N-1}\otimes\id^{A_N^I},
    \end{array}\right.
\end{align}
so that the dual problem, Eq.~\eqref{eq:dual_general_form}, gets here the specific form
\begin{equation}
    \begin{array}{rl}
        \min_{\cal S} & S_0 \\[2mm]
        \text{s.t.} & \left\{\begin{array}{l}
        \forall\, n =1,\ldots,N-1, \\
        \quad S_{n-1}\otimes\id^{A_n^I} \ge \Tr_{A_n^O} S_n, \\[2mm]
        S_{N-1}\otimes\id^{A_N^I} \ge \Tr_{A_N^O}\Omega_\theta(h).
    \end{array}\right.
    \end{array} \label{eq:dual_QCFO_v1}
\end{equation}
To see that Slater's condition for strong duality is satisfied (see above), note that ${\cal W} = \bigoplus_{n=1}^N \frac{\id^{A_{\{1,\ldots,n-1\}}^{IO}A_n^I}}{d_{A_{\{1,\ldots,n\}}^I}}$ for instance satisfies $\Phi({\cal W}) = {\cal B}$ and ${\cal W} > 0$, and that one can always rather trivially find a Hermitian ${\cal S}$ such that $\Phi^\dagger({\cal S}) \ge {\cal A}$.

At this point one may note that it is possible to replace the first $N-1$ inequalities in the constraints of Eq.~\eqref{eq:dual_QCFO_v1} by equalities. Indeed, if ${\cal S}$ is an optimal solution of Eq.~\eqref{eq:dual_QCFO_v1}, then ${\cal S}' = \bigoplus_{n=0}^{N-1} S_n'$, with
\begin{align}
S_n' & = S_0\frac{\id^{A_{\{1,\ldots,n\}}^{IO}}}{d_{A_{\{1,\ldots,n\}}^O}} + \sum_{m=1}^n {}_{[1-A_m^O]}S_m\otimes\frac{\id^{A_{\{m+1,\ldots,n\}}^{IO}}}{d_{A_{\{m+1,\ldots,n\}}^O}} \notag \\
& = S_n + \!\sum_{m=1}^n \!\Big(S_{m-1}\otimes\id^{A_m^I} \notag \\[-7mm]
& \hspace{23mm} - \Tr_{A_m^O} S_m\!\Big)\!\otimes\frac{\!\id^{A_m^OA_{\{m+1,\ldots,n\}}^{IO}}\!}{d_{A_{\{m,\ldots,n\}}^O}}, \label{eq:Sn_prime}
\end{align}
is also an optimal solution (such that $S_0' = S_0$), which satisfies $S_{n-1}'\otimes\id^{A_n^I} = \Tr_{A_n^O} S_n'$ for $n =1,\ldots,N-1$ (as is easily checked using the first form for $S_n'$ given above), and $S_{N-1}'\otimes\id^{A_N^I} \ge S_{N-1}\otimes\id^{A_N^I} \ge \Tr_{A_N^O}\Omega_\theta(h)$ (as is easily checked using the second expression for $S_{N-1}'$, using the constraints of Eq.~\eqref{eq:dual_QCFO_v1} that the $S_n$'s satisfy).

To obtain the optimal QFI for the \textup{\textsf{QC-FO}} class, it remains to insert the above dual problem into Eq.~\eqref{eq_app:min_min}. Using the Schur complement trick introduced above, one can replace the quadratic (in $h$) constraint $S_{N-1}\otimes\id^{A_N^I} \ge \Tr_{A_N^O}\Omega_\theta(h)$ by the linear constraint
\begin{align}
{\mathbf A}_{\text{Schur},N}(S_{N-1},h) \ge 0,
\end{align}
with ${\mathbf A}_{\text{Schur},N}$ as defined in Eq.~\eqref{eq:def_ASchur_kN}.
This thus turns the calculation of the QFI into an SDP problem:
\begin{align}
    & J_\theta^{\textup{\textsf{QC-FO}}}(C_\theta^{\otimes N}) \notag \\
    & \begin{array}{rl}
        = \min_{h,{\cal S}} & S_0 \\[2mm]
        \text{s.t.} & \left\{\begin{array}{l}
        \forall\, n =1,\ldots,N-1, \\
        \quad S_{n-1}\otimes\id^{A_n^I} = \Tr_{A_n^O} S_n, \\[2mm]
        {\mathbf A}_{\text{Schur},N}(S_{N-1},h) \ge 0.
    \end{array}\right.
    \end{array} \label{eq:optim_QFI_QCFO}
\end{align}

As explained in Sec.~\ref{app_subsubsec:reconstruct_Wopt}, the optimal process $\tilde{W}^\text{(opt)}\in\Tr_F\textup{\textsf{QC-FO}}$ is recovered by solving the primal problem of Eq.~\eqref{eq:primal_QCFO} after fixing the optimal value of $h=h^\text{(opt)}$ obtained from Eq.~\eqref{eq:optim_QFI_QCFO} and adding the constraint that ${\mathbf C_\theta^0}{}^\dagger \tilde{W}^T \big( \dot{\mathbf C}_\theta^0 - i \mathbf C_\theta^0 h^\text{(opt)} \big) \in \mathbb{H}_q$, as in Eq.~\eqref{eq:tW_argmax}---and with $\tilde{W} = W_{(N)}\otimes\id^{A_N^O}$ (cf.\ Eq.~\eqref{eq:QC-FO}).

The optimal process with fixed order $W^\text{(opt)} = \ketbra{w}{w}\in\textup{\textsf{QC-FO}}$ is then obtained by purifying $\tilde{W}^\text{(opt)}$.


\subsubsection{SDP problems for the \textup{\textsf{QC-CC}} and \textup{\textsf{QC-QC}} classes}
\label{app:subsubsec_QCCC_QCQC}

The derivation of the SDP problems for the \textup{\textsf{QC-CC}} and \textup{\textsf{QC-QC}} classes follows very similar lines to those for the \textup{\textsf{QC-FO}} class, given in the previous subsection. We will not repeat everything; instead, we just present in the following table  how Eqs.~\eqref{eq:primal_QCFO} to~\eqref{eq:optim_QFI_QCFO} are to be modified for the two new classes we considered in this paper.%
\footnote{For clarification: in the \textup{\textsf{QC-CC}} case, $(k_1,\ldots,k_n)$ for $n=0$ is understood as the empty set, $\emptyset$ (as e.g.\ in Eq.~\eqref{eq:S_QCCC}: ${\cal S} =$ \scalebox{.96}[1]{\mbox{$\bigoplus_{n=0}^{N-1} \!\bigoplus_{(k_1,\ldots,k_n)} \!S_{(k_1,\ldots,k_n)} \!=\! S_\emptyset \!\oplus\! \bigoplus_{n=1}^{N-1} \!\bigoplus_{(k_1,\ldots,k_n)} \!S_{(k_1,\ldots,k_n)}$}}). In the \textup{\textsf{QC-QC}} case, $\mathcal{K}_n\setminus k_n$ is to be understood as $\mathcal{K}_n\setminus \{k_n\}$.}

\begin{widetext}
    \centering
    \begin{longtable}{c|c}
      \textup{\textsf{QC-CC}} & \textup{\textsf{QC-QC}} \\[1mm] \hline
\parbox{.48\columnwidth}{
\vspace{3mm}\underline{Primal problem:} (using Eq.~\eqref{eq:QC-CC})
\begin{align}
    & \hspace{-3mm} \max_{\tilde{W}\in \Tr_F\textup{\textsf{QC-CC}}} \ \Tr\left( \Omega_\theta(h) \, \tilde{W}\right) \notag \\[1mm]
    & \hspace{-3mm} = \max \ \Tr\left( \sum_{(k_1,\ldots,k_N)} \Big( \Tr_{A_{k_N}^O} \Omega_\theta(h) \Big) W_{(k_1,\ldots,k_N)} \right) \notag \\
    & \hspace{-3mm} \quad \text{s.t.} \left\{ \!\begin{array}{l}
        \sum_{k_1} \Tr W_{(k_1)} = 1, \\[2mm]
        \forall\, n = 1, \ldots, N-1, \ \forall\, (k_1, \ldots, k_n), \\[1mm]
        \qquad \sum_{k_{n+1}} \Tr_{\!A_{k_{n+1}}^I} W_{(k_1,\ldots,k_n,k_{n+1})} \\[-1mm]
        \hspace{32mm} = W_{(k_1,\ldots,k_n)} \otimes \id^{A_{k_n}^O}, \\[2mm]
        \forall\, n = 1, \ldots, N, \ \forall\, (k_1, \ldots, k_n), \ W_{(k_1,\ldots,k_n)} \ge 0.
    \end{array} \right. \label{eq:primal_QCCC}
\end{align}
}
&
\parbox{.48\columnwidth}{
\vspace{3mm}\underline{Primal problem:} (using Eq.~\eqref{eq:QC-QC})
\begin{align}
    & \hspace{-3mm} \max_{\tilde{W}\in \Tr_F\textup{\textsf{QC-QC}}} \ \Tr\left( \Omega_\theta(h) \, \tilde{W}\right) \notag \\[1mm]
    & \hspace{-3mm} = \max \ \Tr\left( \sum_{k_N} \Big( \Tr_{A_{k_N}^O} \Omega_\theta(h) \Big) W_{({\cal N} \backslash k_N,k_N)} \right) \notag \\
    & \hspace{-3mm} \quad \text{s.t.} \left\{ \!\begin{array}{l}
        \sum_{k_1} \Tr W_{(\emptyset,k_1)} = 1, \\[2mm]
        \forall\, n = 1, \ldots, N-1, \ \forall\, {\cal K}_n, \\
        \qquad \sum_{k_{n+1} \in {\cal N} \backslash {\cal K}_n} \Tr_{A_{k_{n+1}}^I} W_{({\cal K}_n,k_{n+1})} \\[1mm]
        \hspace{20mm} = \sum_{k_n \in {\cal K}_n} W_{({\cal K}_n \backslash k_n,k_n)}\otimes \id^{A_{k_n}^O}, \\[2mm]
        \forall\, n = 1, \ldots, N, \ \forall\, {\cal K}_{n-1}, k_n, \ W_{({\cal K}_{n-1}, k_n)} \ge 0.
    \end{array} \right. \label{eq:primal_QCQC}
\end{align}
}
\\ \hline
\endfirsthead
\textup{\textsf{QC-CC}} (continued) & \textup{\textsf{QC-QC}} (continued) \\[1mm] \hline
\endhead
\parbox{.48\columnwidth}{
\begin{align}
    {\cal W} & = \bigoplus_{n=1}^N \bigoplus_{(k_1,\ldots,k_n)} W_{(k_1,\ldots,k_n)}, \notag \\
    {\cal A} & = \left(\bigoplus_{n=1}^{N-1} \bigoplus_{(k_1,\ldots,k_n)} \!\!\!\!\!\mathbb{0}^{A_{\{k_1,\ldots,k_{n-1}\}}^{IO}A_{k_n}^I}\! \right) \!\oplus \!\!\bigoplus_{(k_1,\ldots,k_N)}\!\!\!\! \Tr_{A_{k_N}^O}\!\Omega_\theta(h), \notag \\
    \Phi({\cal W}) & = \bigoplus_{n=0}^{N-1} \bigoplus_{(k_1,\ldots,k_n)} \Phi_{(k_1,\ldots,k_n)}({\cal W}) \notag \\
    \text{with } & \left\{\begin{array}{l}
        \Phi_{\emptyset}({\cal W}) = \sum_{k_1} \Tr W_{(k_1)}, \\[2mm]
        \forall\, n =1,\ldots,N-1, \forall\, (k_1,\ldots,k_n), \\[1mm]
        \hspace{2mm} \Phi_{(k_1,\ldots,k_n)}({\cal W}) = \sum_{k_{n+1}} \! \Tr_{A_{k_{n+1}^I}} \!\! W_{(k_1,\ldots,k_n,k_{n+1})} \\ \hspace{30mm} - W_{(k_1,\ldots,k_n)} \otimes \mathbbm{1}^{A_{k_n}^O},
    \end{array}\right. \notag \\
    {\cal B} & = 1 \oplus \bigoplus_{n=1}^{N-1} \bigoplus_{(k_1,\ldots,k_n)} \mathbb{0}^{A_{\{k_1,\ldots,k_n\}}^{IO}}.
\end{align}
}
&
\parbox{.48\columnwidth}{
\begin{align}
    {\cal W} & = \bigoplus_{n=1}^N \bigoplus_{{\cal K}_{n-1},k_n} W_{{\cal K}_{n-1},k_n}, \notag \\
    {\cal A} & = \left(\bigoplus_{n=1}^{N-1} \bigoplus_{{\cal K}_{n-1},k_n} \mathbb{0}^{A_{{\cal K}_{n-1}}^{IO}A_{k_n}^I} \right) \oplus \, \bigoplus_{k_N} \, \Tr_{A_{k_N}^O}\Omega_\theta(h), \notag \\
    \Phi({\cal W}) & = \bigoplus_{n=0}^{N-1} \bigoplus_{{\cal K}_n} \Phi_{{\cal K}_n}({\cal W}) \notag \\
    \text{with } & \left\{\begin{array}{l}
        \Phi_{\emptyset}({\cal W}) = \sum_{k_1} \Tr W_{(\emptyset,k_1)}, \\[2mm]
        \forall\, n =1,\ldots,N-1, \forall\, {\cal K}_n, \\[1mm]
        \quad \Phi_{{\cal K}_n}({\cal W}) = \sum_{k_{n+1}\in \mathcal{N}\backslash\mathcal{K}_n} \Tr_{A_{k_{n+1}}^I}W_{(\mathcal{K}_n,k_{n+1})} \\ \hspace{25mm} - \sum_{k_n \in \mathcal{K}_n} W_{(\mathcal{K}_n\backslash k_n,k_n)}\otimes \mathbbm{1}^{A_{k_n}^O},
    \end{array}\right. \notag \\
    {\cal B} & = 1 \oplus \bigoplus_{n=1}^{N-1} \bigoplus_{{\cal K}_n} \, \mathbb{0}^{A_{{\cal K}_n}^{IO}}.
\end{align}
}
\\ \hline
\parbox{.48\columnwidth}{
\begin{align}
    {\cal S} & = \bigoplus_{n=0}^{N-1} \bigoplus_{(k_1,\ldots,k_n)} \!\!S_{(k_1,\ldots,k_n)} \quad \in \bigoplus_{n=0}^{N-1} \bigoplus_{(k_1,\ldots,k_n)} \!\!{\cal L}({\cal H}^{A_{\{k_1,\ldots,k_n\}}^{IO}}) . \label{eq:S_QCCC}
\end{align}
}
&
\parbox{.48\columnwidth}{
\begin{align}
    {\cal S} & = \bigoplus_{n=0}^{N-1} \bigoplus_{{\cal K}_n} S_{{\cal K}_n} \quad \in \bigoplus_{n=0}^{N-1} \bigoplus_{{\cal K}_n} {\cal L}({\cal H}^{A_{{\cal K}_n}^{IO}}).
\end{align}
}
\\ \hline
\parbox{.48\columnwidth}{
\begin{align}
    \Phi^\dagger({\cal S}) & = \bigoplus_{n=1}^N \bigoplus_{(k_1,\ldots,k_n)} (\Phi^\dagger)_{(k_1,\ldots,k_n)}({\cal S}) \notag \\
    \text{with } & \left\{\begin{array}{l}
        \forall\, n =1,\ldots,N-1, \forall\, (k_1,\ldots,k_n), \\
        \quad (\Phi^\dagger)_{(k_1,\ldots,k_n)}({\cal S}) = S_{(k_1,\ldots,k_{n-1})}\otimes\id^{A_{k_n}^I} \\
        \hspace{35mm} - \Tr_{A_{k_n}^O} S_{(k_1,\ldots,k_n)}, \\[2mm]
        \forall\, (k_1,\ldots,k_N), \\
        \quad (\Phi^\dagger)_{(k_1,\ldots,k_N)}({\cal S}) = S_{(k_1,\ldots,k_{N-1})}\otimes\id^{A_{k_N}^I}.
    \end{array}\right. \hspace{-2mm}
\end{align}
}
&
\parbox{.5\columnwidth}{
\begin{align}
    \Phi^\dagger({\cal S}) & = \bigoplus_{n=1}^N \bigoplus_{{\cal K}_{n-1},k_n} (\Phi^\dagger)_{({\cal K}_{n-1},k_n)}({\cal S}) \notag \\
    \text{with } & \left\{\begin{array}{l}
        \forall\, n =1,\ldots,N-1, \forall\, {\cal K}_n,\forall\, k_n\in{\cal K}_n, \\
        \quad (\Phi^\dagger)_{({\cal K}_n\backslash k_n,k_n)}({\cal S}) = S_{{\cal K}_n\backslash k_n}\otimes\id^{A_{k_n}^I} \\
        \hspace{40mm} - \Tr_{A_{k_n}^O} S_{{\cal K}_n}, \\[2mm]
        \forall\, k_N, (\Phi^\dagger)_{({\cal N}\backslash k_N,k_N)}({\cal S}) = S_{{\cal N}\backslash k_N}\otimes\id^{A_{k_N}^I}.
    \end{array}\right.
\end{align}
}
\\ \hline
\parbox{.48\columnwidth}{
\vspace{3mm}\underline{Dual problem:}
\begin{align}
    \begin{array}{rl}
        \min_{\cal S} & S_\emptyset \\[2mm]
        \text{s.t.} & \left\{\begin{array}{l}
        \forall\, n =1,\ldots,N-1, \forall\, (k_1,\ldots,k_n), \\
        \quad S_{(k_1,\ldots,k_{n-1})}\otimes\id^{A_{k_n}^I} \ge \Tr_{A_{k_n}^O} S_{(k_1,\ldots,k_n)}, \\[2mm]
        \forall\, (k_1,\ldots,k_N), \\
        \quad S_{(k_1,\ldots,k_{N-1})}\otimes\id^{A_{k_N}^I} \ge \Tr_{A_{k_N}^O}\Omega_\theta(h).
    \end{array}\right.
    \end{array} \hspace{-2mm} \label{eq:dual_QCCC_v1}
\end{align}

To verify Slater's condition for strong duality, one can consider ${\cal W} = \bigoplus_{n=1}^N \bigoplus_{(k_1,\ldots,k_n)} \frac{(N-n)!}{N!} \frac{\id^{A_{\{k_1,\ldots,k_{n-1}\}}^{IO}A_{k_n}^I}}{d_{A_{\{k_1,\ldots,k_n\}}^I}}$, which satisfies $\Phi({\cal W}) = {\cal B}$ and ${\cal W} > 0$. (Finding a Hermitian ${\cal S}$ such that $\Phi^\dagger({\cal S}) \ge {\cal A}$ is rather trivial.)

\vspace{1mm}

Furthermore, a very similar construction to that in Eq.~\eqref{eq:Sn_prime} for the \textup{\textsf{QC-FO}} case (essentially, replacing $S_n^{(\prime)}$ by $S_{(k_1,\ldots,k_n)}^{(\prime)}$ and taking the identity operators in the appropriate spaces) then shows that the first inequalities above can without loss of generality be replaced by equalities, $S_{(k_1,\ldots,k_{n-1})}\otimes\id^{A_{k_n}^I} = \Tr_{A_{k_n}^O} S_{(k_1,\ldots,k_n)}$. \vspace{2mm}

}
&
\parbox{.48\columnwidth}{
\vspace{3mm}\underline{Dual problem:}
\begin{align}
    \begin{array}{rl}
        \min_{\cal S} & S_\emptyset \\[2mm]
        \text{s.t.} & \left\{\begin{array}{l}
        \forall\, n =1,\ldots,N-1, \forall\, {\cal K}_n,\forall\, k_n\in{\cal K}_n, \\
        \hspace{12mm} S_{{\cal K}_n\backslash k_n}\otimes\id^{A_{k_n}^I} \ge \Tr_{A_{k_n}^O} S_{{\cal K}_n}, \\[2mm]
        \forall\, k_N, \ S_{{\cal N}\backslash k_N}\otimes\id^{A_{k_N}^I} \ge \Tr_{A_{k_N}^O}\Omega_\theta(h).
    \end{array}\right.
    \end{array} \label{eq:dual_QCQC_v1}
\end{align}
\vspace{1mm}

To verify Slater's condition for strong duality, one can consider ${\cal W} = \bigoplus_{n=1}^N \bigoplus_{{\cal K}_{n-1},k_n} \frac{1}{n\binom{N}{n}} \frac{\id^{A_{{\cal K}_{n-1}}^{IO}A_{k_n}^I}}{d_{A_{{\cal K}_{n-1}\cup\{k_n\}}^I}}$, which satisfies $\Phi({\cal W}) = {\cal B}$ and ${\cal W} > 0$. (Finding a Hermitian ${\cal S}$ such that $\Phi^\dagger({\cal S}) \ge {\cal A}$ is rather trivial.)

\vspace{5mm}

(Note that the construction of Eq.~\eqref{eq:Sn_prime} does not work here: except for the case $n=1$, the inequalities above can in general not be replaced by equalities.)
\vspace{7mm}
}
\\ \hline \\
\parbox{.48\columnwidth}{
$S_{(k_1,\ldots,k_{N-1})}\otimes\id^{A_{k_N}^I} \ge \Tr_{A_{k_N}^O}\Omega_\theta(h)$ in Eq.~\eqref{eq:dual_QCCC_v1} above can be linearized into
\begin{align}
{\mathbf A}_{\text{Schur},k_N}(S_{(k_1,\ldots,k_{N-1})},h) \ge 0.
\end{align}
}
&
\parbox{.48\columnwidth}{
$S_{{\cal N}\backslash k_N}\otimes\id^{A_{k_N}^I} \ge \Tr_{A_{k_N}^O}\Omega_\theta(h)$ in Eq.~\eqref{eq:dual_QCQC_v1} above can be linearized into
\begin{align}
{\mathbf A}_{\text{Schur},k_N}(S_{{\cal N}\backslash k_N},h) \ge 0.
\end{align}
}
\\ \hline
\parbox{.48\columnwidth}{
\vspace{3mm}\underline{Optimal QFI:}
\begin{align}
    & \hspace{-2mm} J_\theta^{\textup{\textsf{QC-CC}}}(C_\theta^{\otimes N}) \notag \\
    & \hspace{-2mm} \begin{array}{rl}
        \!=\!\! & \min_{h,{\cal S}} \ S_\emptyset \\[2mm]
        & \text{s.t.} \left\{\!\begin{array}{l}
        \forall\, n =1,\ldots,N-1, \forall\, (k_1,\ldots,k_n), \\
        \qquad S_{(k_1,\ldots,k_{n-1})}\otimes\id^{A_{k_n}^I} = \Tr_{A_{k_n}^O} S_{(k_1,\ldots,k_n)}, \\[2mm]
        \forall\, (k_1,\ldots,k_N), \, {\mathbf A}_{\text{Schur},k_N}(S_{(k_1,\ldots,k_{N-1})},h) \ge 0.
    \end{array}\right.
    \end{array}\hspace{-2mm} \label{eq:optim_QFI_QCCC}
\end{align}
}
&
\parbox{.48\columnwidth}{
\vspace{3mm}\underline{Optimal QFI:}
\begin{align}
    & J_\theta^{\textup{\textsf{QC-QC}}}(C_\theta^{\otimes N}) \notag \\
    & \begin{array}{rl}
        = & \min_{h,{\cal S}} \ S_\emptyset \\[2mm]
        & \text{s.t.} \ \left\{\begin{array}{l}
        \forall\, n =1,\ldots,N-1, \forall\, {\cal K}_n,\forall\, k_n\in{\cal K}_n, \\
        \hspace{12mm} S_{{\cal K}_n\backslash k_n}\otimes\id^{A_{k_n}^I} \ge \Tr_{A_{k_n}^O} S_{{\cal K}_n}, \\[2mm]
        \forall\, k_N, \ {\mathbf A}_{\text{Schur},k_N}(S_{{\cal N}\backslash k_N},h) \ge 0.
    \end{array}\right.
    \end{array} \label{eq:optim_QFI_QCQC}
\end{align}
}
\\ \hline
\parbox{.48\columnwidth}{
\vspace{3mm}

The optimal process $\tilde{W}^\text{(opt)}\in\Tr_F\textup{\textsf{QC-CC}}$ is recovered by solving Eq.~\eqref{eq:primal_QCCC} after fixing $h=h^\text{(opt)}$ obtained from Eq.~\eqref{eq:optim_QFI_QCCC} and adding the constraint that ${\mathbf C_\theta^0}{}^\dagger \tilde{W}^T \big( \dot{\mathbf C}_\theta^0 - i \mathbf C_\theta^0 h^\text{(opt)} \big) \in \mathbb{H}_q$, as in Eq.~\eqref{eq:tW_argmax}---and with $\tilde{W} = \sum_{(k_1,\ldots,k_N)} W_{(N)}\otimes\id^{A_N^O}$ (see Eq.~\eqref{eq:QC-CC}).

Recall here that the purification $W^\text{(opt)} = \ketbra{w}{w}$ of $\tilde{W}^\text{(opt)}$ is however not guaranteed to be in \textup{\textsf{QC-CC}}. \vspace{3mm}
}
&
\parbox{.48\columnwidth}{
\vspace{3mm}
The optimal process $\tilde{W}^\text{(opt)}\in\Tr_F\textup{\textsf{QC-QC}}$ is recovered by solving Eq.~\eqref{eq:primal_QCQC} after fixing $h=h^\text{(opt)}$ obtained from Eq.~\eqref{eq:optim_QFI_QCQC} and adding the constraint that ${\mathbf C_\theta^0}{}^\dagger \tilde{W}^T \big( \dot{\mathbf C}_\theta^0 - i \mathbf C_\theta^0 h^\text{(opt)} \big) \in \mathbb{H}_q$, as in Eq.~\eqref{eq:tW_argmax}---and with $\tilde{W} = \sum_{k_N \in \mathcal{N}}W_{(\mathcal{N}\setminus k_N,k_N)}\otimes \mathbbm{1}^{A_{k_N}^O}$ (see Eq.~\eqref{eq:QC-QC}).

The optimal process $W^\text{(opt)} = \ketbra{w}{w}\in\textup{\textsf{QC-QC}}$ is then obtained by purifying $\tilde{W}^\text{(opt)}$. \vspace{3mm}
}
\\ \hline
    \end{longtable}
\end{widetext}


\subsubsection{SDP problems for the \textup{\textsf{QC-Par}}, \textup{\textsf{QC-Sup}} and \textup{\textsf{Gen}} classes}

For completeness, let us briefly write the explicit SDP problems one needs to consider for the remaining classes considered in this paper. These can be obtained in a similar fashion to the previous ones, or as in Ref.~\cite{liu} (up to the difference mentioned in Footnote~\ref{ftn:diff_SDPs}).

\medskip

For the \textup{\textsf{QC-Par}} class:
\begin{align}
    \begin{array}{rl}
        J_\theta^{\textup{\textsf{QC-Par}}}(C_\theta^{\otimes N}) = & \min_{h,S_0} \ \ S_0 \\[2mm]
        & \quad \text{s.t. } \ S_0 \,\id^{A_{\cal N}^I} \ge \Tr_{A_{\cal N}^O}\Omega_\theta(h),
    \end{array}
\end{align}
with $S_0 \in \mathbb{R}$, and where the constraint can be linearized into
\begin{align}
    {\mathbf A}_\text{Schur}(S_0 \,\id^{A_{\cal N}^I},{\mathbf C}_{\cal N}(h)) \ge 0
\end{align}
with (similarly to Eq.~\eqref{eq:CkNh}, with clear enough notation)
\begin{align}
    {\mathbf C}_{\cal N}(h) = 2\Big(\bra{j_1,\ldots,j_N}^{A_1^O\cdots A_N^O} \big({\dot{\mathbf C}_\theta^0} - i {\mathbf C_\theta^0} h \big)^{\!*} \Big)_{j_1,\ldots,j_N}.
\end{align}
The optimal $\tilde{W}^\text{(opt)}\in\Tr_F\textup{\textsf{QC-Par}}$ (and $W^\text{(opt)} = \ketbra{w}{w}\in\textup{\textsf{QC-Par}}$, after purification) is recovered by solving Eq.~\eqref{eq:tW_argmax}, translating the constraint $\tilde{W}\in \Tr_F\textup{\textsf{QC-Par}}$ into (see Eq.~\eqref{eq:QC-Par})
\begin{align}
    \left\{ \begin{array}{l}
        \tilde{W} = W_{(I)} \otimes \id^{A_{\cal N}^O}, \\[1mm]
        \Tr W_{(I)} = 1, \\[1mm]
        W_{(I)} \ge 0.
    \end{array} \right.
\end{align}

\medskip

For the \textup{\textsf{QC-Sup}} class:
\begin{align}
    & J_\theta^{\textup{\textsf{QC-Sup}}}(C_\theta^{\otimes N}) \notag \\
    & \begin{array}{rl}
        = & \min_{h,{\cal S}} \ S_0 \\[2mm]
        & \text{s.t.} \ \forall\, \pi, \ \left\{\begin{array}{l}
        S_0 \,\id^{A_{\pi(1)}^I} = \Tr_{A_{\pi(1)}^O} S_{(\pi,1)}, \\[2mm]
        \forall\, n =2,\ldots,N-1, \\
        \quad S_{(\pi,n-1)}\otimes\id^{A_{\pi(n)}^I} = \Tr_{A_{\pi(n)}^O} S_{(\pi,n)}, \\[2mm]
        S_{(\pi,N-1)}\otimes\id^{A_{\pi(N)}^I} \ge \Tr_{A_{\pi(N)}^O}\Omega_\theta(h),
    \end{array}\right.
    \end{array}
\end{align}
where the variables in the optimization (all encapsulated in ${\cal S}$, as done previously) are taken to be $S_0 \in \mathbb{R}$ and $S_{(\pi,n)} \in {\cal L}({\cal H}^{A_{\{\pi(1),\ldots,\pi(n)\}}^{IO}})$ for all $\pi$ and $n =1,\ldots,N-1$;
and where the last constraints (for each $\pi$) can be linearized into
\begin{align}
    {\mathbf A}_{\text{Schur},\pi(N)}(S_{(\pi,N-1)},h) \ge 0.
\end{align}
The optimal $\tilde{W}^\text{(opt)}\in\Tr_F\textup{\textsf{QC-Sup}}$ (and $W^\text{(opt)} = \ketbra{w}{w}\in\textup{\textsf{QC-Sup}}$, after purification) is recovered by solving Eq.~\eqref{eq:tW_argmax}, translating the constraint $\tilde{W}\in \Tr_F\textup{\textsf{QC-Sup}}$ into (see Eq.~\eqref{eq:QC-Sup}, replacing $q_\pi W_\pi$ by $\tilde{W}_\pi$)
\begin{align}
    \left\{ \begin{array}{l}
        \tilde{W} = \sum_\pi \tilde{W}_\pi, \\[1mm]
            \forall\, \pi, \tilde{W}_\pi \in \Tr_F(\text{\textup{\textsf{QC-FO}}}_\pi^\propto), \\[1mm]
            \Tr \tilde{W} = d_{A_{\cal N}^O},
        \end{array} \right.
\end{align}
where $\textup{\textsf{QC-FO}}_\pi^\propto$ denotes the class of process matrices that are proportional (with a nonnegative proportionality factor) to matrices in $\textup{\textsf{QC-FO}}_\pi$, or more explicitly,
\begin{align}
    \left\{ \begin{array}{l}
        \tilde{W} = \sum_\pi W_{(\pi,N)} \otimes \id^{A_{\pi(N)}^O}, \\[1mm]
        \sum_\pi \Tr W_{(\pi,1)} = 1, \\[1mm]
        \forall\, \pi, \forall\, n = 1, \ldots, N-1, \\
        \qquad \Tr_{A_{\pi(n+1)}^I} \! W_{(\pi,n+1)} = W_{(\pi,n)} \otimes \id^{A_{\pi(n)}^O}, \\[1mm]
        \forall\, \pi, \forall\, n = 1, \ldots, N, \ W_{(\pi,n)} \ge 0
    \end{array} \right.
\end{align}
(with each $W_{(\pi,n)} \in {\cal L}({\cal H}^{A_{\{\pi(1),\ldots,\pi(n-1) \}}^{IO}A_{\pi(n)}^I})$).

\medskip

For the \textup{\textsf{Gen}} class, finally:
\begin{align}
    & J_\theta^{\textup{\textsf{Gen}}}(C_\theta^{\otimes N}) \notag \\
    & \begin{array}{rl}
        = & \min_{h,S} \ \Tr[S]/d_{A_{\cal N}^I} \\[2mm]
        & \text{s.t.} \ \left\{\begin{array}{l}
        \forall\, k =1,\ldots,N, \ {}_{[1-A_k^I]}\Tr_{A_k^O}S = 0, \\[2mm]
        S \ge \Omega_\theta(h),
    \end{array}\right.
    \end{array} \label{eq:J_ICO_SDP}
\end{align}
with $S\in{\cal L}({\cal H}^{A_{\cal N}^{IO}})$, and where the last constraint can be linearized into
\begin{align}
    {\mathbf A}_\text{Schur}(S,2 (\dot{\mathbf C}_\theta^0 - i \mathbf C_\theta^0 h )^*) \ge 0.
\end{align}
The optimal $\tilde{W}^\text{(opt)}\in\Tr_F\textup{\textsf{Gen}}$ (and $W^\text{(opt)} = \ketbra{w}{w}\in\textup{\textsf{Gen}}$, after purification) is recovered by solving Eq.~\eqref{eq:tW_argmax}, translating the constraint $\tilde{W}\in \Tr_F\textup{\textsf{Gen}}$ into (see Eq.~\eqref{eq:ICO})
\begin{align}
    \left\{ \begin{array}{l}
        \forall\, \emptyset \subsetneq {\cal K} \subseteq {\cal N}, \ {}_{ \prod_{k \in {\cal K}}[1-A_k^O]} (\Tr_{A_{\mathcal{N}\setminus\mathcal{K}}^{IO}}\tilde{W}) = 0, \\[1mm]
        \Tr \tilde{W} = d_{A_{\cal N}^O}, \\[1mm]
        \tilde{W} \ge 0.
    \end{array} \right.
\end{align}


\subsection{On the importance of imposing Eq.~\eqref{eq:saddle_derivative}}
\label{app:subsec:saddle_condition}

To wrap up this appendix, let us clarify the need for imposing Eq.~\eqref{eq:saddle_derivative} when reconstructing the optimal process (in step~\ref{app_subsubsec:reconstruct_Wopt} above).

Consider some continuous, real-valued function $f(h,w)$, defined over some compact domains for $h$ and $w$.
Fan's minimax theorem~\cite{fan53} says that the optimal values of the $\min_h \max_w$ and $\max_w \min_h$ optimization problems coincide:
\begin{align}
    \min_h \max_w f(h,w) = \max_w \min_h f(h,w). \label{eq:minmaxf_maxminf}
\end{align}
However, the solutions to both problems are in general not the same: a solution to the $\min_h \max_w$ is not necessarily a solution to the $\max_w \min_h$ problem.

Suppose that $(h^\downarrow,w^\downarrow)$ and $(h^\uparrow,w^\uparrow)$ are solutions to the $\min_h \max_w$ and $\max_w \min_h$ problems, respectively. Then one clearly has
\begin{align}
    f(h^\downarrow,w^\downarrow) \ge f(h^\downarrow,w^\uparrow) \ge f(h^\uparrow,w^\uparrow),
\end{align}
and from Eq.~\eqref{eq:minmaxf_maxminf} it follows that both inequalities are in fact equalities~\cite{liu}, so that $(h^\downarrow,w^\uparrow)$ is a solution to both the $\min_h \max_w$ and $\max_w \min_h$ problems: it corresponds to a saddle point.

\begin{figure}[hbtp]
     \centering
         \includegraphics[width=\columnwidth]{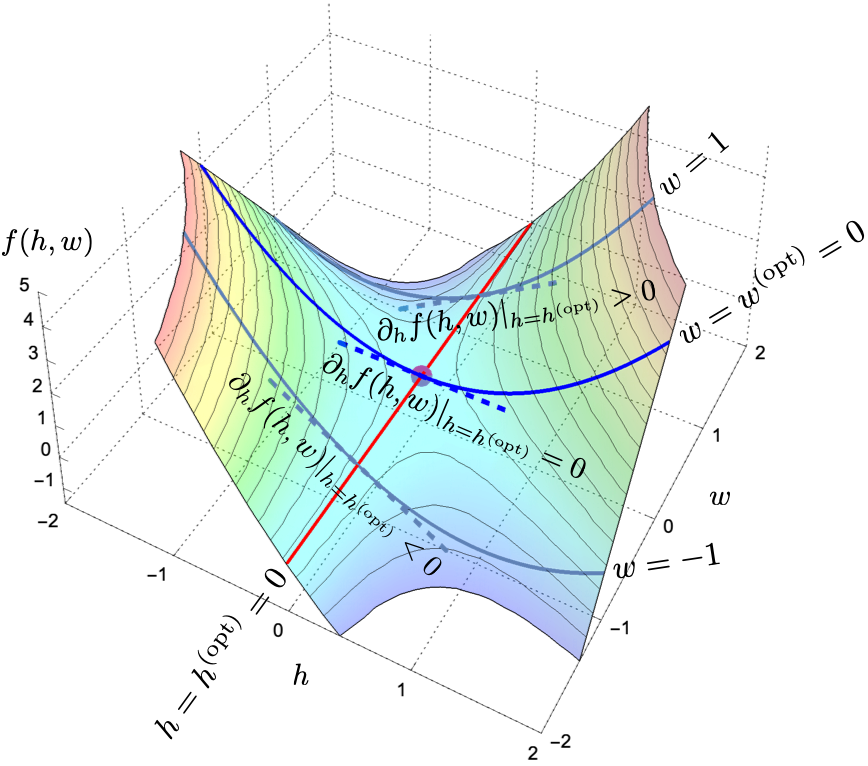}
         \caption{Plot of the function $f(h,w) = h^2+2hw$, as an illustration of a function that is convex in $h$ and concave (even linear) in $w$. $f(h,w)$ has a saddle point for $(h^\text{(opt)},w^\text{(opt)}) = (0,0)$. All points on the red line, for $h = h^\text{(opt)} = 0$, are solutions to the $\min_h \max_w f(h,w)$ problem. However, when fixing $h = h^\text{(opt)} = 0$ the function $f(h^\text{(opt)},w)$ becomes constant in $w$: the saddle point, which is the only one that is also solution to the $\max_w \min_h f(h,w)$ problem, can then only be found after imposing that the derivative of $f(h,w)$ over $h$ must be null, i.e., $\partial_h f(h,w)|_{h=h^\text{(opt)}}= 0$.}
        \label{fig:saddle}
\end{figure}

Hence, once a solution $(h^\downarrow,w^\downarrow)$ of the $\min_h \max_w$ problem is known, we know there exists a solution to the $\max_w \min_h$ problem with $h = h^\downarrow = h^\text{(opt)}$. Simply optimizing $w$ for this fixed value of $h$, however, is not enough, in general, to identify such a solution $(h^\text{(opt)},w^\text{(opt)})$ to the $\max_w \min_h$ problem. Indeed, it could be that several values of $w$, other than the saddle point, give the same value of $f(h^\text{(opt)},w) = f(h^\text{(opt)},w^\text{(opt)})$.
Such a situation is illustrated in Fig.~\ref{fig:saddle}.

To make sure that one indeed finds a solution to the $\max_w \min_h$ problem when optimizing $w$ for the fixed $h = h^\text{(opt)}$, one thus needs to also impose that for the optimal value of $w$, $h^\text{(opt)}$ is indeed a solution to $\min_h f(w,h)$. This can be done by requiring that $\partial_h f(h,w)|_{h=h^\text{(opt)}}= 0$ (which is a defining property for a saddle point). For the function $f$ we consider in this work, this translates into Eq.~\eqref{eq:saddle_derivative}.

\medskip

The situation just depicted for a general (convex-concave) function $f(h,w)$ is indeed commonly encountered in the metrological context of our work. As a simple example, consider for instance the case of $N=1$ qubit depolarizing channel $\mathcal{D}_\theta$ as in Eq.~\eqref{eq:depo_channel}. Since $N=1$, all the classes of processes considered in this work coincide. 
Taking $\{\ket{C_{\theta,j}^0}\}_{j=1}^4$ to be an eigendecomposition of the Choi matrix of $\mathcal{D}_\theta$ (with eigenvectors in the degenerate eigensubspace taken to be independent of $\theta$, except for their normalization) and following the approach described above, we find numerically, for all values of $\theta$, the optimal value of $h$ to be $h^\text{(opt)} = 0$.

Now, for this value one finds that $\Tr_{A^O}\Omega_\theta(h^\text{(opt)}) = \frac{3}{(1-\theta)(1+3\theta)}\id^{A^I}$. Since for all classes, $\tilde{W}$ is of the form $\tilde{W} = \rho^{A^I}\otimes\id^{A^O}$ with $\tr(\rho^{A^I}) = 1$, one then finds that $\Tr(\Omega_\theta(h^\text{(opt)})\,\tilde{W}) = \frac{3}{(1-\theta)(1+3\theta)}$ is independent from $\tilde{W}$. While this indeed gives the correct value for a single qubit depolarizing channel QFI (see Eq.~\eqref{eq:hierarchy_depol}), clearly one could not recover the optimal $\tilde{W}$ (or $\rho^{A^I}$) by directly optimizing $\Tr(\Omega_\theta(h^\text{(opt)})\,\tilde{W})$. Imposing Eq.~\eqref{eq:saddle_derivative} however, enforces $\rho^{A^I}$ to be the maximally mixed state $\frac{\id^{A^I}}{2}$, which indeed (after purification into a maximally entangled 2-qubit state) gives the optimal QFI of the depolarizing channel.%
\footnote{Considering inputting half of a 2-qubit state $\ket{\psi_{\text{in}}} = \alpha\ket{0,0} + \beta\ket{1,1}$ into $\mathcal{D}_\theta$, one finds (e.g., using Eq.~\eqref{eq:QFI_simple}) the QFI of the 2-qubit output state to be $\frac{1+3\theta+8|\alpha|^2|\beta|^2}{(1-\theta^2)(1+3\theta)}$. This indeed clearly depends on the input state, and is optimal for a maximally entangled state.}


\section{Numerical results for the QFI of the composed rotation \& amplitude damping channels $\mathcal{A}_\theta$}
\label{app:hierarchy_AD_data}

\begin{figure}[htbp]
      \centering
          \includegraphics[width=0.47\textwidth]{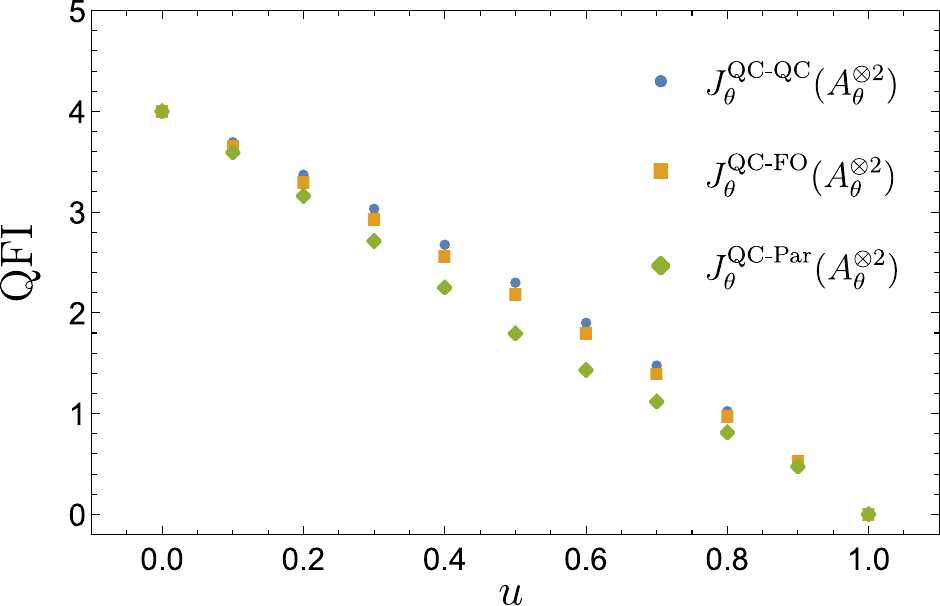}
          \caption{QFI for $N=2$ uses of the composed rotation \& amplitude damping channel $\mathcal{A}_\theta$, as in Table~\ref{tab:AD_N_2}.}
         \label{fig:QFI_AD_2.pdf}
 \end{figure}
\begin{figure}[htbp]
      \centering
          \includegraphics[width=0.47\textwidth]{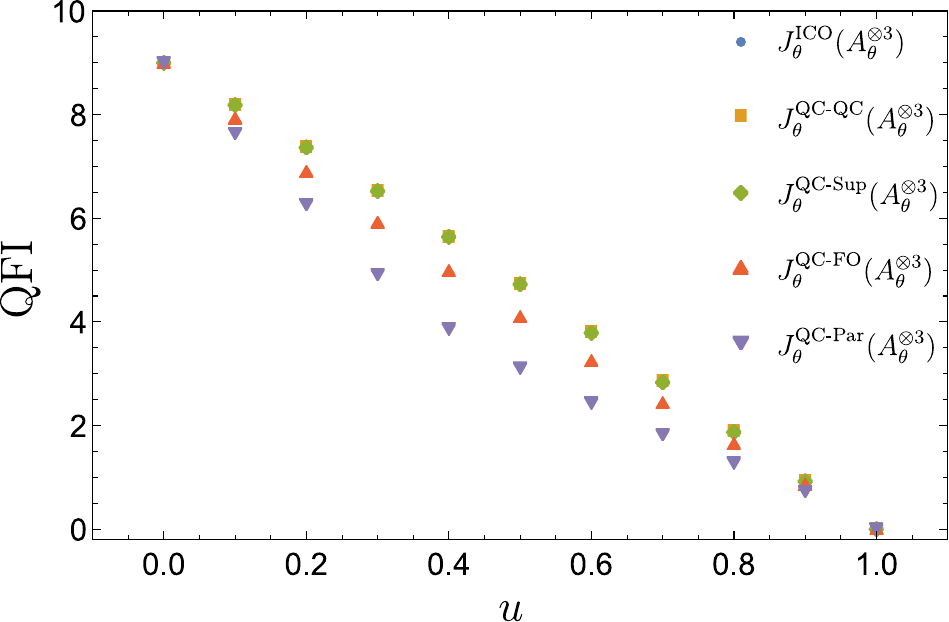}
          \caption{QFI for $N=3$ uses of the composed rotation \& amplitude damping channel $\mathcal{A}_\theta$, as in Table~\ref{tab:AD_N_3}.}
         \label{fig:QFI_AD_3.pdf}
 \end{figure}

We report in Tables~\ref{tab:AD_N_2} and~\ref{tab:AD_N_3}, and in Figures~\ref{fig:QFI_AD_2.pdf} and~\ref{fig:QFI_AD_3.pdf}, the optimal values of the QFI obtained numerically (by solving the SDP) for the different classes, in the case of $N=2$ and $N=3$ queries (respectively) to the composed rotation \& amplitude damping channel $\mathcal{A}_\theta$ defined in Eq.~\eqref{eq:def_Atheta}. As mentioned in the main text, the QFI does not depend on the specific value of $\theta$ for these channels. 
We recover the hierarchies obtained in Eqs.~\eqref{eq:hierarchy_N2_AD} and~\eqref{eq:hierarchy_N3_AD}, respectively.

\setcounter{table}{0}

\begin{table}[h]
  \centering
\begin{tabular}{|c || c | c | c |}
\hline $u$ & $J_\theta^{\textup{\textsf{QC-Par}}}(A_\theta^{\otimes 2})$ & $\!\begin{array}{r}
    \\[-8pt] J_\theta^{\textup{\textsf{QC-FO}}}(A_\theta^{\otimes 2}), \\[1mm]
    J_\theta^{\textup{\textsf{QC-CC}}}(A_\theta^{\otimes 2})
\end{array}\!$ & $\!\begin{array}{r}
    \\[-8pt] J^{\textup{\textsf{QC-CC}}}_{\theta,\textup{purif}}(A_\theta^{\otimes 2}), \, J_\theta^{\textup{\textsf{QC-Sup}}}(A_\theta^{\otimes 2}), \\[1mm]
    J_\theta^{\textup{\textsf{QC-QC}}}(A_\theta^{\otimes 2}), \, J_\theta^{\textup{\textsf{Gen}}}(A_\theta^{\otimes 2})
\end{array}\!$ \\[10pt]
\hline  0 & 4 & 4  & 4 \\
\hline  0.1 & 3.5900 & 3.6477  & 3.6929  \\
\hline  0.2 &  3.1605 &3.2904 &  3.3709  \\
\hline  0.3 & 2.7128
 &2.9273 & 3.0324\\
\hline  0.4 &  2.2500 &2.5574 & 2.6760\\
\hline  0.5 &  1.7946 &2.1793 & 2.2995  \\
\hline  0.6 & 1.4314 &1.7911 & 1.9005 \\
\hline  0.7 &  1.1177 &1.3900 & 1.4764\\
\hline  0.8 & 0.8117 &0.9714 & 1.0240 \\
\hline  0.9 &  0.4734 &0.5260 & 0.5410  \\
\hline  1 & 0 & 0  & 0 \\ 
\hline
\end{tabular}
\caption{QFI for $N=2$ uses of the composed rotation \& amplitude damping channel $\mathcal{A}_\theta$, for the different classes under consideration, for $0 \leq u \leq 1$ (and any fixed value of $\theta$). Notice that here (except for $u=0$ and $u=1$) $J^{\textup{\textsf{QC-CC}}}_{\theta,\textup{purif}}(A_\theta^{\otimes 2})$ is a strict upper bound on $J^{\textup{\textsf{QC-CC}}}_{\theta}(A_\theta^{\otimes 2})$---which, in the case of $N=2$, we know is equal to $J_\theta^{\textup{\textsf{QC-FO}}}(A_\theta^{\otimes 2})$; cf.\ Eq.~\eqref{eq:hierarchy_N2_pur}.}
\label{tab:AD_N_2}
\end{table}

\begin{table}[h]
  \centering
\resizebox{\columnwidth}{!}{\begin{tabular}{|c || c | c | c | c | c |}
\hline $u$ & $J_\theta^{\textup{\textsf{QC-Par}}}(A_\theta^{\!\otimes 3})$ & $J_\theta^{\textup{\textsf{QC-FO}}}(A_\theta^{\!\otimes 3})$ & $\!\!\begin{array}{r}
    \\[-8pt] J^{\textup{\textsf{QC-CC}}}_{\theta,\textup{purif}}(A_\theta^{\!\otimes 3}), \\[1mm]
    J_\theta^{\textup{\textsf{QC-Sup}}}(A_\theta^{\!\otimes 3})
\end{array}\!\!$ &  $J_\theta^{\textup{\textsf{QC-QC}}}(A_\theta^{\!\otimes 3})$ & $J_\theta^{\textup{\textsf{Gen}}}(A_\theta^{\!\otimes 3})$ \\[10pt]
\hline  0 & 9 & 9  & 9  & 9 & 9\\
\hline  0.1 & 7.6366 & 7.9238 &  8.1848 & 8.1986 & 8.2001  \\
\hline  0.2 & 6.2628 & 6.8951 & 7.3643 & 7.3741 & 7.3751  \\
\hline  0.3 & 4.9111 & 5.9140 &  6.5231 & 6.5242 & 6.5242  \\
\hline  0.4 & 3.8614 & 4.9797 & 5.6417 & 5.6461 & 5.6468  \\
\hline  0.5 & 3.1082 & 4.0905 & 4.7247 & 4.7373 & 4.7431  \\
\hline  0.6 & 2.4344 & 3.2427 & 3.7859 & 3.8048 & 3.8150  \\
\hline  0.7 & 1.8217 & 2.4300 & 2.8317 & 2.8564 & 2.8698  \\
\hline  0.8 & 1.2786 & 1.6412 & 1.8710 & 1.8957 & 1.9086  \\
\hline  0.9 & 0.7255 & 0.8553 & 0.9177 & 0.9260 & 0.9299  \\
\hline  1 & 0 & 0  & 0  & 0 & 0\\
\hline
\end{tabular}}
\caption{QFI for $N=3$ uses of the composed rotation \& amplitude damping channel $\mathcal{A}_\theta$, for the different classes under consideration, for $0 \leq u \leq 1$ (and any fixed value of $\theta$). Recall that $J^{\textup{\textsf{QC-CC}}}_{\theta,\textup{purif}}(A_\theta^{\otimes 3})$ is only an upper bound on $J^{\textup{\textsf{QC-CC}}}_{\theta}(A_\theta^{\otimes 3})$, which our approach does not allow us to access directly; cf.\ Eq.~\eqref{eq:hierarchy_N3_dual}.}
\label{tab:AD_N_3}
\end{table}

Recall that our SDPs allow us, for each class, to reconstruct the processes that give the optimal values of the QFI. Except in the simplest cases%
\footnote{For instance, for $N=2$, for the class of parallel strategies \textup{\textsf{QC-Par}} and for $u \lesssim 0.45$, the SDP returns optimal processes of the form $\tilde W = \Tr_F W = (a \ketbra{00}{00} + b \ketbra{11}{11})^{A_1^IA_2^I}\otimes\id^{A_1^OA_2^O}$, for some (nonnegative, $u$-dependent) values of $a,b$. These can be purified in the form $W = \ketbra{w}{w}$ with $\ket{w} = (\sqrt{a}\ket{000}+\sqrt{b}\ket{111})^{A_1^IA_2^IF_I}\otimes|\id\rangle\!\rangle^{A_1^OF_1}\otimes|\id\rangle\!\rangle^{A_2^OF_2}$; the QFI of the output state can then be calculated analytically, and optimized (for some fixed $u$) over the values of $a,b$---which leads to $J^{\textup{\textsf{QC-Par}}}(A_\theta^{\otimes 2}) = \frac{4(1-u)^2}{(1-u/2)^2}$. For $u \gtrsim 0.45$ the optimal processes are found to be of the form $\tilde W = \Tr_F W = (a \ketbra{00}{00} + b \ketbra{11}{11} + c \ketbra{01}{01} + c \ketbra{10}{10})^{A_1^IA_2^I}\otimes\id^{A_1^OA_2^O}$, rather, and expressing the corresponding optimal QFI analytically is already much more involved.}
it is, however, difficult to gain much insight or intuition from the form of the processes thus obtained numerically.


\bibliography{biblio_draft}

\end{document}